\documentclass[a4paper,UKenglish]{lipics-v2019}

\usepackage{amssymb}
\usepackage{afterpage}
\usepackage{changepage}
\usepackage{etoolbox}
\usepackage{macros}
\usepackage{stmaryrd}
\usepackage{tikz}
\undef\second
\usepackage{turnstile}
\usepackage{xspace}

\usepackage{prftree}
\usepackage{pbox}

\hideLIPIcs

\usetikzlibrary{arrows}

\newcommand\logicname{Wasm Logic\xspace}

\newtheoremstyle{break}
  {\topsep}{\topsep}%
  {\itshape}{}%
  {\bfseries\sffamily}{}%
  {\newline}{}%
\theoremstyle{break}
\newtheorem{theorem1}{Theorem}

\newcommand\void[1]{}

\newcommand\x[1]{\ensuremath{{\mathit{#1}}}\xspace}
\newcommand\f[1]{\ensuremath{\mathop{\operatorname{\mathrm{#1\null}}}\nolimits}\xspace}

\newcommand\K[1]{\ensuremath{\textsf{\sf#1}}}
\newcommand\KK[1]{\ensuremath{\K{\textbf{#1}}}}

\newcommand\Kto{\ensuremath{\rightarrow}\xspace}

\definecolor{hilite}{rgb}{0.8,0,0}

\definecolor{hiliteb}{rgb}{0,0,0.9}

\definecolor{shade}{rgb}{0.6,0.6,0.6}
\newcommand{\shade}[1]{\color{shade}#1\color{black}}

\definecolor{hishade}{rgb}{1,0.4,0.4}

\newcommand{\Witype}{\ensuremath{\K{i32}}}
\newcommand{\Wiconst}[1]{\ensuremath{(\KK{i32.const}~{#1})}\xspace}

\newcommand{\Wiadd}{\ensuremath{(\KK{i32.add})}\xspace}
\newcommand{\Wisub}{\ensuremath{(\KK{i32.sub})}\xspace}
\newcommand{\Wimul}{\ensuremath{(\KK{i32.mul})}\xspace}

\newcommand{\Wieq}{\ensuremath{(\KK{i32.eq})}\xspace}
\newcommand{\Wine}{\ensuremath{(\KK{i32.ne})}\xspace}
\newcommand{\Wigt}{\ensuremath{(\KK{i32.gt})}\xspace}
\newcommand{\Wilt}{\ensuremath{(\KK{i32.lt})}\xspace}

\newcommand{\Wile}{\ensuremath{(\KK{i32.le})}\xspace}
\newcommand{\Wbr}[1]{\ensuremath{(\KK{br}~{#1})}\xspace}
\newcommand{\Wbrif}[1]{\ensuremath{(\KK{br\textnormal{\_}if}~{#1})}\xspace}

\newcommand{\Wgetlocal}[1]{\ensuremath{(\KK{get\_local}~{#1})}\xspace}
\newcommand{\Wsetlocal}[1]{\ensuremath{(\KK{set\_local}~{#1})}\xspace}
\newcommand{\Wteelocal}[1]{\ensuremath{(\KK{tee\_local}~{#1})}\xspace}

\newcommand{\Wsetglobal}[1]{\ensuremath{(\KK{set\_global}~{#1})}\xspace}

\newcommand{\Wiload}{\ensuremath{(\KK{i32.load})}\xspace}
\newcommand{\Wistore}{\ensuremath{(\KK{i32.store})}\xspace}

\newcommand{\Wiloadoff}[1]{\ensuremath{(\KK{i32.load}~\K{offset=}{#1})}\xspace}
\newcommand{\Wistoreoff}[1]{\ensuremath{(\KK{i32.store}~\K{offset=}{#1})}\xspace}

\newcommand{\Wgrowmemory}{\ensuremath{(\KK{mem.grow})}\xspace}
\newcommand{\Wsizememory}{\ensuremath{(\KK{mem.size})}\xspace}

\newcommand{\Wdrop}{\ensuremath{(\KK{drop})}}

\newcommand{\Wfunc}[2]{\ensuremath{(\KK{func}~\f{#1}~{#2}}\xspace}
\newcommand{\Wblock}[1]{\ensuremath{(\KK{block}~{#1}}\xspace}
\newcommand{\Wloop}[1]{\ensuremath{(\KK{loop}~{#1}}\xspace}
\newcommand{\Wif}{\ensuremath{(\KK{if}}\xspace}

\newcommand{\Welse}{\ensuremath{\KK{else}}\xspace}

\newcommand{\Wend}{\ensuremath{\KK{end}})\xspace}

\newcommand{\Wcall}[1]{\ensuremath{(\KK{call}} #1)\xspace}

\newcommand{\Wsize}[1]{\ensuremath{\KK{size}(#1)}\xspace}

\newenvironment{Wfunce}[3][]
{
\Wfunc{#2}{#3}
{#1}
\begin{adjustwidth}{4\fontdimen2\font}{}
}
{
\end{adjustwidth}
\Wend
}

\newenvironment{Wblocke}[2][]
{
\Wblock{#2}
{\color{purple}#1}
\begin{adjustwidth}{4\fontdimen2\font}{}
}
{
\end{adjustwidth}
\Wend
}
\newenvironment{Wloope}[2][]
{
\Wloop{#2}
{\color{purple}#1}
\begin{adjustwidth}{4\fontdimen2\font}{}
}
{
\end{adjustwidth}
\Wend
}
\newenvironment{Wife}[2][]
{
\Wif{#2}
{\color{purple} #1}
\begin{adjustwidth}{4\fontdimen2\font}{}
}
{
\end{adjustwidth}
\Wend
}

\newcommand{\ba}[3]{\pred{BA}(#1, #2, #3)}
\newcommand{\oba}[3]{\pred{OBA}(#1, #2, #3)}
\newcommand{\oband}[3]{\pred{OBA_{nd}}(#1, #2, #3)}
\newcommand{\obaseg}[2]{\pred{Aseg}(#1, #2)}
\newcommand{\ordered}[1]{\pred{Ordered}(#1)}
\newcommand{\listToSet}[1]{\pred{ToSet}(#1)}
\newcommand{\setToList}[1]{\pred{ToList}(#1)}
\newcommand{\sublist}[3]{\pred{SubList}(#1, #2, #3)}

\newcommand{\card}[1]{\pred{card}(#1)}

\newcommand{\page}[1]{\pred{Page}(#1)}
\newcommand{\meta}[4]{\pred{Meta}(#1, #2, #3, #4)}
\newcommand{\free}[1]{\pred{Free}(#1)}
\newcommand{\node}[4]{\pred{Node}(#1, #2, #3, #4)}
\newcommand{\nkeys}[2]{\pred{Keys}(#1, #2)}
\newcommand{\nptrs}[2]{\pred{Ptrs}(#1, #2)}

\newcommand{\btree}[2]{\pred{BTree}(#1, #2)}
\newcommand{\btreerec}[4]{\pred{BTreeRec}^{t, r, \mu}(#1, #2, #3, #4)}

\newcommand\stackr[2]{\mbox{$[#2]$}}
\newcommand\stacks{,}

\newcommand\interp[1]{\llbracket{#1}\rrbracket}

\title{A Program Logic for First-Order Encapsulated WebAssembly}

\author{Conrad Watt}{University of Cambridge, UK}{conrad.watt@cl.cam.ac.uk}{}{}

\author{Petar Maksimovi\'c}{Imperial College London, UK; Mathematical Institute SASA, Serbia}{p.maksimovic@imperial.ac.uk}{}{}

\author{Neelakantan R. Krishnaswami}{University of Cambridge, UK}{nk480@cl.cam.ac.uk}{}{}

\author{Philippa Gardner}{Imperial College London, UK}{p.gardner@imperial.ac.uk}{}{}

\authorrunning{C. Watt, P. Maksimovi\'c, N. R. Krishnaswami, P. Gardner}

\Copyright{Conrad Watt, Petar Maksimovi\'c, Neelakantan R. Krishnaswami, Philippa Gardner}



\acknowledgements{
We would like to thank the reviewers, whose comments were valuable in improving the paper.
All authors were supported by the EPSRC Programme Grant `REMS: Rigorous Engineering for Mainstream Systems' (EP/K008528/1).
In addition: Watt was supported by an EPSRC DTP award (EP/N509620/1);
Maksimovi\'{c} was supported by the Serbian Ministry of Education and Science through the Mathematical Institute SASA, projects ON174026 and III44006;
Krishnaswami was supported by the EPSRC Programme Grant `Semantic Foundations for Interactive Programs' (EP/N02706X/2);
and Gardner was supported by the EPSRC Fellowship `VeTSpec: Verified Trustworthy Software Specification' (EP/R034567/1).
}

\EventEditors{Alastair F. Donaldson}
\EventNoEds{1}
\EventLongTitle{33rd European Conference on Object-Oriented Programming (ECOOP 2019)}
\EventShortTitle{ECOOP 2019}
\EventAcronym{ECOOP}
\EventYear{2019}
\EventDate{July 15--19, 2019}
\EventLocation{London, United Kingdom}
\EventLogo{}
\SeriesVolume{134}
\ArticleNo{10}
\nolinenumbers 

\begin{document}
\maketitle

\begin{abstract}
We introduce \logicname, a sound program logic for first-order, encapsulated WebAssembly. 
%
%
We design a novel assertion syntax, tailored to WebAssembly's stack-based semantics and the strong guarantees given by WebAssembly's type system, and show how to adapt the standard separation logic triple and proof rules in a principled way to capture WebAssembly's uncommon structured control flow.
Using \logicname, we specify and verify a simple WebAssembly B-tree library, giving abstract specifications independent of the underlying implementation.
We mechanise \logicname and its soundness proof in full in Isabelle/HOL. As part of the soundness proof, we formalise and fully mechanise a novel, big-step semantics of WebAssembly, which we prove equivalent, up to transitive closure, to the original WebAssembly small-step semantics. \logicname is the first program logic for WebAssembly, and represents a first step towards the creation of static analysis tools for WebAssembly.
 \end{abstract}


\section{Introduction}

WebAssembly~\cite{Haas:2017:BWU:3062341.3062363} is a stack-based, statically typed bytecode language. It is the first new language to be natively supported on the Web in nearly 25 years, following JavaScript (JS).
It was created to act as the safe, fast, portable low-level code of the Web, in answer to the growing sophisticated, computationally intensive demands of the Internet of today, such as 3D visualisation, audio/video processing, and games.
For years, developers wishing to execute calculation-heavy programs written in C/C++ on the Web had to compile them to asm.js~\cite{asmjs_spec}, a subset of JS.
In time, such code has become widespread~\cite{Zakai:2011:ELC:2048147.2048224, asmjsunreal,asmjsunity}, but the fundamental limitations of JS as a compilation target have become too detrimental to ignore.
WebAssembly is designed from the ground up to be an efficient, Web-compatible compilation target, obsoleting asm.js and other similar endeavours, such as Native Client~\cite{yee2009native}.
All major browser vendors, including Google, Microsoft, Apple, and Mozilla, have pledged to support WebAssembly, and the past two years have seen a flurry of implementation activity~\cite{wasm_roadmap}.

These facts alone would be enough to motivate that WebAssembly will be an important technology, and a worthy target for formal methods. The designers of WebAssembly have anticipated this, and have specified WebAssembly using a precise formal small-step semantics, combined with a sound type system. Moreover, WebAssembly's semantics, type system, and soundness have already been fully mechanised~\cite{Watt:2018:MVW:3176245.3167082}, and the WebAssembly Working Group requires any further additions to WebAssembly to be formally specified. 

The main use case for WebAssembly is to inter-operate with JS in creating content for the Web. More precisely, WebAssembly functions can be grouped into \textit{modules}, which provide interfaces through which users can call WebAssembly code, and self-contained (encapsulated) modules can be used as drop-in replacements for their existing JS counterparts and already constitute a major design pattern in WebAssembly. 
We believe that having a formalism for describing and reasoning about WebAssembly modules and their interfaces is essential, in line with WebAssembly's emphasis on formal methods.
Thus far, very little work has been done on static analysis for WebAssembly (cf.~\S\ref{sec:relwork}).

We present \logicname, a sound program logic for reasoning about first-order, encapsulated WebAssembly modules, such as data structure libraries. 
Enabled by the strong~guarantees of WebAssembly's type system, we design a novel assertion syntax, tailored~to~WebAssembly's stack-based semantics. We further adapt the standard separation logic triple and proof rules in a principled way to capture WebAssembly's uncommon structured control flow.

Having a program logic for WebAssembly is valuable for several reasons.
First, as WebAssembly programs are distributed without their originating source code, any client-side verification would have to rely on a WebAssembly-level logic.
Similarly, verification techniques such as proof-transforming compilation~\cite{DBLP:conf/tools/NordioMM08,Bannwart:2005:PLB:1705549.1706041, Muller:2007:PCP:1292316.1292321,Saabas:2006:CNS:1706641.1706946} rely on the existence of a program logic for the target language.
Finally, some fundamental data structure libraries are expected to be implemented directly in WebAssembly for efficiency reasons.
For example, the structure of B-trees strongly aligns with the way in which WebAssembly memory is managed (cf.~\S\ref{subsec:btrees}). 

To demonstrate the usability of \logicname, we implement, specify, and verify a simple WebAssembly B-tree library.  In doing so, we discuss how the new and adapted \logicname proof rules can be used in practice. The specifications that we obtain are abstract, in that they do not reveal any details about the underlying implementation.

We mechanise \logicname and its soundness proof in full in Isabelle/HOL, 
building on a previous WebAssembly mechanisation of Watt~\cite{Watt:2018:MVW:3176245.3167082}.
We prove \logicname sound against a novel, big-step semantics of WebAssembly, and also mechanise a proof of equivalence between the transitive closure of the original small-step semantics and our big-step semantics. Our mechanisation totals \textasciitilde10,400 lines of non-comment, non-whitespace Isabelle code, not counting code inherited from the existing mechanisation.


\section{A Brief Overview of WebAssembly}

We give the syntax and an informal description of the semantics of WebAssembly.
A precise account of its semantics is given through our program logic in \S\ref{sec:proglog} and also through our big-step semantics, introduced in \S\ref{sec:soundness} and presented in full in~\cite{wasmlogic:techrep}. 

\subsection{WebAssembly Syntax}
\label{sec:wsyn}

WebAssembly has a human-readable \textit{text format} based on s-expressions, which we use throughout.
The abstract syntax of WebAssembly programs~\cite{Haas:2017:BWU:3062341.3062363}, is given in full in Figure~\ref{fig:syntax}. 
As we consider first-order, encapsulated modules, we grey out the remaining, non-relevant syntax.
We describe the semantics of the instructions informally in \S\ref{sec:wasminstr}, and additional syntax as it arises in the paper. A full description of WebAssembly can be found in~\cite{Haas:2017:BWU:3062341.3062363}.


\begin{figure}[!h]
\vspace*{-0.5cm}
\fontsize{8pt}{8pt}
$$
\begin{array}{@{}l@{}l@{}}
\begin{array}{@{}l@{}}
\begin{array}{@{}l@{~}r@{~}c@{~}l@{}}
\text{(constants)} & k &::=& \dots \\
\text{(immediates)} & \x{im} &::=& i,a,o \in \x{nat} \\
\text{(packed types)} & \x{pt} &::=&
  \K{i8} ~|~
  \K{i16} ~|~
  \K{i32} \\
\text{(value types)} & t &::=&
  \K{i32} ~|~
  \K{i64}  ~|~
  \K{f32} ~|~
  \K{f64} \\
\text{(function types)} & \x{ft} &::=&
  t^\ast \Kto t^\ast \\
\text{(global types)} & \x{gt} &::=&
  \K{mut}^?~t \\
\end{array}
\\~\vspace{-2ex}\\
\begin{array}{@{}l@{~}r@{~}c@{~}l@{}}
\multicolumn{2}{@{}r}{\x{unop}_{\K{i}N}} & ::= &
  \KK{clz} ~|~
  \KK{ctz} ~|~
  \KK{popcnt} \\
\multicolumn{2}{@{}r}{\x{unop}_{\K{f}N}} & ::= &
  \KK{neg} ~|~
  \KK{abs} ~|~
  \KK{ceil} ~|~
  \KK{floor} ~|~ 
  \KK{trunc} ~|~
  \KK{nearest} ~|~
  \KK{sqrt} \\
\multicolumn{2}{@{}r}{\x{binop}_{\K{i}N}} & ::= &
  \KK{add} ~|~
  \KK{sub} ~|~
  \KK{mul} ~|~
  \KK{div\textunderscore}\x{sx} ~|~ 
  \KK{rem\textunderscore}\x{sx} ~|~
  \KK{and} ~|~ \\&&&
  \KK{or} ~|~
  \KK{xor} ~|~ 
  \KK{shl} ~|~
  \KK{shr\textunderscore}\x{sx} ~|~
  \KK{rotl} ~|~
  \KK{rotr} \\
\multicolumn{2}{@{}r}{\x{binop}_{\K{f}N}} & ::= &
  \KK{add} ~|~
  \KK{sub} ~|~
  \KK{mul} ~|~
  \KK{div} ~|~ \\&&&
  \KK{min} ~|~
  \KK{max} ~|~
  \KK{copysign} \\
\multicolumn{2}{@{}r}{\x{testop}_{\K{i}N}} & ::= &
  \KK{eqz} \\
\multicolumn{2}{@{}r}{\x{relop}_{\K{i}N}} & ::= &
  \KK{eq} ~|~
  \KK{ne} ~|~
  \KK{lt\textunderscore}\x{sx} ~|~
  \KK{gt\textunderscore}\x{sx} ~|~ \\&&&
  \KK{le\textunderscore}\x{sx} ~|~
  \KK{ge\textunderscore}\x{sx} \\
\multicolumn{2}{@{}r}{\x{relop}_{\K{f}N}} & ::= &
  \KK{eq} ~|~
  \KK{ne} ~|~
  \KK{lt} ~|~
  \KK{gt} ~|~
  \KK{le} ~|~
  \KK{ge} \\
\multicolumn{2}{@{}r}{\x{cvtop}} & ::= &
  \KK{convert} ~|~
  \KK{reinterpret} \\
\multicolumn{2}{@{}r}{\x{sx}} & ::= &
  \KK{s} ~|~
  \KK{u} \\
\end{array}
\end{array}
\hspace{-15.5ex}
\begin{array}{@{}l@{}}
\begin{array}{@{}l@{~~}r@{~}c@{~}l@{}}
\text{(instructions)} & e &::=&
  t\KK{.const}~k ~|~
  \KK{drop} ~|~
  \KK{nop} ~|~
  \KK{select}~|~\KK{unreachable}~| \\&&&
  t\KK{.}\x{unop}_t ~|~
  t\KK{.}\x{binop}_t ~|~ 
  t\KK{.}\x{testop}_t ~|~
  t\KK{.}\x{relop}_t ~|~ \\&&&
  t\KK{.}\x{cvtop}_{t\KK{\textunderscore}\x{sx}^?} ~|~ 
  \KK{get\textunderscore{}local}~\x{im} ~|~
  \KK{set\textunderscore{}local}~\x{im} ~|~ \\&&&
  \KK{tee\textunderscore{}local}~\x{im} ~|~
  \KK{get\textunderscore{}global}~\x{im} ~|~ \\&&&
  \KK{set\textunderscore{}global}~\x{im} ~|~ 
  t\KK{.store}~\x{pt}^?~a~o ~|~\\&&&
  t\KK{.load}~(\x{pt}\KK{\textunderscore}\x{sx})^?~a~o ~|~ 
  \KK{mem.size} ~|~
  \KK{mem.grow} ~|~ \\&&&
  \KK{block}~\x{ft}~e^\ast~\KK{end} ~|~
  \KK{loop}~\x{ft}~e^\ast~\KK{end} ~|~ \\&&&
  \KK{if}~\x{ft}~e^\ast~\KK{else}~e^\ast~\KK{end} ~|~ \\&&&
  \KK{br}~\x{im} ~|~
  \KK{br\textunderscore{}if}~\x{im} ~|~
  \KK{br\textunderscore{}table}~\x{im}^+ ~|~ \\&&&
  \KK{return} ~|~
  \KK{call}~\x{im} ~|~
  \shade{\KK{call\textunderscore{}indirect}~\x{ft} }
\end{array}
\\~\\[-1ex]\hspace{0ex}
\begin{array}{@{}l@{~}r@{~}c@{~}l@{}}
\text{(functions)} & \x{func} &::=&
  \shade{\x{ex}^\ast}~\KK{func}~\x{ft}~\KK{local}~\x{t}^\ast~e^\ast ~|~ 
  \shade{\x{ex}^\ast~\KK{func}~\x{ft}~\x{imp}} \\
\text{(globals)} & \x{glob} &::=&
  \shade{\x{ex}^\ast}~\KK{global}~\x{gt}~e^\ast ~|~
  \shade{\x{ex}^\ast~\KK{global}~\x{gt}~\x{imp}} \\
\text{\shade{(tables)}} & \x{\shade{tab}} & \shade{::=} &
  \shade{\x{ex}^\ast}~\KK{\shade{table}}~\x{\shade{n}}~\x{\shade{im}}^\ast ~|~
  \shade{\x{ex}^\ast~\KK{table}~\x{n}~\x{imp}} \\
\text{(memories)} & \x{mem} &::=&
  \shade{\x{ex}^\ast}~\KK{memory}~\x{n}~|~
  \shade{\x{ex}^\ast~\KK{memory}~\x{n}~\x{imp}} \\
\shade{\text{(imports)}} & \shade{\x{imp}} &\shade{::=}&
  \shade{\KK{import}~\text{``\x{name}''}~\text{``\x{name}''}} \\
\shade{\text{(exports)}} & \shade{\x{ex}} & \shade{::=}&
  \shade{\KK{export}~\text{``\x{name}''}} \\
\text{(modules)} & \x{mod} &::=&
  \KK{module}~\x{func}^\ast~\x{glob}^\ast~\shade{\x{tab}^?}~\x{mem}^? \\ 
\end{array}
\end{array}
\end{array}
$$

\vspace*{-0.1cm}
{\small {\bfseries Note}: we denote lists with a $*$ superscript: for example, $t^\ast$ denotes a list of types.}

\caption{WebAssembly Abstract Syntax of~\cite{Haas:2017:BWU:3062341.3062363}, with aspects not relevant to this work greyed~out.}
\label{fig:syntax}
\end{figure}

\subsection{The WebAssembly Memory Model}
\label{sec:wasmmm}

\subparagraph{Values} WebAssembly values, $v$, may have one of four \textit{value types}, representing 32- and 64-bit IEEE-754 integers and floating-point numbers: \K{i32}, \K{i64}, \K{f32}, or \K{f64}. We denote values using their type: for example, a 32-bit representation of the integer 42 is denoted $42_{\K{i32}}$. If the type of a value is not given, it is assumed to be {\K{i32}} by default.

\subparagraph{Local and Global Variables} 
WebAssembly programs have access to statically declared variables, which may be \textit{local} or \textit{global}. 
Local variables are declared per-function. They live in local variable stores, which exist only in the body of their declaring function. They include function arguments, followed by a number of ``scratch'' local variables initialised to zero when the function is called.
Global variables are declared by the enclosing module. They live in a global variable store, are initialised to zero at the beginning of the execution, and are accessible by all of the functions of the module.

In contrast to most standard programming languages, WebAssembly variables cannot be referenced by name.
Instead, both the global and local variable stores are designed as mappings from natural numbers to WebAssembly values, and variables are referenced by their index in the corresponding variable store, as shown in \S\ref{sec:wasminstr}.

\subparagraph{Stack}
WebAssembly computation is based on a stack machine: all instructions pop their arguments from and push their results onto a stack of WebAssembly values. 
By convention, stack concatenation is implicit and the top of the stack is written on the right-hand side: for example, a stack with a 32-bit 0 at its top followed by $m$ WebAssembly values would be denoted as~$v^m~0$. Note that the type system of WebAssembly allows us to statically know both the number of elements on the stack and their types at every point of program~execution.

\subparagraph{Memory}
WebAssembly has a linear memory model. A WebAssembly memory is an array of bytes, indexed by \K{i32} values, which are interpreted as offsets.
Memory is allocated in units of \textit{pages}, and each page is exactly 64k bytes in size.

\subsection{WebAssembly Instructions}
\label{sec:wasminstr}

\begin{figure}[!t]
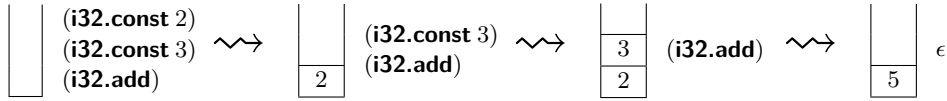

\centering
\begin{tabular}{lclclcl}
\begin{tabular}{|c|}
\\
\\
\\
\hline
\end{tabular}
\hspace{0.1em}
\begin{minipage}{0.11\textwidth}
\small
\Wiconst 2\\
\Wiconst 3\\
\Wiadd
\end{minipage}
&
\huge
$\leadsto$
&
\begin{tabular}{|c|}
\\

\\
\hline
2 \\
\hline
\end{tabular}
\hspace{0.1em}
\begin{minipage}{0.11\textwidth}
\small
\Wiconst 3\\
\Wiadd
\end{minipage}
&
\huge
$\leadsto$
&
\begin{tabular}{|c|}
\\
\hline
3\\
\hline
2\\
\hline
\end{tabular}
\hspace{0.1em}
\begin{minipage}{0.08\textwidth}
\small
\Wiadd
\end{minipage}
&
\huge
$\leadsto$
&
\begin{tabular}{|c|}
\\
\\
\hline
5\\
\hline
\end{tabular}
\hspace{0.1em}
\begin{minipage}{0.0\textwidth}
$\epsilon$
\end{minipage}
\end{tabular}
\caption{Addition in WebAssembly.}
\label{fig:add}
\end{figure}

WebAssembly has a wide array of instructions, which we divide into: basic instructions, variable management instructions, memory management instructions, function-related instructions, and control flow instructions, all of which we discuss below. Every instruction consumes its arguments from the stack, carries out its operation, and pushes any resulting value back onto the stack. Moreover, every instruction is typed, with its type describing the types of its arguments and result. We illustrate how this works in Figure~\ref{fig:add}, which describes WebAssembly addition of two 32-bit integers starting from an empty stack. In particular, the \KK{i32.const} command, whose type is $[] \rightarrow [\K{i32}]$, does not require any arguments and puts the given value on the stack, whereas the \KK{i32.add} instruction, whose type is $[\K{i32},\K{i32}] \rightarrow [\K{i32}]$, takes two arguments from the stack and returns their sum.

WebAssembly gives two official, equivalent, semantics: a semi-formal prose semantics and an entirely formal small-step semantics~\cite{wasm_spec}. In this paper, we introduce an additional, equivalent, big-step semantics as part of the soundness proof of our logic.
Most of our diagrams and explanatory text throughout the paper follow the style of the prose semantics, as its treatment of the value stack is most useful in explaining the behaviour of the logic.
%
%
We denote prose-style execution steps using $\leadsto$, and introduce the other semantics as necessary.


\subparagraph{Basic Instructions}
WebAssembly values can be declared using the $t$.\KK{const} command, typed $[] \rightarrow [t]$, in the style of $\Wiconst{2}$ of Figure~\ref{fig:add}. The (\KK{drop}) command, typed $[t] \rightarrow []$, pops and discards the top stack item, while (\KK{nop}), typed $[] \rightarrow []$ has no effect. The (\KK{select})  instruction, typed $[t, t, \Witype] \rightarrow []$, takes three values from the stack, $v_1$, $v_2$, and $c$. If $c$ is non-zero, $v_1$ is pushed back onto the stack, and $v_2$ otherwise. The (\KK{unreachable}) instruction, typed $[] \rightarrow t^\ast$, causes the program to halt with a runtime error, which is represented in WebAssembly by a special \KK{Trap} execution result~(cf.~\S\ref{sec:abrupt_completions}).

WebAssembly also provides a variety of (type-annotated) unary and binary arithmetic operations (Figure~\ref{fig:syntax}, \textit{unop} and \textit{binop}, respectively), unary and binary logical operations (Figure~\ref{fig:syntax}, \textit{testop} and \textit{relop}, respectively), and casting operations (Figure~\ref{fig:syntax}, \textit{cvtop}). Some of these operations can cause a \KK{Trap}: for example, if we attempt division by zero or try to convert a floating-point number to an integer when the result is not representable. Their meaning is detailed in~\cite{Haas:2017:BWU:3062341.3062363}, and we address them in this paper by need.

%

\subparagraph{Variable Management Instructions}
Local and global variables can be read from and written to using the appropriate \KK{get} and \KK{set} instructions, and all variable accesses are performed using static indexes. For example, $\Wgetlocal i$, typed $[] \rightarrow [t]$ (where $t$ is the statically known type of the $i$-th local variable), will push the value of the $i$-th declared local variable of the current function onto the stack, and $\Wsetglobal i$, typed $[t] \rightarrow []$, will set the value of the $i$-th declared global variable to the value at the top of the stack, which is consumed in the process. It is also possible to set a local variable without consuming this value from the stack by using the \KK{tee\_local} instruction, typed $[t] \rightarrow [t]$.

\subparagraph{Memory Management Instructions} 
Stack values may be serialised and copied into the appropriate number of bytes in memory through the type-annotated \KK{store} instruction.
The $(t.\KK{store})$ instruction, typed $[\Witype, t] \rightarrow []$, interprets its $\Witype$ argument as an index into the memory, while the second is serialised into the appropriate number of bytes to be stored sequentially, starting from the indexed memory location.

Conversely, the type-annotated \KK{load} instruction reads bytes from the memory and produces the appropriate stack value.
$(t.\KK{load})$, typed $[\Witype] \rightarrow [t]$, will consume a single \K{i32} value (the address), and then read the appropriate number of bytes starting from that address, leaving the corresponding value of type $t$ on the top of the stack.
WebAssembly specifies that every value can be serialised, and every byte sequence of the appropriate length can be interpreted as a value; there are no trap representations for values.

The size of the memory can be inspected by executing the (\KK{mem.size}) instruction, typed $[] \rightarrow [\Witype]$, which returns an 32-bit integer denoting the current memory size in pages.
The WebAssembly memory may also be grown by executing the (\KK{mem.grow}) instruction, typed $[\Witype] \rightarrow [\Witype]$, which takes a single \K{i32} value from the top of the stack and attempts to grow the memory by that many pages, returning the previous size of the memory, in pages, as a 32-bit integer if successful.
(\KK{mem.grow}) is always allowed to fail non-deterministically, to represent some memory limitation of the host environment.
In this case, the memory is not altered, and the value ${-1}_{\K{i32}}$ is returned.

\subparagraph{Control Flow Instructions}
Most WebAssembly features have many similarities to other bytecodes, such as that of the Java Virtual Machine~\cite{jvm}.
WebAssembly's approach to control flow, however, is uncommon.
WebAssembly does not allow unstructured control flow in the style of a \KK{goto} instruction.
Instead, it has three control constructs that implement structured control flow: $(\KK{block}~ft~e^\ast~\KK{end})$, $(\KK{loop}~ft~e^\ast~\KK{end})$, and $(\KK{if}~ft~e^\ast~\KK{else}~e^\ast~\KK{end})$.
%
%
Each of these control constructs is annotated with a function type $ft$ of the form $t^m \rightarrow t^n$, meaning that its body, $e^\ast$, requires $m$ elements from the stack and places back $n$ elements onto the stack on exit.
The semantics guarantees that this type precisely describes the effect the construct will have on the stack after it/its body terminates.
For example, a $(\KK{loop}~(t^m \rightarrow t^n)~e^\ast~\KK{end})$, no matter the behaviour of its body, will always leave precisely $n$ additional values on the stack upon termination.
Control constructs may be nested within each other in the intuitive way.
The execution of a control construct consists of executing its body to termination.

\begin{figure}[!t]
\centering
\begin{tikzpicture}[x=\baselineskip,y=\baselineskip]
\node[anchor=west] (1) at (0,2) {\Wloop{\x{tf}}};
\node[anchor=west] (2) at (1,1) {\Wif{~$\x{tf}'$}~\Wbr{0}\Welse\Wbr{1}\Wend};
\node[anchor=west] (3) at (0,0) {\Wend};

\path [->] (2.200) edge[bend left = -40, line width=0.75mm] (2.350);
\path [->] (2.25) edge[bend left = -40, line width=0.75mm] (1.140);

\end{tikzpicture}
\caption{Example of WebAssembly control flow. Executing \Wbr{0} jumps to the end of the \KK{if}, while \Wbr{1} jumps to the \textit{start} of the \KK{loop}.}
\label{fig:loopbr}
\end{figure}
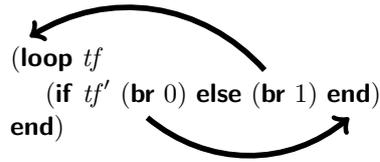

Within the body of a control construct, a break instruction, \Wbr i, may be executed. As control constructs can be nested, \KK{br} is parameterised by a static index $i$, indicating the control construct that it targets (indexing inner to outer).
The behaviour of \KK{br} depends on the type of its target.
When targeting a \KK{block} or an \KK{if}, \KK{br} acts as a ``break'' statement of a high-level language, which transfers control to the matching \KK{end} opcode, jumping out of all intervening constructs.
When targeting a \KK{loop}, the break instruction acts like a ``continue'' statement, transferring control back to the beginning of the loop.
If the body of a \KK{loop} terminates without executing a~\KK{br}, the loop terminates with the result of the body.
The \KK{br} instruction is, therefore, required for loop iteration.
We illustrate this in Figure~\ref{fig:loopbr}. The first break, \Wbr 0, targets its enclosing \KK{if} instruction, meaning that control should be transferred to the end of that \KK{if} instruction. The second break, \Wbr 1, targets the outer \KK{loop} instruction, meaning that control should be transferred to the beginning of that~loop. 

WebAssembly also has two instructions for conditional breaking: \KK{br\_if} and \KK{br\_table}. The $(\KK{br\_if}~i)$ instruction takes one \Witype~value off the stack and, if this value is not equal to zero, behaves as $(\KK{br}~i)$, and as (\KK{nop}) otherwise.
On the other hand, the ($\KK{br\_table}~i_0 \ldots i_n~i$) instruction acts like a switch statement. It takes one \Witype~value $v$ off the stack and then: if $0 \leq v \leq n$, it behaves as $(\KK{br}~i_v)$; otherwise, it behaves as $(\KK{br}~i)$.


\subparagraph{Function-related Instructions}
WebAssembly supports two types of functions. 
First, the host environment wil supply \textit{import} functions for use by the WebAssembly module.
These functions may be JavaScript \textit{host} functions or may come from other WebAssembly modules.
Second, the module itself will define its own native WebAssembly functions.

Functions are called using the $(\KK{call}~i)$ instruction, which executes the $i$-th function, indexing imports first, followed by module-native functions in order of declaration. As WebAssembly functions are declared with a precise type annotation, $(\KK{call}~i)$ also takes the type of the $i$-th function. WebAssembly also provides a mechanism for dynamic dispatch through the \KK{call\_indirect} instruction. 

Our core logic does not support imported functions, as well as the \KK{call\_indirect} dynamic dispatch, as all of these features require JavaScript intervention for non-trivial use.
Without \KK{call\_indirect}, WebAssembly provides no mechanism for higher-order code---this is why we characterise our logic as supporting ``first-order, encapsulated WebAssembly''.
We view these features as part of further work on JavaScript/WebAssembly interoperability and discuss the ramifications of providing support for them in \S\ref{futureconclusions}.

Finally, the (\KK{return}) instruction is analogous to \KK{br}, except that it breaks out of \textit{all} enclosing constructs, concluding the execution of the function.

\subparagraph{Modules}
A WebAssembly program is represented as a module, which consists of: a list of functions; a list of global variables; the (optional) \KK{call\_indirect} table; and the (optional) linear memory.
Formally, this is written as \mbox{$\KK{module}~\textit{func}^\ast~\textit{glob}^\ast~\textit{tab}^?~\textit{mem}^?$}.
Functions are made of a function type \x{ft}, a series of typed local variable declarations $t^\ast$, and a function body~$e^\ast$.
Globals are made up of a type declaration \x{gt} (including an optional immutable flag for declaring constants) and an initializer expression $\x{e}^\ast$.
Tables collect a list of function indexes for use by the \KK{call\_indirect} instruction.
Memories declare their initial size measured in pages.
Functions, globals, tables, and memories may be shared between modules through a system of imports and exports, but we do not support this in our current logic, in large part because WebAssembly modules cannot satisfy each other's imports natively, but must currently rely on JavaScript ``glue code'' to compose together.

\subsection{WebAssembly Semantics}
\label{sec:abrupt_completions}

WebAssembly's official specification \cite{Haas:2017:BWU:3062341.3062363} provides a formal small-step semantics, mechanised in Isabelle/HOL by Watt~\cite{Watt:2018:MVW:3176245.3167082}. As part of the soundness proof of our program logic, we define and mechanise in Isabelle/HOL a WebAssembly big-step semantics that we formally prove equivalent, up to transitive closure, to the mechanised small-step semantics of~\cite{Watt:2018:MVW:3176245.3167082}. 
We introduce a fine-grained semantics of the \KK{br} and \KK{return} instructions, which is independent of the style of semantics chosen and streamlines formal reasoning.

\subparagraph{Execution Results}
%
WebAssembly executions terminate with one of the following~results:
\begin{itemize}
\item $\KK{Normal}~v^\ast$, representing standard termination with a list of values $v^\ast$ (in future, we often elide the $\KK{Normal}$ constructor and consider it to be the default result type);
\item $\KK{Trap}$, representing a runtime error (cf. \S\ref{sec:wasminstr} for examples of instructions that can trap);
\item $\KK{Break}~n~v^\ast$, describing an in-progress \KK{br} instruction;
\item $\KK{Return}~v^\ast$, describing an in-progress \KK{return} instruction.
\end{itemize}

Whereas the first two types of results are introduced by Haas et al. in \cite{Haas:2017:BWU:3062341.3062363}, the last two are introduced by us in this paper.
The reason for this is that the WebAssembly formal semantics of~\cite{Haas:2017:BWU:3062341.3062363} gives a very coarse-grained semantics to the \KK{br} and \KK{return} instructions.
A \KK{br} instruction targetting a control construct is defined as breaking to it immediately in a single step, discarding everything in between, including all other nested control constructs.

This complicates inductive proofs over the semantics, impairing formal reasoning~\cite{Watt:2018:MVW:3176245.3167082}.
In fact, this semantics is too coarse-grained for our proof system
and we need to introduce the notions of ``in-progress'' \KK{br} and  \KK{return} instructions as explicit execution results.


\begin{figure}[!b]
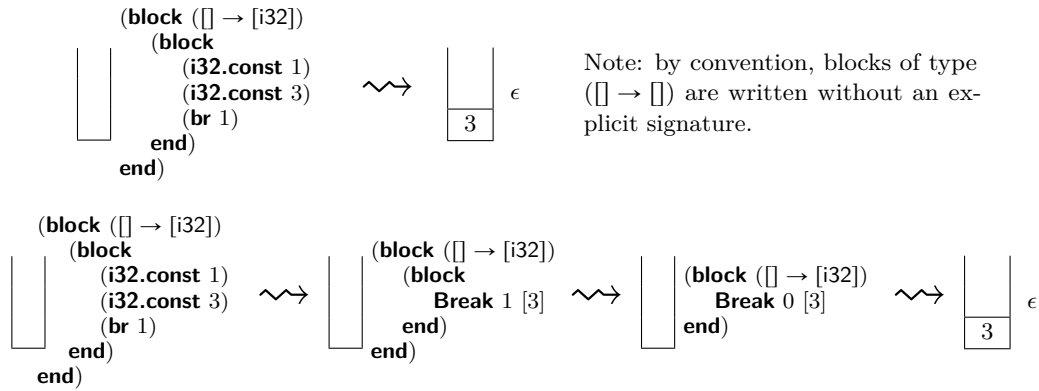



\begin{minipage}{\textwidth}
\centering
\small
\begin{tabular}{lclcl}
\begin{tabular}{|c|}
\\
\\
\\
\hline
\end{tabular}
\begin{minipage}{0.23\textwidth}
\footnotesize
\begin{Wblocke}{([] \rightarrow [\Witype])}
    \begin{Wblocke}{}
      \Wiconst{1} \\
      \Wiconst{3} \\
      \Wbr{1}
    \end{Wblocke}
\end{Wblocke}
\end{minipage}
&
\huge
\hspace{-0.05\textwidth}
$\leadsto$
&
\begin{tabular}{|c|}
\\
\\
\hline
3\\
\hline
\end{tabular}~
\begin{minipage}{0.0\textwidth}
$\epsilon$
\end{minipage}
\end{tabular}
\qquad
\begin{minipage}{0.37\textwidth}
Note: by convention, blocks of type \mbox{$([] \rightarrow [])$} are written without an explicit signature.
\end{minipage}
\end{minipage}
\\[0.4cm]

\begin{minipage}{\textwidth}
\centering
\begin{tabular}{lclclcl}
\begin{tabular}{|c|}
\\
\\
\\
\hline
\end{tabular}
\hspace{-1em}
\begin{minipage}{0.21\textwidth}
\footnotesize
\begin{Wblocke}{([] \rightarrow [\Witype])}
    \begin{Wblocke}{}
      \Wiconst{1} \\
      \Wiconst{3} \\
      \Wbr{1}
    \end{Wblocke}
\end{Wblocke}
\end{minipage}
&
\huge
\hspace{-0.05\textwidth}
$\leadsto$
\hspace{-0.03\textwidth}
&
\begin{tabular}{|c|}
\\
\\
\\
\hline
\end{tabular}
\begin{minipage}{0.19\textwidth}
\footnotesize
\begin{Wblocke}{([] \rightarrow [\Witype])}
    \begin{Wblocke}{}
      \KK{Break}~1~[3]
    \end{Wblocke}
\end{Wblocke}
\end{minipage}
&
\huge
\hspace{-0.05\textwidth}
$\leadsto$
\hspace{-0.03\textwidth}
&
\begin{tabular}{|c|}
\\
\\
\\
\hline
\end{tabular}
\begin{minipage}{0.19\textwidth}
\footnotesize
\begin{Wblocke}{([] \rightarrow [\Witype])}
      \KK{Break}~0~[3]
\end{Wblocke}
\end{minipage}
&
\huge
\hspace{-0.04\textwidth}
$\leadsto$
\hspace{-0.03\textwidth}
&
\begin{tabular}{|c|}
\\
\\
\hline
3\\
\hline
\end{tabular}~
\begin{minipage}{0.0\textwidth}
$\epsilon$
\end{minipage}
\end{tabular}
\end{minipage}
\caption{Granularity of \KK{br} executions: Haas et al.~\cite{Haas:2017:BWU:3062341.3062363}~(top); our approach (bottom).}\label{fig:abrupt_completion}
\end{figure}

We illustrate the difference between the approach of Haas et al.~\cite{Haas:2017:BWU:3062341.3062363} and our approach in Figure~\ref{fig:abrupt_completion}.
The top reduction follows the official semantics of~\cite{Haas:2017:BWU:3062341.3062363}.
There, $\Wbr 1$ breaks out of two blocks in a single step, transferring exactly one value out of the \KK{block}, in order to satisfy the targeted block's type signature.
We make this semantics more granular by introducing an auxiliary \KK{Break} result type. Concretely, \mbox{$\KK{Break}~n~v^\ast$} denotes an in-progress \KK{br} instruction, with $n$ remaining contexts to break out of, in the process of transferring $v^\ast$ values to the target context, as shown in the bottom reduction of Figure~\ref{fig:abrupt_completion}.
Similarly, \mbox{$\KK{Return}~v^\ast$} represents an in-progress \KK{return} instruction, with the only difference being that \KK{Return} does not require a remaining context count, as it breaks out of all enclosing constructs.

\subparagraph{Big-Step Semantic Judgement}
The judgement of our big-step semantics is of the~form
$$
(\x{s}, \x{loc}^\ast, \x{v}_\x{e}^\ast \x{e}^\ast) \Downarrow_{inst}^{\x{labs,ret}} (\x{s}', \x{loc}'^\ast, \x{res}).
$$
On the left-hand side of the judgement, we have \emph{configurations} of the form \mbox{$(\x{s}, \x{loc}^\ast, \x{v}_\x{e}^\ast \x{e}^\ast)$}, where $s$ is a \emph{store} containing whole-program runtime information (e.g.~global variables and the memory), $\x{loc}^\ast$ is the list of current local variables, and $\x{v}_e^\ast$ is a value stack $\x{v}^\ast$ lifted to \KK{const} instructions, which is then directly concatenated with $\x{e}^\ast$, the list of instructions to execute.\footnote{This treatment of the value stack is a key difference between the official prose and formal semantics.
In the prose semantics, the stack is represented as a list of values $v^\ast$, together with an executing list of instructions $e^\ast$, which modifies the stack.
In the formal semantics, the value stack is represented as a list of \KK{const} instructions, and directly concatenated with the executing list of instructions to form a single list.
Reduction rules are defined between configurations, pattern-matching between \KK{const} instructions and other instructions, such as \KK{add}, without ever explicitly manipulating a separate value stack.}
Configuration execution yields an updated store $s'$, updated local variables~$loc'^*$, and a result~\x{res}, which has one of the four above-mentioned result types.

Additionally, execution is defined with respect to a subscript \x{inst}. This is the \textit{run-time instance}, a record which keeps track of which elements of $s$ have been allocated by the current program.
In the case of the encapsulated modules that we consider, its role in the formalism is trivial, but its full role is described in the official specification~\cite{Haas:2017:BWU:3062341.3062363}, and we give a full definition in~\cite{wasmlogic:techrep}, along with our big-step semantics.

Finally, execution is also defined with respect to a list of break label arities \x{labs} (a \x{nat~list}), and a return label arity \x{ret} (a single \x{nat}).
As depicted in Fig.~\ref{fig:abrupt_completion}, \KK{Break} and \KK{Return} results must transfer precisely the correct number of values to satisfy the type of the context it is targeting. The \x{labs} and \x{ret} parameters keep track of the number of values required, so that, for example, if \x{res} is of the form \mbox{$\KK{Break}~k~v^n$}, then $\x{labs}!k = n$.
Similarly, if \x{res} is of the form \mbox{$\KK{Return}~v^n$}, then $\x{ret} = n$.

\subparagraph{Equivalence Result} We recall the original formal small-step semantic judgement of~\cite{Haas:2017:BWU:3062341.3062363}, which is of the form $(\x{s}, \x{loc}^\ast, v_e^\ast\x{e}^\ast) \hookrightarrow_\x{inst} (\x{s}', \x{loc}'^\ast, \x{v}_e'^\ast e'^\ast)$. This judgement does not include our break or return labels.

We state our equivalence result in Theorem~\ref{thm:equiv} and mechanise its proof in Isabelle/HOL. We denote the transitive closure of the small-step semantics by $\hookrightarrow^\ast$. Both $\hookrightarrow$ and $\Downarrow$ are subscripted by the instance $\x{inst}$, the big-step derivation starts with empty \x{labs} ([]) and empty \x{ret} ($\epsilon$) components, and $v'^\ast$ denotes the list of values obtained from $v_e'^\ast$ by removing their leading \KK{const}s.

\begin{theorem1}[reduce\_trans\_equiv\_reduce\_to]
\label{thm:equiv}
\mbox{$(\x{s}, \x{loc}^\ast, v_e^\ast\x{e}^\ast) \hookrightarrow_\x{inst}^\ast (\x{s}', \x{loc}'^\ast, \x{v}_e'^\ast)$} $\Longleftrightarrow$ \mbox{$(\x{s}, \x{loc}^\ast, v_e^\ast\x{e}^\ast) \Downarrow_\x{inst}^{[],\epsilon} (\x{s}', \x{loc}'^\ast, \KK{Normal}~\x{v'}^\ast)~\wedge$}\\
\mbox{$(\x{s}, \x{loc}^\ast, v_e^\ast\x{e}^\ast) \hookrightarrow_\x{inst}^\ast (\x{s}', \x{loc}'^\ast, [\KK{trap}])$} $\Longleftrightarrow$ \mbox{$(\x{s}, \x{loc}^\ast, v_e^\ast\x{e}^\ast) \Downarrow_\x{inst}^{[],\epsilon} (\x{s}', \x{loc}'^\ast, \KK{Trap})$}
\end{theorem1}

Theorem~\ref{thm:equiv} relates terminal states (values $\x{e}^\ast$ or a trap result $[\KK{trap}]$) in the small step semantics with execution results in the big-step semantics.
In particular, it shows that the small- and big-step semantics give equivalent results for all terminating programs.
The proof also requires auxiliary lemmas about how the big-step \KK{Break} and \KK{Return} execution results correspond to behaviours in the small-step semantics.
These lemmas are not included here for space, but can be found in the mechanisation.

\section{\logicname}
\label{sec:proglog}

We present \logicname, a program logic for first-order, encapsulated WebAssembly modules.
We define a novel assertion syntax, with a highly structured stack assertion which takes advantage of WebAssembly's strict type system.
Our proof rules for the WebAssembly \KK{br} and \KK{return} instructions are inspired by a foundational proof rule for ``structured \KK{goto}'' by Clint and Hoare~\cite{Clint1972}, and extend their work to the world of separation logic~\cite{Reynolds:2002:SLL:645683.664578}.
We fully mechanise and prove soundness of \logicname in Isabelle/HOL, as detailed in \S\ref{sec:soundness}.

\subsection{Assertion Language}
\label{ass:lang}

\logicname assertions encode information about WebAssembly runtime states.
Their semantic interpretation is formally described in \S\ref{sec:soundness}, in the context of our soundness result.

In many programming languages, program state is made up of the values stored in \textit{variables} and the values stored in the \textit{heap}.
In this case, it is natural for assertions to be expressed using a \textit{separation logic}, which extends predicate logic with connectives for reasoning about resource separation, and is useful for modular client reasoning~\cite{Reynolds:2002:SLL:645683.664578}.

WebAssembly, however, also allows values to be stored in the stack.
Given how the WebAssembly's type system provides static knowledge of the stack size and of the types of each of its elements at every program point, we believe that reasoning about the WebAssembly stack \textit{should} be simple: that is, it should not result in proofs more complicated than those of traditional separation logic. We manage to achieve this thanks to our structured stack assertion and the associated proof rules.
While one's first instinct could be to treat assertions about stack values like assertions about local variables, such a system would require substantial bookkeeping, since the stack changes shape during execution.
Benton~\cite{Benton:2005:TCL:2099708.2099741} uses this approach for a language with a similar typed-value stack, but ends up describing the resulting proofs as ``fussily baroque'' and ``extremely tedious to construct by hand''.
%


\begin{figure}[!h]
$$
\begin{array}{ll@{~~}c@{~~}l}
\text{constants} & c \in \consts &::=& \text{c}_{\text{i32}}~\pmb{|}~\text{c}_{\text{i64}}~\pmb{|}~\text{c}_{\text{f32}}~\pmb{|}~\text{c}_{\text{f64}}  \\
\text{variables (logical/local/global)} & \x{\nu} \in \vars &::=& \textit{x}~ \pmb{|}~l_i   ~ \pmb{|}~g_i, \text{where } i \in \mathbb{N} \\
\text{terms}  & \x{\tau} \in \terms &::=& c~\pmb{|}~\x{\nu}~\pmb{|}~f(\x{\tau}_1 \ldots \x{\tau}_n) \\
\text{heap assertions}  & \ha, \ha' \in \hasrts &::=& \bot~\pmb{|}~\neg \ha~\pmb{|}~\ha \wedge \ha'~\pmb{|}\\
&&& \exists \x{x}.~\ha~\pmb{|}~  p(\x{\tau}_1 \ldots \x{\tau}_n)  ~ \pmb{|} \\
&&& \textbf{emp} ~\pmb{|}~ \ha * \ha' ~\pmb{|} \Star{\tau_1 < \textit{ x }< \tau_2}\ha~\pmb{|} \\
&&& \x{\tau}_1 \mapsto \x{\tau}_2~\pmb{|}~\textbf{size}(\x{\tau}) \\
\text{stack assertions} & \sa \in \sasrts &::=& []~~\pmb{|}~~\sa :: \x{\tau}  \\
\text{assertions} & P,Q \in \asrts &::=& (\sa ~|~ \ha)~~\pmb{|}~~ \exists \x{x}.~P
\end{array}
$$

$$
[\exists \overrightarrow{x}.~ (\sa~|~H)] \otimes H_f ~ \triangleq ~ \exists \overrightarrow{x}.~(\sa~|~H * H_f) \quad \text{iff $\textit{fv}(H_f) \cap \overrightarrow{x} = \emptyset$}
$$
\caption{Syntax of \logicname assertions.}
\label{fig:ass_syntax}
\end{figure}

The syntax of \logicname assertions is defined in Fig.~\ref{fig:ass_syntax}.
Constants, $c$, can have one of the four WebAssembly value types.
Next we have logical, local, and global variables, with local/global variables having dedicated variable names, $l_i$/$g_i$, where $i \in \mathbb N$.
Terms can either be constants, or variables, or functions (for example,~unary and binary operators).

Heap assertions are mostly familiar from traditional separation logic~\cite{Reynolds:2002:SLL:645683.664578}.
First, we have the pure assertions of predicate logic, including predicates $p(\x{\tau}_1 \ldots \x{\tau}_n)$ over terms (for example, term equality).
We also have the standard spatial assertions: $\emp$ describes an empty heap, $\ha * \ha'$ is the separating conjunction (star), and the iterated star operator,~$\bigoasterisk$, aggregates assertions composed by $\slstar$ in the same way that $\sum$ aggregates arithmetic expressions composed by $+$.
Finally, we have two WebAssembly-specific spatial assertions: the cell assertion $\x{\tau}_1 \mapsto \x{\tau}_2$ describes a single heap cell at address denoted by $\x{\tau}_1$ with contents denoted by $\x{\tau}_2$, and the \Wsize{\tau} assertion states that the number of pages currently allocated is denoted by $\x{\tau}$.

A stack assertion, denoted by $\sa$, is a list of terms, each of which represents the value of the corresponding stack position in the value stack.
This is possible due to the size of the WebAssembly stack always being precisely known statically.
Were this not true, the stack assertion would need to be able to represent that the stack may have multiple sizes, and could not be represented purely as a single list of terms.
The list appends on the right, to match the conventions of the WebAssembly type system.

Finally, a \logicname assertion is a two-part, possibly existentially quantified assertion consisting of a stack assertion $\sa$, and a pure/heap assertion $\ha$.
We define an operator, $\otimes$, for distributing heap frames through \logicname assertions, which will be used later in~\S\ref{sec:stackops} to define our frame rule.
The notation $\exists\overrightarrow{x}$ is a shorthand for some set of outer existentially quantified variables, while $\textit{fv}(H_f)$ returns the set of \textit{free variables} in the heap assertion $H_f$.

\subparagraph{Notation}
For clarity of presentation, we introduce the following notational conventions:
\begin{itemize}
\item (Stack Length) We denote by $P_n$ an assertion whose stack part is of length $n$.
\item{ (Type Annotations in Cell Assertions)} The cell assertion $\tau_1 \mapsto \tau_2$ encodes the value of a single byte in memory. As WebAssembly values normally take up either four or eight bytes, it is convenient for us to define the corresponding shorthand, which we do by annotating the arrow with the appropriate type: $\tau_1 \mapsto_t \tau_2$. For example, we have that
$\tau_1 \mapsto_{i32} \tau_2 \triangleq \tau_1 \mapsto b_0 * (\tau_1+1) \mapsto b_1 * (\tau_1+2) \mapsto b_2 * (\tau_1+3) \mapsto b_3$,
\noindent where $b_k$ denotes the $k^\text{th}$ least significant byte of the 32-bit representation of $\tau_2$.

\item{ (Operator Domain)} To avoid clutter, we overload all mathematical operators (e.g., $+$, $\cdot$, $\leq$, $\ldots$) instead of explicitly stating their domain ($\K{i32}$, $\K{i64}$, $\K{f32}$, $\K{f64}$, $\mathbb{N}$, $\mathbb{Z}$, or $\mathbb{R}$) on each use. When required, we state the domain either of a single operator (e.g., $\ittplus$, $\isfplus$, $\ldots$) or of a parenthesised expression (e.g., $(3.14 - 2.71 \cdot x)_{\K{f64}}$)), in which case the domain applies to all operators and operands of the expression. The default domain is~$\K{i32}$.
\end{itemize}

%

\subsection{\logicname Triple}
\label{sec:proof_system}
We define a program logic for first-order, encapsulated WebAssembly modules. We base our encoding of program behaviour on \textit{Hoare triples}~\cite{Hoare:1969:ABC:363235.363259}. \logicname triples are of the form%
$$\Gamma \vdash \speclineold{\{P\}}~e^\ast~\speclineold{\{Q\}}$$
\noindent where $e^\ast$ is the WebAssembly program to be executed, $P$ is its \textit{pre-condition}, $Q$ is its \textit{post-condition}, and $\Gamma$ represents the context in which the program is executed.

Before giving the interpretation of the \logicname triple, we have to explain the context~$\Gamma$ in detail. A context contains four fields: 
\dtag{1} the \emph{functions} field, $F$, containing a list of all function definitions of the module;
\dtag{2} the \emph{assumptions} field, $A$, containing a set of assertions of the form $\speclineold{\{ P \}}~\KK{call}~i~\speclineold{\{Q \}}$, used by the [call] rule to correctly capture mutually recursive functions;
\dtag{3} the \emph{labels} field, $L$, containing a list of assertions used to describe the behaviour of the \KK{br} instruction; and
\dtag{4} the \emph{return} field, $R$, containing an optional return assertion, used to describe the behaviour of the \KK{return} instruction.
A context may be alternatively presented as $(F,A,L,R)$, and any of its fields may be referenced directly: for example, $\Gamma.F$ refers to the functions field of the context.
We use $P;\Gamma$ as syntactic shorthand for $\Gamma$ with $P$ appended to the head of its labels field, since this pattern occurs commonly.

\subparagraph{Interpretation of \logicname Triples} The meaning of the triple \mbox{$\Gamma \vdash \speclineold{\{P\}}~e^\ast~\speclineold{\{Q\}}$} is, informally, as follows.
Let $e^\ast$ be executed from a state satisfying $P$. Then: if $e^\ast$ terminates normally, it will terminate in a state satisfying $Q$;
if it terminates with a $\KK{Return}~v^\ast$ result, the resulting state must satisfy $\Gamma.R$; and
if it terminates with a $\KK{Break}~i~v^\ast$ result, the resulting state must satisfy the $i$-th assertion of $\Gamma.L$.
A formal definition is given in~\S\ref{sec:soundness}.

\subsection{Proof Rules}
\label{sec:stackops}

\subparagraph{Basic Instructions}
The proof rules for basic instructions are given in Figure~\ref{fig:basinstrs}.
These rules manipulate only the stack and pure logical assertions, and can be intuitively motivated by their effects on the stack.
In particular, the effect of the [{select] rule is conditional on the value of $\tau_3$: we know that it has placed exactly one value on the stack, but whether it is $\tau_1$ or $\tau_2$ depends on whether or not $\tau_3 \neq 0$. These rules, despite manipulating the WebAssembly stack, appear very standard: this is precisely due to our structured stack assertions.


\begin{figure}[!b]
{
\footnotesize
$$
\prftree[r]{[const]}
{
\Gamma \vdash \speclineold{\{\stackr{0}{}~|~\textbf{emp}\}}~\textit{t}.\KK{const}~c~\speclineold{\{\stackr{0}{{c}}~|~\textbf{emp}\}}
}
\qquad
\prftree[r]{[unreachable]}
{
\Gamma \vdash \speclineold{\{\stackr{0}{}~|~\bot\}}~\KK{unreachable}~\speclineold{\{ Q\}}
}
$$

$$
\prftree[r]{[nop]}
{
\Gamma \vdash \speclineold{\{\stackr{0}{}~|~\textbf{emp}\}}~\KK{nop}~\speclineold{\{\stackr{0}{}~|~\textbf{emp}\}}
}
\qquad
\prftree[r]{[drop]}
{
\Gamma \vdash \speclineold{\{\stackr{0}{\tau}~|~\textbf{emp}\}}~\KK{drop}~\speclineold{\{\stackr{0}{}~|~\textbf{emp}\}}
}
$$

$$
\prftree[r]{[select]}
{
\Gamma \vdash \speclineold{\{ [\tau_1, \tau_2, \tau_3]~|~\textbf{emp}\}}~\KK{select}~\speclineold{\{ \exists x.~[x]~|~\textbf{emp} \wedge (\tau_3 \neq 0 \rightarrow x = \tau_1) \wedge (\tau_3 = 0 \rightarrow x = \tau_2) \}}
}
$$

$$
\hspace*{-0.15cm}
\prftree[r]{[unop]}
{
\Gamma \vdash \speclineold{\{\stackr{0}{\tau}~|~\textbf{emp}\}}~\textit{t}.\textit{unop}~\speclineold{\{\stackr{0}{\textit{unop}(\tau)}~|~\textbf{emp}\}}
}
\quad
\prftree[r]{[testop]}
{
\Gamma \vdash \speclineold{\{\stackr{0}{\tau}~|~\textbf{emp}\}}~\textit{t}.\textit{testop}~\speclineold{\{\stackr{0}{\textit{testop}(\tau)}~|~\textbf{emp}\}}
}
$$

$$
\prftree[r]{[binop]}
{
\Gamma \vdash \speclineold{\{\stackr{0}{\tau_1 \stacks \tau_2}~|~\text{defined}{(\textit{binop}, \tau_1, \tau_2)} \wedge \textbf{emp}\}}~\textit{t}.\textit{binop}~\speclineold{\{\stackr{0}{\textit{binop}(\tau_1, \tau_2)}~|~\textbf{emp}\}}
}
$$

$$
\prftree[r]{[relop]}
{
\Gamma \vdash \speclineold{\{\stackr{0}{\tau_1 \stacks \tau_2}~|~\textbf{emp}\}}~\textit{t}.\textit{relop}~\speclineold{\{\stackr{0}{\textit{relop}(\tau_1, \tau_2)}~|~\textbf{emp}\}}
}
$$

$$
\prftree[r]{[cvtop]}
{
\Gamma \vdash \speclineold{\{\stackr{0}{\tau}~|~\text{defined}{(\textit{cvtop}, \tau)} \wedge \textbf{emp}\}}~\textit{t}.\textit{cvtop}~\speclineold{\{\stackr{0}{\textit{cvtop}( \tau)}~|~\textbf{emp}\}}
}
$$

\medskip
{\bfseries Note}: The $\text{defined}{(\textit{binop}, \tau_1, \tau_2)}$ and $\text{defined}{(\textit{cvtop}, \tau)}$ predicates describe conditions sufficient for binary and conversion operators to be non-trapping.
}
\caption{Proof Rules: Basic Instructions.}
\label{fig:basinstrs}
\end{figure}

\subparagraph{Variable Management Instructions} We give the proof rules for variable management instructions in Figure~\ref{fig:varinstrs} (left). Just like the rules for basic instructions, these also require an empty heap. By observing these rules, we can understand how the dedicated local/global variable names are manipulated. For example, $\Wgetlocal i$ simply puts the variable $l_i$ on the top of the stack.
On the other hand, $\Wsetglobal{i}$ requires one value from the value stack in the pre-condition, and in the post-condition has consumed it, and guarantees that $g_i$, the $i$-th global variable, holds this value.

In Figure~\ref{fig:varinstrs} (right), we give a proof sketch of a simple WebAssembly program that uses basic and variable management instructions, illustrating how stack assertions behave.
We start from the pre-condition \mbox{$\specline{\stackheap{}{\x{l}_1 = 2 \wedge \emp}}$}, which tells us that the stack and the heap are empty and that the first local variable, $l_1$, equals $2$.
Executing \mbox{$\Wgetlocal{1}$} adds $l_1$ to the stack, which we can immediately replace with $2$ due to our pure knowledge that \mbox{$\x{l}_1 = 2$}.
The second line of the program pushes the constant $3$ onto the stack (the top of the stack is on the \textit{right-hand side} of the assertion).
Finally, the two values are added together, and the resulting stack holds a single value, 5.


\begin{figure}[!t]

\begin{minipage}{0.64\textwidth}
{\footnotesize
$$
\prftree[r]{[get\_local]}
{
\text{isDeclaredLocal } i
}
{
\Gamma \vdash \speclineold{\{\stackr{0}{}~|~\textbf{emp}\}}~\KK{get\_local}~\textit{i}~\speclineold{\{ \stackr{0}{l_i}~|~\textbf{emp}\}}
}
$$

$$
\prftree[r]{[set\_local]}
{
\text{isDeclaredLocal } i
}
{
\Gamma \vdash \speclineold{\{\stackr{0}{x}~|~\textbf{emp}\}}~\KK{set\_local}~\textit{i}~\speclineold{\{\stackr{0}{} ~|~\textbf{emp} \wedge l_i = x\}}
}
$$

$$
\prftree[r]{[tee\_local]}
{
\text{isDeclaredLocal } i
}
{
\Gamma \vdash \speclineold{\{\stackr{0}{x}~|~\textbf{emp}\}}~\KK{tee\_local}~\textit{i}~\speclineold{\{\stackr{0}{x}~|~\textbf{emp}\wedge l_i = x\}}
}
$$

$$
\prftree[r]{[get\_global]}
{
\text{isDeclaredGlobal } i
}
{
\Gamma \vdash \speclineold{\{\stackr{0}{}~|~\textbf{emp}\}}~\KK{get\_global}~\textit{i}~\speclineold{\{ \stackr{0}{g_i}~|~\textbf{emp}\}}
}
$$

$$
\prftree[r]{[set\_global]}
{
\text{isDeclaredGlobal } i
}
{
\Gamma \vdash \speclineold{\{\stackr{0}{x}~|~\textbf{emp}\}}~\KK{set\_global}~\textit{i}~\speclineold{\{\stackr{0}{} ~|~\textbf{emp} \wedge g_i = x\}}
}
$$}
\end{minipage}~
\begin{minipage}{0.32\textwidth}
\small
$\specline{\stackheap{}{\x{l}_1 = 2 \wedge \emp}}$ \\
\Wgetlocal{1}\\
$\specline{\stackheap{\x{l}_1}{\x{l}_1 = 2 \wedge \emp}}$ \\
$\specline{\stackheap{2}{\x{l}_1 = 2 \wedge \emp}}$ \\
\Wiconst{3}\\
$\specline{\stackheap{2, 3}{\x{l}_1 = 2 \wedge \emp}}$ \\
\Wiadd \\
$\specline{\stackheap{5}{\x{l}_1 = 2 \wedge \emp}}$
\end{minipage}
{\footnotesize
{\bfseries Note}: The $(\text{isDeclaredLocal } i)$ and $(\text{isDeclaredGlobal } i)$ predicates are an internal detail of the meta-theory ensuring that $l_i$ and $g_i$ do not refer to local/global variables that are not declared by the module. They always hold for any well-typed WebAssembly program.}
\caption{Proof Rules: Variable Management Instructions (left); Simple Proof Sketch (right)}
\label{fig:varinstrs}
\end{figure}

\subparagraph{Memory Management Instructions}
Proof rules for instructions that interact with the WebAssembly memory are given in Figure~\ref{fig:memoryops}.
The $(t.\KK{load})$ and $(t.\KK{store})$ proof rules are similar to standard separation heap rules, except that they are annotated with the type of the value in the heap, which determines the number of bytes that this value occupies, and also a static offset, which is added to the given address. 


\begin{figure}[!b]
{
\footnotesize
\vspace*{-0.3cm}
$$
\hspace*{-0.25cm}
\begin{array}{c}
\prftree[r]{[load]}
{
\Gamma \vdash \speclineold{\{\stackr{0}{\tau_1}~|~(\tau_1 + \textit{off}) \mapsto_t \tau_2 \}}~\textit{t}.\KK{load } \textit{off}~\speclineold{\{\stackr{0}{\tau_2}~|~(\tau_1 + \textit{off}) \mapsto_t \tau_2 \}}
}
\\[0.25cm]
\prftree[r]{[store]}
{
\Gamma \vdash \speclineold{\{\stackr{0}{\tau_1 \stacks \tau_2}~|~(\tau_1 + \textit{off}) \mapsto_t - \}}~\textit{t}.\KK{store } \textit{off}~\speclineold{\{\stackr{0}{}~|~(\tau_1 + \textit{off}) \mapsto_t \tau_2\}}
}
\\[0.25cm]
\prftree[r]{[mem.size]}
{
\Gamma \vdash \speclineold{\{ \stackr{0}{}~|~\textbf{size}(\tau) \}}~\KK{mem.size}~\speclineold{\{ \stackr{0}{\tau}~|~\textbf{size}(\tau) \}}
}
\\[0.25cm]
\prftree[r]{[mem.grow]}
{
\Gamma \vdash \speclineold{\{ \stackr{0}{\tau_1}~|~\textbf{size}(\tau_2) \}}~\KK{mem.grow}~\speclineold{\left \{\exists v.~\stackr{0}{v}\left |~\pbox{0.7\textwidth}{$\left ( \pbox{0.7\textwidth}{\hspace{-0.6cm}$\Star{\hspace{0.6cm}\tau_2~\leq~i/64k~<~(\tau_2+ \tau_1)} i \mapsto 0 \slstar~\textbf{size}(\tau_2 + \tau_1)$ \\ 
$\wedge~v = \tau_2 \wedge~((\tau_2 + \tau_1) \leq 2^{16})_{\mathbb{N}}$}\!\!\!\right )$\hspace*{-0.3cm} \\ $\vee~(\textbf{size}(\tau_2) \wedge v = -1)$} ~ \right.~\right \}}
}
\end{array}
$$
  \caption{Proof Rules: Memory Management Instructions.}
  \label{fig:memoryops}
}
\end{figure}

As discussed, the $\Wsizememory$ and $\Wgrowmemory$ instructions allow WebAssembly to alter the memory size.
The ``permission'' to observe the memory size is encoded using the $\textbf{size}(\tau)$ assertion, which states that the memory is currently $\tau$ pages long.
This permission, however, does not imply permission to access in-bounds locations; the logic still requires $x \mapsto_t n$ to be held in order to access the location $x$, even if $x$ is known to be in-bounds because \textbf{size} is held.
Growing the memory using the $\Wgrowmemory$ instruction confers ownership of all newly created locations, and leaves the index of the first newly allocated location on the stack.


\subparagraph{Control Flow Instructions}
%
%
%
%
The proof rules for WebAssembly control constructs are given in Figure~\ref{fig:controlops}.
%
%
These rules illustrate how the labels ($L$) and return ($R$) fields of the context are used in practice. 
In particular, $L$ contains a list of assertions, and the $i$-th assertion describes the state that has to hold if we break out of $i$ enclosing contexts. Similarly, the $R$ assertion describes the state that has to hold if we execute a function return. 


\begin{figure}[!ht]
\footnotesize
$$
\prftree[r]{[br]}
{
L!i = P
}
{
F,A,L,R \vdash \speclineold{\{ P \}}~\KK{br } \textit{i}~\speclineold{\{Q\}}
}
\qquad
\prftree[r]{[return]}
{
F,A,L,R \vdash \speclineold{\{ R\}}~\KK{return}~\speclineold{\{Q\}}
}
$$

$$
\prftree[r]{[block]}
{
Q_m~;\Gamma \vdash \speclineold{\{ P_n\}}~e^\ast~\speclineold{\{ Q_m\}}
}
{
\Gamma \vdash  \speclineold{\{P_n\}}~\KK{block } \textit{$t^n$ $\rightarrow$ $t^m$ } e^\ast~\KK{end}~\speclineold{\{ Q_m\}}
}
\qquad
\prftree[r]{[loop]}
{
P_n~;\Gamma \vdash \speclineold{\{ P_n\}}~e^\ast~\speclineold{\{ Q_m\}}
}
{
\Gamma \vdash  \speclineold{\{P_n\}}~\KK{loop } \textit{$t^n$ $\rightarrow$ $t^m$ } e^\ast~\KK{end}~\speclineold{\{ Q_m\}}
}
$$

$$
\prftree[r]{[if]}
{
\pbox{0.45\textwidth}{
\centering
$\Gamma \vdash \speclineold{\{\sa~|~H \wedge \tau \neq 0_{i32}\}}~\KK{block } \textit{tf } e^\ast_1~\KK{end}~\speclineold{\{Q\}}$
$\Gamma \vdash \speclineold{\{\sa~|~H \wedge \tau = 0_{i32}\}}~\KK{block } \textit{tf } e^\ast_2~\KK{end}~\speclineold{\{Q\}}$
}
}
{
\Gamma \vdash \speclineold{\{\sa :: \tau~|~H\}}~\KK{if } \textit{tf } e_1^\ast \KK{ else } e_2^\ast~ \KK{ end}~\speclineold{\{Q\}}
}
\quad
\prftree[r]{[br\_if]}
{
\Gamma \vdash \speclineold{\{ \sa~|~H \wedge \tau \neq 0_{i32}\}}~\KK{br } \textit{i}~\speclineold{\{Q\}}
}
{
\Gamma \vdash \speclineold{\{ \sa ::  \tau~|~H\}}~\KK{br\_if } \textit{i}~\speclineold{\{ \sa~|~H \wedge \tau = 0_{i32}\}}
}
$$

$$
\prftree[r]{[br\_table]}
{
\pbox{0.66\textwidth}{
\centering
$\frall k 0 \leq k < \llen {i^\ast} \rightarrow \Gamma \vdash \speclineold{\{\sa~|~H \wedge \tau = k_{\K{i32}}\}}~\KK{br } (i^\ast \lidx k)~\speclineold{\{Q\}}$
$\Gamma \vdash \speclineold{\{\sa~|~H \wedge \lnot(0 \leq \tau < \llen{i^\ast})_{\K{i32}} \}}~\KK{br } \textit{i}~\speclineold{\{Q\}}$
}
}
{
\Gamma \vdash \speclineold{\{\sa :: \tau~|~H\}}~\KK{br\_table } i^\ast~i~\speclineold{\{Q\}}
}
$$
\caption{Proof Rules: Control Flow Instructions.}
\label{fig:controlops}
\end{figure}

In line with this, the precondition of $\Wbr i$ in the [br] rule equals the $i$-th assertions of~$L$. On the other hand, its post-condition is arbitrary, which is justified by the fact that any code following a \KK{br} instruction in the same block of code cannot be reached due to the structured control flow of WebAssembly. Analogously, the precondition of a (\KK{return}) statement in the [return] rule equals the return field of the context, and its post-condition is arbitrary. Observe the clear analogy between the role of \x{labs} and \x{ret} in the semantics and the role of $L$ and $R$ in the proof rules for \KK{br} and \KK{return}, respectively.

The main aspect of the [block] and [loop] rules is how they interact with the context. Concretely, in the [block] rule, the labels field is extended with the post-condition of the block, whereas in the [loop] rule, it is extended with its pre-condition. Bearing in mind the [br] rule, this precisely captures the WebAssembly control flow: when we break to a block, we exit the block, and when we break to a loop, we continue with the next iteration and the pre-condition of the loop acts as its invariant.

This approach is inspired by the proof rule for ``structured'' \KK{goto} statements of Clint and Hoare~\cite{Clint1972}, as WebAssembly's \KK{block} and \KK{br} opcodes replicate the structural conditions imposed by~\cite{Clint1972} on the use of \KK{goto}.
Note also that the explicit type annotations of [block] and [loop], combined with the guarantees of the WebAssembly type system, allow the rules to precisely fix the size of the stack in both the pre- and post-condition.

Next, the [if] rule branches depending on the value that is on the top of the stack. If this value is non-zero, the \KK{then} branch is taken, and the \KK{else} branch otherwise. As is commonplace, the post-conditions of the two \KK{if} branches have to match.

The [br\_if] rule is a conditional break. If the break is taken, the value on the top of the stack is popped, and known to be non-zero, and the instruction functions identically to \KK{br}. The post-condition represents the case where the break is not taken: the value on the top of the stack is popped, and known to be $0$.

Finally, the \KK{br\_table} instruction acts like the switch statement of modern languages, breaking to the appropriate label depending on the value on the top of the stack.

\subparagraph{Structural Proof Rules}
Structural proof rules, shown in Figure~\ref{fig:structureops} and demonstrated in practice throughout \S\ref{sec:verification}, are needed to compose proofs together.
The [seq] rule for program concatenation is inherited from standard separation logic, whereas the others are either new or require adjustment for \logicname.

The existential elimination rule, [exists], has to eliminate the existential from all assertions in $L$ and also the $R$. If we were only to eliminate the existential from the pre- and post-condition, as is standard, the rule would be unsound, as we could derive the following:
$$
\footnotesize
\prftree[r]{[unsound exists]}
{
-,-,[(\stackheap{}{l_0 = k})], -  \vdash \speclineold{\{\stackheap{}{l_0 = k}\}}~\Wbr 0~\speclineold{\{Q\}}
}
{
-,-,[(\stackheap{}{l_0 = k})], -  \vdash \speclineold{\{\exists k'.~\stackheap{}{l_0 = k'}\}}~\Wbr 0~\speclineold{\{\exists x.~Q\}}
}
$$
\noindent which does not correspond to the intended meaning of the context, as the pre-condition of the break no longer implies its matching assertion in $L$.  For similar reasons, the [frame] rule must frame off from all assertions in $L$ and also the $R$. As shown in~\S\ref{sec:verification}, we can derive simpler proof rules for straight-line code that do not require irrelevant manipulation of $L$ and $R$.


\begin{figure}[!t]
{
\footnotesize
$$
\prftree[r]{[seq]}
{
\Gamma \vdash \speclineold{\{P\}}~e^\ast_1~\speclineold{\{Q\}}
}
{
\Gamma \vdash \speclineold{\{Q\}}~e^\ast_2~\speclineold{\{R\}}
}
{
\Gamma \vdash \speclineold{\{P\}}~e^\ast_1~e^\ast_2~\speclineold{\{R\}}
}
\qquad
\prftree[r]{[exists]}
{
F,A,L,R^?  \vdash \speclineold{\{P\}}~e^\ast~\speclineold{\{Q\}}
}
{
F,A,(map~(\exists x.)~L),(\exists x.~R)^?  \vdash \speclineold{\{\exists x.~P\}}~e^\ast~\speclineold{\{\exists x.~Q\}}
}
$$

$$
\prftree[r]{[frame]}
{
F,A,L,R^?  \vdash \speclineold{\{P\}}~e^\ast~\speclineold{\{Q\}}
}
{
\textit{fv}(H) \cap \textit{mv}(e^\ast) = \emptyset
}
{
F,A,(map~(\otimes~H)~L) ,(R \otimes H)^?  \vdash \speclineold{\{P \otimes H\}}~e^\ast~\speclineold{\{Q \otimes H\}}
}
$$

$$
\prftree[r]{[consequence]}
{
\pbox{0.85\textwidth}{
\mbox{$F,A,L',R'_{n'} \vdash \speclineold{\{P'\}}~e^\ast~\speclineold{\{Q'\}}$} \qquad ~
\mbox{$P \Rightarrow P'$} \qquad\hspace*{0.03cm}
\mbox{$Q' \Rightarrow Q$} \qquad 
\mbox{$\llen{L} = \llen{L'}$} \\[0.1cm]
\mbox{$\forall i < \llen L. \exists L_{n}~L'_{n'}.~L!i = L_n \wedge L'!i = L'_{n'} \wedge n' \leq n \wedge L'_{n'} \Rightarrow L_n$} \quad $n' \leq n \wedge R'_{n'} \Rightarrow R_n$
}
}
{
F,A,L,R_n \vdash \speclineold{\{P\}}~e^\ast~\speclineold{\{Q\}}
}
$$

$$
\prftree[r]{[extension]}
{
\Gamma \vdash \speclineold{\{\exists \overrightarrow{x}.~(\sa_p~|~ H)\}}~e^\ast~\speclineold{\{\exists \overrightarrow{y}.~(\sa_q~|~ H')\}}\quad
}
{
 fv(\sa_k) \cap (\textit{mv}(e^\ast) \cup \overrightarrow{x}  \cup \overrightarrow{y}) = \emptyset
}
{
\Gamma \vdash \speclineold{\{\exists \overrightarrow{x}.~(\sa_k;\sa_p~|~H)\}}~e^\ast~\speclineold{\{\exists \overrightarrow{y}.~(\sa_k;\sa_q~|~H')\}}
}
$$

$$
\prftree[r]{[context]}
{
F,A,L,R^? \vdash \speclineold{\{P\}}~e^\ast~\speclineold{\{Q\}}
}
{
F,A,(L;L_f),R \vdash \speclineold{\{P\}}~e^\ast~\speclineold{\{Q\}}
}
$$
}

\vspace*{-0.2cm}
{
\small 
{\bfseries Note}: $\textit{mv}(e^\ast)$ denotes the set of local and global variables modified by the execution of $e^\ast$.
}
\caption{Proof Rules: Structural}
\label{fig:structureops}
\end{figure}

In addition to the standard strengthening of the pre-condition and weakening of the post-condition, the [consequence] rule allows us to weaken the assertions in $L$ and also the~$R$. This weakening comes with a side condition that we are not allowed to increase the number of elements on the corresponding stack, which comes from the intuition that breaking out carrying $n$ values does not necessarily imply that we can break out with $n+1$ values. The [consequence] rule uses the entailment relation of \logicname, denoted by $P \Rightarrow Q$ and defined in the standard way in \S\ref{sec:soundness}, Figure~\ref{fig:interpretation}.

The two new rules introduced for \logicname are [extension] and [context]. The [extension] rule is the analog of [frame] for stacks, and it allows us to arbitrarily extend the ``bottom'' of the stack. This, in turn, enables the proof rules of Figures~\ref{fig:basinstrs}, \ref{fig:varinstrs}, and \ref{fig:memoryops} to be generalised to arbitrary stacks, with the rules modifying only the head. The [context] rule allows us to remove unneeded assertions from $L$ and also, potentially, $R$. This rule is sound because the triple encodes that $e^\ast$, when executed, will only jump to targets in $L$, so it is trivially correct for $L$ to be further enlarged.


\begin{figure}[!t]
{
\footnotesize
\vspace*{-0.3cm}
$$
\prftree[r]{[function]}
{
\pbox{0.85\textwidth}{
$\textit{func} = \KK{func}~t^n \rightarrow t^m ~\KK{local}~t^k~e^\ast \quad \sa_n = [x_0..x_{n-1}] \quad \forall i.~l_i \notin fv(\sa_n) \cup fv(H) \cup fv(Q_m)$\\[0.1cm]
$(F,A,[\speclineold{Q_m}],\speclineold{Q_m}) \vdash \speclineold{\{\stackr{0}{}~|~H \wedge \bigwedge_{0 \leq i < n}(l_i = x_i) \wedge \bigwedge_{n \leq i < n + k}(l_{i} = 0)\}}~e^\ast~\speclineold{\{Q_m\}}$
}
}
{
F,A,L,R \vdash \speclineold{\{\sa_n~|~H\}}~\KK{callcl}~\textit{func}~\speclineold{\{Q_m\}}
}
$$

$$
\prftree[r]{[call]}
{
\speclineold{\{P\}}~\KK{call}~i~\speclineold{\{Q\}} \in A(\Gamma)
}
{
i < \llen{F(\Gamma)}
}
{
\Gamma \vdash \speclineold{\{P\}}~\KK{call}~\textit{i}~\speclineold{\{Q\}}
}
$$

$$
\prftree[r]{[specsI]}
{
\forall (\speclineold{\{P\}}~e^\ast~\speclineold{\{Q\}}) \in \textit{specs}.~\Gamma \vdash \speclineold{\{P\}}~e^\ast~\speclineold{\{Q\}}
}
{
\Gamma \Vdash \textit{specs}
}
\qquad
\prftree[r]{[specsE]}
{
\Gamma \Vdash \textit{specs}
}
{
(\speclineold{\{P\}}~e^\ast~\speclineold{\{Q\}}) \in \textit{specs}
}
{
\Gamma \vdash \speclineold{\{P\}}~e^\ast~\speclineold{\{Q\}}
}
$$

$$
\prftree[r]{[module]}
{
\pbox{0.7\textwidth}{
\mbox{$\forall \textit{spec} \in \textit{specs}.~\textit{spec} = \speclineold{\{ \_ \}}~\KK{call}~\_\speclineold{~\{\_\}}$}\\[0.3em]
\mbox{$F,\textit{specs},[],[] \Vdash \{~\speclineold{\{P\}}~\KK{callcl}~(F!j)~\speclineold{\{Q\}}~|~ \speclineold{\{P\}}~\KK{call}~j~\speclineold{\{Q\}} \in \textit{specs}~\}$}
}
}
{
F,[],[],[] \Vdash \textit{specs}
}
$$
\caption{Proof Rules: Function-Related Instructions, Modules}
\label{fig:callops}
}
\end{figure}

\subparagraph{Function-Related Instructions, Modules}
The proof rules for function-related instructions and modules are given in Fig.~\ref{fig:callops}.
We give a unified semantics to function calls in WebAssembly through the auxiliary \KK{callcl} instruction and the corresponding [function] rule, which we now explain in detail.
First, when inside a function body, if we execute \Wbr 0 at top-level or $(\KK{return})$ anywhere, the function terminates. For this reason, the context from which we start proving a function body has the labels and the return field set to the post-condition of the function $Q_m$.
Next, as previously described, the function arguments are taken from the stack. Therefore, we require the length of the stack to match the number of function parameters, $n$, given in the function definition.
Next, the $n$ arguments themselves are transferred into the first $n$ local variables ($l_0$ through $l_{n-1}$), whereas the remaining declared local variables ($l_n$ through $l_{n+k}$) are set to~0.
Finally, as local variables are declared per-function, we forbid function pre- and post-conditions from talking about local variables altogether in order to avoid name clashes.
Note that, as with [block] and [loop], the function's explicit type annotation allows us to precisely fix the stack size of both the pre- and post-condition.

At the top level, we have rules for proving specifications for sets of mutually recursive functions.
We follow the strategy described by Oheimb~\cite{Oheimb:1999:HLM:646837.708364} and Nipkow~\cite{10.1007/3-540-45793-3_8}. 
There, each individual function body is initially proven while assuming the specifications of all other functions (the [function] rule), recursive calls and calls to other functions only use the assumptions (the [call] rule), and from this, it can be concluded that all function specifications are correct without any assumptions (the [module] rule).


\section{Using \logicname: A Verified B-Tree Library}
\label{sec:verification}

We demonstrate the applicability of \logicname by specifying and verifying a simple WebAssembly B-tree library. B-trees are one of the data structures that we expect to be implemented directly in WebAssembly for efficiency reasons. In particular, a B-tree node commonly occupies an entire page of secondary storage (for example, a hard drive) and WebAssembly memory is allocated in pages.
Our B-tree implementation is underpinned by the ordered, bounded array data structure, which we use to demonstrate in detail how \logicname rules can be used in practice~(\S\ref{subsec:obas}). We focus on the two non-standard aspects of the logic: stack manipulation and the interplay between structural rules (framing, existential variable elimination, and consequence) and WebAssembly's control flow. We further describe the structure of the B-trees that we implement and present abstract specifications for some of the main B-tree operations~(\S\ref{subsec:btrees}). The full details of our B-tree implementation are available in the accompanying technical report~\cite{wasmlogic:techrep}.

\subparagraph{Additional Notation (Lists/Sets)} We denote: the empty list by $\lemp$; the list resulting from prepending an element $a$ to a list $\alpha$ by $a \lcons \alpha$; concatenation of two lists $\alpha$ and $\beta$ by~$\alpha \lcat \beta$; the length of a list $\alpha$ by $\llen{\alpha}$; the $n$-th element of a list $\alpha$ by $\alpha \lidx n$; the sublist of a list $\alpha$ starting from index $k$ and containing $n$ elements by $\sublist\alpha k n$; and the set corresponding to a list $\alpha$ by $\listToSet\alpha$. We also denote the number of elements of a set $X$ by $\card{X}$. 

\subsection{Ordered, Bounded Arrays in WebAssembly}
\label{subsec:obas}

An ordered, bounded array (OBA) is an array whose elements are ordered and which has a fixed upper bound on the number of elements it can contain. We have found OBAs to be an appropriate data structure for representing B-tree nodes, as discussed in detail in~\cite{wasmlogic:techrep}.

In separation logic, it is commonplace to describe data structures using \textit{abstract predicates} in order to abstract their implementation and simplify the textual representation of the associated proofs.\footnote{In some separation logics, abstract predicates are distinct formal entities, but in \logicname they are simply a syntactic shorthand for some particular assertion.} 
We define the abstract predicate for a 32-bit OBA at address $x$, with maximum size $n$ and contents $\alpha$, written $\oba{x}{n}{\alpha}$. Informally, the layout of OBAs in  memory, illustrated below, is as follows: the first 32-bit cell holds the length of the list $\alpha$; the next $\llen\alpha$ 32-bit cells hold the contents of the list $\alpha$; and the remaining $(n - \llen\alpha)$ 32-bit cells constitute over-allocated~space.
\begin{center}
\includegraphics[width=0.7\textwidth]{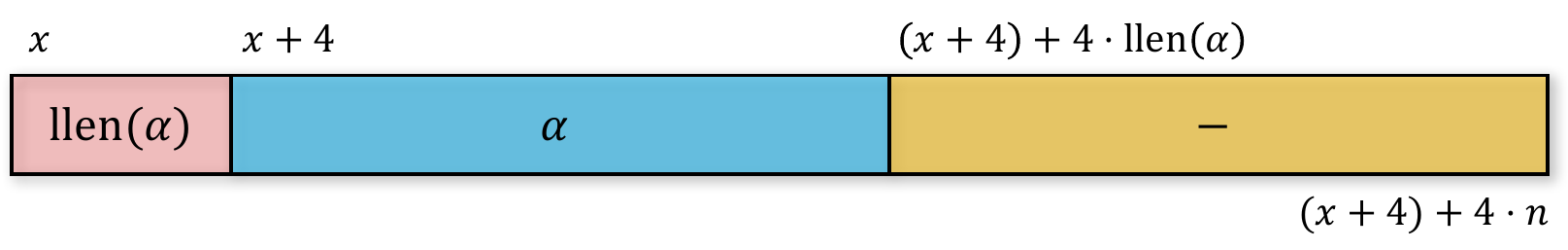}
\end{center}

Formally, the definition of the $\oba{x}{n}{\alpha}$ predicate is:
\begin{align*}
\oba{x}{n}{\alpha} :=~ & (\ittcell{x}{\llen{\alpha}} \slstar \obaseg{x + 4}{\alpha} \slstar \Star{\llen{\alpha} \leq i < n}(\ittcell{x + 4 + 4 \cdot i}{-}))\wedge \\ 
& \qquad (\ordered{\alpha} \wedge \llen{\alpha} \leq n \wedge (x + 4 + 4 \cdot n) \leq \ittmax)_\mathbb{N}),
\end{align*}
where: the predicate $\obaseg{x}{\alpha}$ describes the contents as an array segment: 
\begin{align*}
\obaseg{x}{\alpha} := \Star{0 \leq i < \llen \alpha}(\ittcell{x + 4 \cdot i}{\alpha \lidx i});
\end{align*}
the predicate $\ordered{\alpha}$ denotes that $\alpha$ is ordered in ascending order:
\begin{align*}
\ordered{\alpha} := \frall{i} 0 < i < \llen{\alpha} \Rightarrow \alpha \lidx (i - 1) \leq \alpha \lidx i;
\end{align*}
and $\ittmax$ denotes the maximal positive integer of $\K{i32}$. 
Additionally, we require that the length of the list be bounded $(\llen\alpha \leq n)$. Finally, since we are working in $\K{i32}$, we have to explicitly prevent overflow by stating that $(x + 4 + 4 \cdot n \leq \ittmax)_\mathbb{N}$. 

\subparagraph{Straight-Line Code: OBAGet}
We demonstrate the basics of proof sketches in \logicname using the example of the $\pred{OBAGet}(x, k)$ function,  specified and verified in Figure~\ref{fig:OBAgetMain}. $\pred{OBAGet}$ takes two parameters: $x$, denoting the memory address at which the OBA starts; and $k$, denoting the (non-negative) index of the OBA element to be retrieved. Assuming that $k$ does not exceed the current OBA length, the function returns the $k$-th element of the OBA.

\begin{figure}[!t]
\footnotesize
$\specline{\stackheap{x, k}{\oba{x}{n}{\alpha} \wedge 0 \leq k < \llen \alpha}}$ \\
\begin{Wfunce}[]{\pred{OBAGet}}{[\Witype, \Witype] \rightarrow [\Witype]}
  $\specline{\stackheap{}{\oba{x}{n}{\alpha} \wedge 0 \leq k < \llen \alpha \wedge l_0 = x \wedge l_1 = k}}$ \\
  $\begin{leftvruled}{frame}
  \specline{\stackheap{}{\emp}} \\
  \Wgetlocal{0} \\
  \specline{\stackheap{l_0}{\emp}} \\
  \begin{leftvruled}{extension}
  \specline{\stackheap{}{\emp}} \\
  \Wgetlocal{1} \\
  \specline{\stackheap{l_1}{\emp}} \\
  \end{leftvruled} \\
  \specline{\stackheap{l_0, l_1}{\emp}} \\
  \Wiconst{4} \\
  \specline{\stackheap{l_0, l_1, 4}{\emp}} \\
  \Wimul 
  \Wiadd \\
  \specline{\stackheap{l_0 + 4 \cdot l_1}{\emp}} 
  \end{leftvruled}$ \\
  $\specline{\stackheap{l_0 + 4 \cdot l_1}{\oba{x}{n}{\alpha} \wedge 0 \leq k < \llen \alpha \wedge l_0 = x \wedge l_1 = k}}$ \\
  $\specline{\stackheap{x + 4 \cdot k}{\oba{x}{n}{\alpha} \wedge 0 \leq k < \llen \alpha}}$ \annot{(by consequence)} \\
  \annot{[[ Unfold $\oba{x}{n}{\alpha}$ ]]} \\
  $\specline{\stackheap{x + 4 \cdot k}{(\ittcell{x}{\llen{\alpha}} \slstar \Star{0 \leq i < \llen \alpha}(\ittcell{x + 4 + 4 \cdot i}{\alpha \lidx i}) \slstar \Star{\llen{\alpha} \leq i < n}(\ittcell{x + 4 + 4 \cdot i}{-}))\wedge \\ \hspace*{1.6cm}(\ordered{\alpha} \wedge \llen{\alpha} \leq n~\wedge (x + 4 + 4 \cdot n \leq \ittmax)_\mathbb{N}) \wedge 0 \leq k < \llen \alpha}}$ \\
  $\begin{leftvruled}{frame}
  \specline{\stackheap{x + 4 \cdot k}{\ittcell{x + 4 + 4 \cdot k}{\alpha \lidx k}}} \\
  \Wiloadoff{4} \\
  \specline{\stackheap{\alpha \lidx k}{\ittcell{x + 4 + 4 \cdot k}{\alpha \lidx k}}} \\
  \end{leftvruled}$ \\
  $\specline{\stackheap{\alpha \lidx k}{(\ittcell{x}{\llen{\alpha}} \slstar \Star{0 \leq i < \llen \alpha}(\ittcell{x + 4 + 4 \cdot i}{\alpha \lidx i}) \slstar \Star{\llen{\alpha} \leq i < n}(\ittcell{x + 4 + 4 \cdot i}{-}))\wedge \\ \hspace*{0.95cm}(\ordered{\alpha} \wedge \llen{\alpha} \leq n~\wedge (x + 4 + 4 \cdot n \leq \ittmax)_\mathbb{N}) \wedge 0 \leq k < \llen \alpha}}$ \\
  \annot{[[ Fold $\oba{x}{n}{\alpha}$ ]]} \\
  $\specline{\stackheap{\alpha \lidx k}{\oba{x}{n}{\alpha} \wedge 0 \leq k < \llen \alpha}}$ 
\end{Wfunce} \\
$\specline{\stackheap{\alpha \lidx k}{\oba{x}{n}{\alpha} \wedge 0 \leq k < \llen \alpha}}$
\caption{$\pred{OBAGet}$: Specification and Verification: }
\label{fig:OBAgetMain}
\end{figure}

This example illustrates the following aspects of \logicname: the interaction between function parameters, the stack, and the local variables; basic stack and heap manipulation; basic use of the frame, extension, and consequence rules; and predicate unfolding and folding. 

In Wasm, function inputs are taken from and function outputs are put onto the stack, as specified in the pre- and post-conditions. When verifying the function body, the values of the function parameters are introduced as local variables (here, $l_0$ and $l_1$), which are propagated throughout the proof and are forgotten in the post-condition (cf. the [function] rule).

When the code being verified is straight-line, i.e.~when the labels and the return fields of the context are empty, the [frame] and [consequence] rules can be used as in standard separation logic. On the other hand, the [extension] rule, which manipulates the stack analogously to [frame] manipulating the heap, can always be applied independently of the context (to limit clutter, in Figure~\ref{fig:OBAgetMain}, we show only one use of the [extension] rule and do not show the context $\Gamma$, since it is not relevant for this particular proof). 

Predicate unfolding and folding in \logicname is standard. For example, in Figure~\ref{fig:OBAgetMain}, we have to unfold the $\pred{OBA}$ predicate and frame off the excess resource in order to isolate the $k$-th element of the OBA in the heap, perform the lookup according to the [load] rule, and then frame the resource back on and fold the predicate.

\afterpage{%
\begin{figure}
\footnotesize
$\specline{\stackheap{x, e}{\oba{x}{n}{\alpha}}}$ \\
\begin{Wfunce}[\\(\KK{locals} \Witype)]{OBAFind}{[\Witype, \Witype] \rightarrow [\Witype]}
  $\specline{\stackheap{}{\oba{x}{n}{\alpha} \wedge l_0 = x \wedge l_1 = e \wedge l_2 = 0}}$ \\
  \annot{$P_{inv} : \oba{x}{n}{\alpha} \wedge l_0 = x \wedge l_1 = e \wedge 0 \leq l_2 \leq \llen\alpha \wedge \left(\frall{j} 0 \leq j < l_2 \Rightarrow \alpha \lidx j < e \right)$ \\
  $\specline{\stackheap{}{P_{inv}}}$} \annot{(by consequence)} \\
  \begin{Wloope}[\\$(\stackheap{}{P_{inv}}) \vdash$]{}
    $\specline{\stackheap{}{P_{inv}}}$ \\
    $\begin{leftvruled}{frame}
    {\color{purple} (\stackheap{}{\emp}) \vdash}~\specline{\stackheap{}{\emp}} \\
    \Wgetlocal{2} \\
    {\color{purple} (\stackheap{}{\emp}) \vdash}~\specline{\stackheap{l_2}{\emp}} \\
    \end{leftvruled}$ \\
    $\specline{\stackheap{l_2}{P_{inv}}}$ \\
    \Wgetlocal{0} 
    \Wiload \\
    $\specline{\stackheap{l_2, \llen \alpha}{P_{inv}}}$ \\
    \Wilt \\
    \annot{$C_1 : (v = 0 \Rightarrow l_2 = \llen{\alpha}) \wedge (v \neq 0 \Rightarrow l_2 < \llen{\alpha})$} \\
    $\specline{\exsts{v}\stackheap{v}{P_{inv} \wedge C_1}}$ \\
    ${\color{purple} (\exsts v \stackheap{}{P_{inv}}) \vdash~} \specline{\exsts{v}\stackheap{v}{P_{inv} \wedge C_1}}$ \annot{(by consequence)}\\
      $\begin{leftvruled}{exists}
      \begin{minipage}{0.7\textwidth}
    ${\color{purple} (\stackheap{}{P_{inv}}) \vdash~} \specline{\stackheap{v}{P_{inv} \wedge C_1}}$ \\
    \begin{Wife}[\\$(\stackheap{}{P_{inv} \wedge C_2}), (\stackheap{}{P_{inv}}) \vdash$]{}
      $\specline{\stackheap{}{P_{inv} \wedge l_2 < \llen{\alpha}}}$ \\
      \Wgetlocal{0}
      \Wgetlocal{2} \\
      $\specline{\stackheap{x, l_2}{P_{inv} \wedge l_2 < \llen{\alpha}}}$ \\
      $\begin{leftvruled}{(S2)}
      \begin{minipage}{0.5\textwidth}
      ${\color{purple} \vdash}~\specline{\stackheap{x, l_2}{\oba{x}{n}{\alpha} \wedge 0 \leq l_2 < \llen \alpha}}$ \\
      $\Wcall{\text{OBAGet}}$ \\
      ${\color{purple} \vdash}~\specline{\stackheap{\alpha!l_2}{\oba{x}{n}{\alpha} \wedge 0 \leq l_2 < \llen \alpha}}$ 
      \end{minipage}
      \end{leftvruled}$ \\
      ${\color{purple} (\stackheap{}{P_{inv} \wedge C_2}), (\stackheap{}{P_{inv}}) \vdash}~\specline{\stackheap{\alpha!l_2}{P_{inv} \wedge l_2 < \llen{\alpha}}}$ \\
      \Wgetlocal{1} 
      \Wilt \\
      $\specline{\exsts{v}\stackheap{v}{P_{inv} \wedge l_2 < \llen{\alpha} \wedge (v = 0 \Rightarrow \alpha \lidx l_2 \geq e) \wedge (v \neq 0 \Rightarrow \alpha \lidx l_2 < e)}}$ \\
    \begin{Wife}[\\
    \annot{$C_2 : ((l_2 < \llen{\alpha} \wedge \alpha \lidx l_2 \geq e) \vee l_2 = \llen\alpha)$} 
     \\$(\stackheap{}{P_{inv} \wedge C_2}), (\stackheap{}{P_{inv} \wedge C_2}), (\stackheap{}{P_{inv}})  \vdash$]{}
         $\specline{\stackheap{}{P_{inv} \wedge l_2 < \llen{\alpha} \wedge \alpha \lidx l_2 < e}}$ \\
        \Wgetlocal{2} 
        \Wiconst{1} 
        \Wiadd \\
        $\specline{\stackheap{l_2 + 1}{P_{inv} \wedge l_2 < \llen{\alpha} \wedge \alpha \lidx l_2 < e}}$ \\
        \Wsetlocal{2} \\
        $\specline{\stackheap{}{\oba{x}{n}{\alpha} \wedge l_0 = x \wedge l_1 = e \wedge l_2{-}1 < \llen{\alpha}~\wedge \\ \tab~~ \left(\frall{j} 0 \leq j < l_2{-}1 \Rightarrow \alpha \lidx j < e \right) \wedge \alpha \lidx (l_2{-}1) < e}}$ \\
        $\specline{\stackheap{}{\oba{x}{n}{\alpha} \wedge l_0 = x \wedge l_1 = e \wedge \left(\frall{j} 0 \leq j < l_2 \Rightarrow \alpha \lidx j < e \right) \wedge l_2 \leq \llen{\alpha}}}$ \\
        $\specline{\stackheap{}{P_{inv}}}$ \\
        \Wbr{2} \\
        $\specline{\stackheap{}{P_{inv} \wedge C_2}}$ 
      \end{Wife} \\
      $\specline{\stackheap{}{P_{inv} \wedge C_2}}$ 
    \end{Wife} \\
    $\specline{\stackheap{}{P_{inv} \wedge C_2}}$ 
    \end{minipage} 
    \end{leftvruled}$ \\
    ${\color{purple} (\exsts v \stackheap{}{P_{inv}}) \vdash~} \specline{\exsts v \stackheap{}{P_{inv} \wedge C_2}}$ \\
    ${\color{purple} (\stackheap{}{P_{inv}}) \vdash~} \specline{\stackheap{}{P_{inv} \wedge C_2}}$ \annot{(by consequence)} 
  \end{Wloope} \\
  $\specline{\stackheap{}{P_{inv} \wedge C_2}}$ \\
  \Wgetlocal{2} \\
  $\specline{\stackheap{l_2}{\oba{x}{n}{\alpha} \wedge l_0 = x \wedge l_1 = e \wedge 0 \leq l_2 \leq \llen\alpha \wedge \left(\frall{j} 0 \leq j < l_2 \Rightarrow \alpha \lidx j < e \right) \wedge C_2}}$ \\
  $\specline{\exsts{i} \stackheap{i}{\oba{x}{n}{\alpha} \wedge l_0 = x \wedge l_1 = e \wedge l_2 = i \wedge 0 \leq i \leq \llen\alpha~\wedge \\ \tab\tab\tab~\left(\frall{j} 0 \leq j < i \Rightarrow \alpha \lidx j < e \right) \wedge \left(\frall{j} i \leq j < \llen{\alpha}\Rightarrow e \leq \alpha \lidx j\right)}}$ \\
\end{Wfunce} \\
$\specline{\exsts{i}\stackheap{i}{\oba{x}{n}{\alpha} \wedge 0 \leq i \leq \llen\alpha \wedge \left(\frall{j} 0 \leq j < i \Rightarrow \alpha \lidx j < e \right) \wedge \left(\frall{j} i \leq j < \llen{\alpha}\Rightarrow e \leq \alpha \lidx j\right) }}$ 
\caption{OBAFind: Specification and Verification}
\label{fig:OBAfindMain}
\end{figure}%
\clearpage%
}

\subparagraph{Conditionals and Loops: OBAFind}
We demonstrate how to reason about WebAssembly conditionals and loops in \logicname using the example of the $\pred{OBAFind}(x, e)$ function, specified and verified in Figure \ref{fig:OBAfindMain}. $\pred{OBAFind}$ takes two parameters: $x$, denoting the memory address at which the OBA starts; and $e$, a 32-bit integer. The function returns the index $i$ of the first element of the OBA that is not smaller than $e$, or $\llen \alpha$ if such an element does not exist. The index~$i$ effectively tells us the position in the OBA at which either $e$ appears for the first time or would be inserted.

This example addresses, among other things, the following features of \logicname: interaction between conditionals, loops, and the break statement; advanced use of the frame, existential elimination, and consequence rules; and function calls. To focus on these features, we elide previously discussed details, such as predicate management, from the proof sketch.

First, observe how local variables are initialised. The function itself expects two parameters, as given by the type of the function (cf. the [function] rule). These form the first two local variables. The explicitly declared local variables, starting from index 2, are initialised to zero.

The body of the function is a loop that uses the local variable $l_2$ to iterate over the OBA and find its first element that is not smaller than $e$. First, the loop checks if $l_2$ is smaller than the length of the OBA. If it is, the loop terminates (by reaching the loop end), and we know that all of the elements of the OBA are smaller than $e$. Otherwise, it checks if the $l_2$-nd element of the OBA is smaller than $e$. If it is, the loop terminates, and we know that we have found an element not smaller than $e$ in the OBA. Otherwise, $l_2$ is incremented and the loop restarts (by executing the break instruction).

For the loop construct, we establish the appropriate invariant, $(\stackheap{}{P_{inv}})$, using the [consequence] rule in the standard way. This invariant essentially states that all of the previously examined elements are smaller than $e$. Then, following the [loop] rule, we verify the body of the loop while extending the labels field of the context with the invariant. We explicitly state modifications to the context at the point at which they first occur.

As soon as the labels or the return field of the context is not empty, the use of the frame and existential elimination becomes more involved. For example, when framing off, we have to frame off not only from the current state, but also from all of the labels, as well as from the return assertion. We illustrate this in Figure \ref{fig:OBAfindMain}, using the first instruction of the loop body, \Wgetlocal 2, where we have to frame off $P_{inv}$ both from the state and the labels of the context in order to apply the [get\_local] rule. 

In the general case, however, the label assertions, the return assertion, and the state need not match in resource, meaning that the [frame] rule may be unable to manipulate the label/return context. In practice, we have identified two strategies for handling this issue: \dtag{S1} specialising ``falsey'' labels/return via the [consequence] rule; or \dtag{S2} adjusting the context via the [context] rule. 

We illustrate the first strategy using the following derivation tree:
$$
\footnotesize
\prftree[r]{[cons]}
{
	\begin{minipage}{0.45\textwidth}
	\prftree[r]{[frame]}
	{-, -, [(S_1 \mid \bot), (S_2, \bot)], (S_R, \bot) \vdash \triple {S_P \mid P} {e^\ast} {S_Q \mid Q}}
	{\begin{array}{l}{-,-,[(S_1 \mid \bot \slstar F), (S_2 \mid \bot \slstar F)], (S_R \mid \bot \slstar F) \vdash} \\ {\qquad\qquad\qquad\qquad\qquad\triple{S_P \mid {P \slstar F}}{e^\ast}{S_Q \mid {Q \slstar F}}}\end{array}}
	\end{minipage}
	\quad
	\begin{minipage}{0.34\textwidth}
	$(S_1 \mid \bot \slstar F) \Rightarrow (S_1 \mid H_1)$ \\
	$(S_2 \mid \bot \slstar F) \Rightarrow (S_2 \mid H_2)$ \\
	$(S_R \mid \bot \slstar F) \Rightarrow (S_R \mid H_R)$
	\end{minipage}
}
{
-,-,[(S_1 \mid H_1), (S_2 \mid H_2)], (S_R \mid H_R) \vdash \triple{{S_P \mid P \slstar F}}{e^\ast}{{S_Q \mid  Q \slstar F}}
}
$$

This strategy takes advantage of the fact that if $e^\ast$ never actually executes (for example) $\Wbr{n}$, then $L!n$ can have a $\bot$ component, allowing the manufacturing of any frame through application of the [consequence] rule.

An example of the second strategy works as follows:
$$
\footnotesize
\prftree[r]{[context]}
{
	\begin{minipage}{0.5\textwidth}
	\prftree[r]{[frame]}
	{-, -, [], \mathtt{None} \vdash \triple {S_P \mid P} {e^\ast} {S_Q \mid Q}}
	{-,-,[], \mathtt{None} \vdash \triple{S_P \mid {P \slstar F}}{e^\ast} {S_Q \mid {Q \slstar F}}}
	\end{minipage}
}
{
-,-,[L_1, L_2], R \vdash \triple{{S_P \mid P \slstar F}}{e^\ast} {{S_Q \mid  Q \slstar F}}
}
$$
Here, we use the [context] rule to temporarily remove all of the labels and the return, allowing us to frame off only from the state. This strategy be seen in action immediately before the function call to OBAGet in Figure~\ref{fig:OBAfindMain}.

Both  strategies can normally be applied before any non-break, non-return instruction, although the second strategy is preferred.
However, there are occasions where the first strategy must be used. For example, if $e^\ast$ executes $\Wbr{1}$, then $L \lidx 0$ can no longer be removed by [context].
However, it can still be falsified, allowing the first approach.


Existential elimination is another fundamental separation logic rule that needs to consider the context in \logicname and can only be applied if all of the labels, the return, and the state have the same leading existential variable(s). This requirement can normally be established via the [consequence] rule and can be used regardless of the context and the position in the code. For example, consider the following part of the proof derivation for the first $\KK{if}$ statement of OBAFind (cf. Figure~\ref{fig:OBAfindMain} for more details):
$$
\footnotesize
\prftree[r]{[cons]}
{
	\begin{minipage}{0.5\textwidth}
	\prftree[r]{[exists]}
	{-,-,[(\stackheap{}{P_{inv}})], \mathtt{None} \vdash \triple{\stackheap{v}{P_{inv} \wedge C_1}}{(\KK{if} \ldots \KK{end})}{\stackheap{}{P_{inv} \wedge C_2}}}
	{-,-,[(\exsts v \stackheap{}{P_{inv}})], \mathtt{None} \vdash \triple{\exsts v \stackheap{v}{P_{inv} \wedge C_1}}{(\KK{if} \ldots \KK{end})}{\exsts v \stackheap{}{P_{inv} \wedge C_2}}}
	\end{minipage}
}
{
-,-,[(\stackheap{}{P_{inv}})], \mathtt{None} \vdash \triple{\exsts v \stackheap{v}{P_{inv} \wedge C_1}}{(\KK{if} \ldots \KK{end})}{\stackheap{}{P_{inv} \wedge C_2}}
}
$$ 

Here, we use [consequence] to add the existential~$v$ directly to the label (possible because $v$ is not featured in $P_{inv}$) and remove it from the obtained post-condition (possible because $v$ is not featured in $R_2$). In cases where this direct approach would lead to variable capture, we would have an additional first step of renaming the existentials appropriately.

In the first $\KK{if}$ statement of OBAFind, we also encounter a call to the OBAGet function.  In \logicname, function calls are handled in the standard way, meaning that frame and consequence are used first to isolate the appropriate pre-condition from the current state and then to massage the obtained post-condition into a desired form. For simplicity, in the code we call the functions by name, rather than by index.

Finally, we comment on the treatment of break statements, using the example of the \Wbr {2} statement seen in OBAFind. Given the [br] rule, the pre-condition of that break statement must match the loop invariant $(\stackheap{}{P_{inv}})$, which we establish. The post-condition, however, is left free in the [br] rule, and has to be chosen correctly so that the subsequent derivation makes sense. Observe that, due to the design of WebAssembly, any code found between a break statement and the end of the block of code in which it is found is dead code. In our case, this means that we never reach the exit of that $\KK{if}$ branch---instead, we unconditionally jump to the head of the main loop. The only way to reach the end of that $\KK{if}$ statement is if the test of that $\KK{if}$ yields zero, in which case our state would be $(\stackheap{}{P_{inv} \wedge C_2})$. Now, since the [if] rule requires the final states from both branches to be the same, we can choose precisely $(\stackheap{}{P_{inv} \wedge C_2})$ to be the post-condition of the break statement. More generally, a safe option is to always choose the post-condition of a break statement to be $(\stackheap{}{\bot})$, and from there derive any required assertion using the [consequence] rule.

\subparagraph{Additional OBA Functions}
In order to support basic B-tree operations, we also need to be able to insert/delete elements into/from an OBA. Moreover, as B-tree keys are unique (cf.~\S\ref{subsec:btrees}), we strengthen the OBA predicate to enforce non-duplication of elements:
$$
\oband{x}{n}{\alpha} := \oba{x}{n}{\alpha} \wedge \llen\alpha = \card{\listToSet\alpha}.
$$
Note that the previously presented OBA functions, $\pred{OBAGet}$ and $\pred{OBAFind}$, can also be used with an $\pred{OBA_{nd}}$. We give the specifications of $\pred{OBAInsert}$ and $\pred{OBADelete}$ in Figure~\ref{fig:btreespecs} (left). Their corresponding proof sketches are available in~\cite{wasmlogic:techrep}.

\subsection{B-Trees in WebAssembly}
\label{subsec:btrees}

\begin{figure}[t]
\begin{minipage}[t]{0.48\textwidth}
\small
$\specline{\stackheap{x, e}{\oband{x}{n}{\alpha} \wedge \llen \alpha < n}}$ \\
\mbox{(\KK{func}~OBAInsert~$[\Witype, \Witype] \rightarrow []~\ldots~$ \KK{end})} \\
$\specline{\exsts{\alpha'} \stackheap{}{\oband{x}{n}{\alpha'}~\wedge \\ \hspace{1.15cm}\listToSet{\alpha'} = \listToSet{\alpha} \cup \{e\}}}$ \\[1em]

$\specline{\stackheap{x, e}{\oband{x}{n}{\alpha}}}$ \\
\mbox{(\KK{func}~OBADelete~$[\Witype, \Witype] \rightarrow []~\ldots~$ \KK{end})} \\
$\specline{\exsts{\alpha'} \stackheap{}{\oband{x}{n}{\alpha'}~\wedge \\ \hspace{1.15cm}\listToSet{\alpha'} = \listToSet{\alpha} \setminus \{e\}}}$
\end{minipage}
~
\begin{minipage}[t]{0.48\textwidth}
\small
$\specline{\stackheap{t}{\Wsize 0} \wedge 2 \leq t \leq 4095}$ \\
\mbox{(\KK{func}~$\pred{BTreeCreate}~[\Witype] \rightarrow [] \ldots$ \KK{end})}  \\
$\specline{\stackheap{}{\btree t \emptyset} \wedge 2 \leq t \leq 4095 }$ \\[1em]
$\specline{\stackheap{k}{\btree t \kappa}}$ \\
\mbox{(\KK{func}~$\pred{BTreeSearch}~[\Witype] \rightarrow [\Witype] \ldots$ \KK{end})} \\
$\specline{\exsts {b} \stackheap{b}{\btree t \kappa}~\wedge \\ \qquad\quad~(k \in \kappa \Rightarrow b = 1) \wedge (k \notin \kappa \Rightarrow b = 0) }$ \\[1em]
$\specline{\stackheap{k}{\btree t \kappa}}$ \\
\mbox{(\KK{func}~$\pred{BTreeInsert}~[\Witype] \rightarrow [\Witype] \ldots$ \KK{end})} \\
$\specline{\stackheap{}{\btree t {\kappa \cup \{ k \}}} }$
\end{minipage}
\caption{Specifications of: OBAInsert/OBADelete (left); B-Tree operations (right)}
\label{fig:btreespecs}
\end{figure}

B-trees are self-balancing tree data structures that allow search, sequential access, insertion, and deletion in logarithmic time. They generalise binary search trees in that a node of a B-tree can have more than two children. B-trees are particularly well-suited for storage systems that manipulate large blocks of data, such as hard drives, and are commonly used in databases and file systems~\cite{algorithms}. 

Every node $x$ of a B-tree contains: an indicator denoting whether or not it is a leaf,~$\lambda$; the number of keys that it holds, $n$; and the $n$ keys themselves, $\kappa_1, \ldots \kappa_{n}$. Additionally, each non-leaf node contains $n + 1$ pointers to its children, $\pi_1, \ldots, \pi_{n+1}$. 

The number of keys that a B-tree node may have is bounded. These bounds are expressed in terms of a fixed integer $t \geq 2$, called \emph{the branching factor} of the B-tree. In particular, every node except the root must have at least $t-1$ keys, and every node must have \emph{at most} $2t - 1$ keys. Moreover, if a B-tree is non-empty, the root must have at least one key. Finally, all of the leaves of the B-tree have the same depth.

The keys of a B-tree are ordered, in the sense that the keys of every node are ordered (for us, in ascending order), and that every key of a non-leaf node is greater than all of the keys of its left child and smaller than all of the keys of its right child.

As an illustrative example, in Figure~\ref{fig:btreeexample} we show a B-tree with branching factor $t = 2$ that contains all prime numbers between 1 and 100. It has 25 keys distributed over 12 nodes, with every node having at least $t-1 = 1$ and at most $2t-1 = 3$ keys. 

\begin{figure}[!h]
\centering
\includegraphics[width=0.95\textwidth]{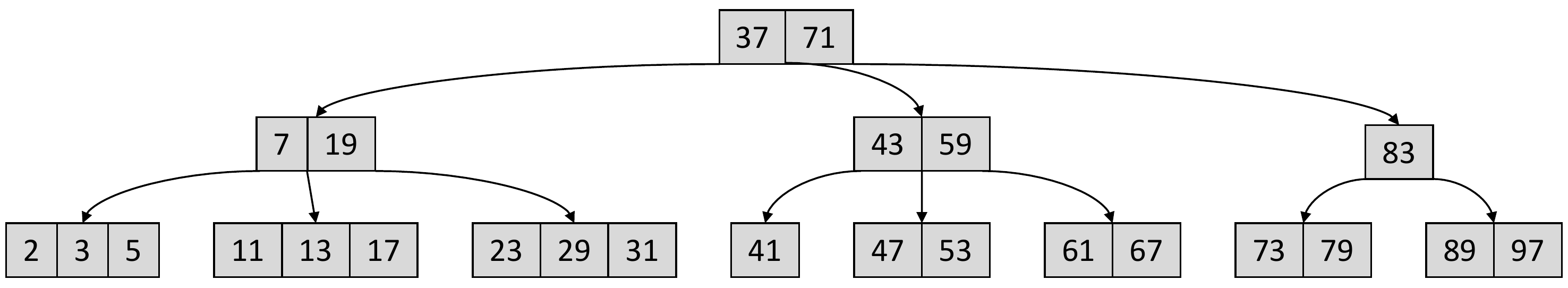}
\caption{Prime numbers from 1 to 100 in a B-tree of branching factor two, with the $\lambda$ and $n$ parameters of the nodes elided.}
\label{fig:btreeexample}
\end{figure}

Onward, we describe the layout of a B-tree in WebAssembly memory, define the associated predicates, and show the specifications for B-tree creation, search, and insertion, implemented based on the algorithms and auxiliary functions  in~\cite{algorithms}. 
The implementations are available, together with their accompanying proof sketches, in full in~\cite{wasmlogic:techrep}. 

\subparagraph{B-Tree Metadata Page}
The first page of memory is reserved for keeping track of information about the state of the module.
For example, one aspect of module state are the addresses of ``free'' pages where nodes can be allocated, and another is the root node address.

We first define what it means to be a page in memory with (non-negative integer) index~$n$:
\begin{align*}
\page{n} := \Star{n \cdot 64k \leq i < (n+1) \cdot 64k}(\ittcell{i}{-}) \land 0 \leq n \land ((n+1) \cdot 64k \leq \ittmax)_\mathbb{N}
\end{align*}
Next, we define the predicate capturing the free pages, $\free{\varphi}$, which stores the list of free pages, $\varphi$, in an $\pred{OBA_{nd}}$, and confers ownership of all of the pages in $\varphi$.
The $\pred{OBA_{nd}}$ length $(64k/4-3 = 16381)$ is chosen to ensure that it can never overflow over the bounds of the metadata page, taking into account the two first elements of the page as well as the length of the array itself that is stored in the $\pred{OBA_{nd}}$.
\begin{align*}
\free{\varphi} := \oband{8}{16381}{\varphi} \Star{0 \leq i < \llen \varphi}(\page{\varphi \lidx i});
\end{align*}
The full metadata predicate, $\meta t r l \varphi$, describes the metadata page layout: $t$ denotes the branching factor of the B-tree; $r$ denotes the address of its root; $\mu$ denotes the current memory size in pages; and $\varphi$ denotes the list of free pages. 
\begin{align*}
\meta t r \mu \varphi := \ittcell{0}{t} \slstar \ittcell{4}{r} \slstar \Wsize{\mu} \slstar \free{\varphi}.
\end{align*}

\subparagraph{B-Tree Nodes} We next show the definition of the abstract predicate $\node x \lambda \kappa \pi$, which captures a B-tree node at page $x$, with leaf indicator $\lambda$, keys $\kappa$, and pointers $\pi$.
A B-tree node takes up an entire WebAssembly page in memory, which can hold 16384 32-bit integers. The first 32-bit integer of the page is the leaf indicator (non-zero means non-leaf); the next 8191 32-bit integers hold information about the node keys; and the last 8192 32-bit integers hold information about the node pointers. The associated predicates are defined as follows:
\begin{align*}
\nkeys{x}{\kappa} :=~& \oband{x \cdot 64k + 4}{8090}{\kappa}; \\
\nptrs{x}{\pi} :=~& \ba{x \cdot 64k + 32k}{8091}{\pi}; \\
\node x \lambda \kappa \pi :=~& \ittcell{x \cdot 64k}{\lambda} \slstar \nkeys{x}{\kappa} \slstar \nptrs{x}{\pi}.
\end{align*}

Note that, since the pointers need not be ordered, we describe them using use a simpler bounded array predicate, $\ba x n \alpha$, whose definition is the same as that of the $\pred{OBA}$ predicate given in \S\ref{subsec:obas}, but without the ordering requirement. Recall also that the OBAs and BAs come with a leading 32-bit integer capturing their length, meaning that the maximum number of keys/pointers our B-tree node can hold is 8090/8091 and that the maximal branching factor of our B-trees is 4095.
%
%
%
%

\subparagraph{B-Tree Definition and Operations} Finally, we define an abstract predicate, $\btree t \kappa$, capturing a WebAssembly B-Tree with branching factor $t$ and set of keys $\kappa$:
\begin{align*}
\btree t \kappa \triangleq~& \exsts{r, \mu, \varphi, \lambda, \phi} \meta t r \mu \varphi \slstar \btreerec r \kappa \lambda \phi.
\end{align*}
Due to lack of space, the full definition of the $\pred{BTreeRec}$ predicate is shown and explained in detail in~\cite{wasmlogic:techrep}. Informally, $\btreerec {r'} \kappa \lambda \phi$ captures a subtree of a B-tree with branching factor $t$, root $r$, in a memory of size $\mu$. This subtree has root $r'$ and set of keys $\kappa$. Additionally, the B-tree node at $r'$ is a leaf iff $\lambda \neq 0$ and is full iff $\phi \neq 0$.

In Figure~\ref{fig:btreespecs} (right), we give the specifications of WebAssembly functions for basic B-tree operations: creation; search; and insertion. The specifications are abstract, in that they do not reveal any detail of the underlying implementations. 


\section{Soundness}
\label{sec:soundness}

The semantic interpretation of our triple and the accompanying soundness proof are informed by the approaches of de Bruin~\cite{deBruin1981} and Oheimb~\cite{Oheimb:1999:HLM:646837.708364}.
The former gives us a semantics for \KK{goto} which we use as the foundation for WebAssembly's \KK{br} and \KK{return} instructions.
The latter gives us a strategy for handling mutual recursion.

Interpretation is defined against an \emph{abstract variable store}, $\store \in \stores$. Abstract variable stores are finite partial mappings from variables to constants: $\stores \equiv \vars \rightharpoonup \consts$. 

\begin{figure}[!t]
\small
\begin{minipage}{0.49\textwidth}
$$
\begin{array}{r@{~}c@{~}l}
\multicolumn{3}{c}{\textbf{Interpretation of terms} \vspace*{0.2cm}} \\
\multicolumn{3}{c}{\interp{\cdot} ::  \terms \Rightarrow \stores \Rightarrow \consts \vspace*{0.2cm}} \\
\interp{c}(\store) & \triangleq & c \\
\interp{\nu}(\store) & \triangleq & \store(\nu) \\
\interp{f(\tau_1, \ldots, \tau_n)}(\store) & \triangleq & f(\interp{\tau_1}(\store), \ldots, \interp{\tau_n}(\store)) 
\end{array}
$$
\end{minipage}
~
\begin{minipage}{0.49\textwidth}
$$
\begin{array}{r@{~}c@{~}l@{}}
\multicolumn{3}{c}{\textbf{Interpretation of stack assertions}\vspace*{0.2cm}} \\
\multicolumn{3}{c}{\interp{\cdot} :: \terms \text{ list} \Rightarrow \stores \Rightarrow \consts \text{ list} \vspace*{0.2cm}} \\
\interp{~[]~}(\store) & \triangleq & [] \\
 \interp{~\mathcal{S} :: \tau~}(\store) & \triangleq & \interp{~\mathcal{S}~}(\store) :: \interp{\tau}(\store) 
\end{array}
$$
\end{minipage}

\begin{minipage}{0.38\textwidth}
$$
\begin{array}{r@{~}c@{~}l}
\multicolumn{3}{c}{\textbf{Abstract heaps}\vspace*{0.15cm}} \\
\multicolumn{3}{c}{\textit{size} ::= \bullet~\pmb{|}~\textit{i32}} \\[0.2cm]
\multicolumn{3}{c}{\heaps ::= (\textit{i32} \rightharpoonup \textit{byte}) \times \textit{size}}\\
(\heaplet, \bullet) \uplusbullet (\heaplet', \bullet) &\triangleq& (\heaplet \uplus \heaplet', \bullet) \\
(\heaplet, \bullet) \uplusbullet (\heaplet', n) &\triangleq& (\heaplet \uplus \heaplet', n) \\
(\heaplet, n) \uplusbullet (\heaplet', \bullet) &\triangleq& (\heaplet \uplus \heaplet', n) \\
\end{array}
$$

\smallskip
{\bfseries Note}: the two last cases require that 
\vspace*{-0.2cm}
$$\forall i \in dom(\heaplet) \uplus dom(\heaplet').~i < n * 64k$$
\end{minipage}
\begin{minipage}{0.61\textwidth}
$$
\begin{array}{r@{~}c@{~}l}
\multicolumn{3}{c}{\textbf{Interpretation of pure/heap assertions}\vspace*{0.2cm}} \\
\interp{\cdot} & :: & \hasrts \Rightarrow \stores \Rightarrow \heaps\text{ set} \vspace*{0.2cm} \\
\interp{\bot}(\store) & \triangleq & \emptyset \\
\interp{\tau_1 = \tau_2}(\store) & \triangleq & \{~h~|~\interp{\tau_1}(\store) = \interp{\tau_2}(\store)~\} \\
\interp{\tau_1 \mapsto \tau_2}(\store) & \triangleq & \{~(\interp{\tau_1}(\store) \mapsto \interp{\tau_2}(\store), \bullet)~\} \\
\interp{\tau_1 \wedge \tau_2}(\store) & \triangleq & \interp{\tau_1}(\store) \cap \interp{\tau_2}(\store) \\
\interp{\neg H}(\store) & \triangleq & (\interp{H}(\store))^c\\
\interp{\exists x.~ H}(\store) & \triangleq & \{~h~|~\exists c.~h \in \interp{H}(\store[x \mapsto c])~\}\\
\interp{p(\tau_1, \ldots, \tau_n)}(\store) & \triangleq & \{~h~|~p(~\interp{\tau_1}(\store), \ldots, \interp{\tau_n}(\store)~)~\} \\
\interp{H * H'}(\store) & \triangleq & \{~h_1 \uplusbullet h_2~|~h_1 \in \interp{H}(\store),~h_2 \in \interp{H'}(\store)~\}\\
\interp{\textbf{size}(\tau)}(\store) & \triangleq & \{~(\emptyset, \interp{\tau}(\store))~\}
\end{array}
$$
\end{minipage}

\begin{minipage}{0.59\textwidth}
$$
\begin{array}{r@{~}c@{~}l}
\multicolumn{3}{c}{\textbf{Interpretation of assertions}\vspace*{0.2cm}} \\[0.2cm]
\interp{\cdot} & :: & \asrts \Rightarrow \stores \Rightarrow (\consts \text{ list} \times \heaps)~\text{set} \\
\interp{~\mathcal{S}~|~H~}(\store) & \triangleq & \{(v^\ast,h)~|~ v^\ast = \interp{\mathcal{S}}(\store), h \in \interp{H}(\store)\} \\
\interp{~\exists x.~P~}(\store) & \triangleq & \{(v^\ast,h)~|~ \exists x.~(v^\ast,h) \in \interp{P}(\store[x \mapsto c])\} \\
\end{array}
$$
\end{minipage}
~
\begin{minipage}{0.4\textwidth}
$$
\begin{array}{r@{~}c@{~}l}
\multicolumn{3}{c}{\textbf{Entailment}\vspace*{0.2cm}} \\
P \Rightarrow Q & \triangleq & \frall \store \interp{P}(\store) \subseteq \interp{Q}(\store)
\end{array}
$$
\end{minipage}
\caption{Interpretations of Terms and Assertions}
\label{fig:interpretation}
\end{figure}

Defining interpretation for terms and stack assertions is straightforward. On the other hand, interpretation of heap assertions is more involved. In traditional separation logic~\cite{Reynolds:2002:SLL:645683.664578}, ownership and existence of memory locations are conflated to simplify the soundness proof. This, however, cannot be done for WebAssembly: in the concrete WebAssembly linear memory, the existence of the addressable location $x+1$ implies that the addressable location~$x$ also exists. However, asserting ownership of location $x+1$ should not imply ownership of~$x$.

To address this, we define a two-stage interpretation of heap assertions. We first define their interpretation into a set of abstract heaps, $\heaps$. An abstract heap, $\heap \in \heaps$, is a map from locations to bytes that additionally keeps track of the memory size, which may be fixed by ownership of the \textbf{size} resource.
The \textbf{size} resource can be thought of as tracking the state of memory allocation, with ownership of \textbf{size} implying permission to perform allocations through \KK{mem.grow}, similarly to the ``free set'' resource of~\cite{10.1007/978-3-540-78499-9_15}.
Each abstract heap that is a member of the assertion interpretation represents a possible set of owned locations. Our separation algebra is defined over abstract heaps, as shown in Figure~\ref{fig:interpretation}.

Before describing the second, \emph{reification} stage, we recall the definition of \emph{instances} and \emph{WebAssembly stores} as defined in the official WebAssembly specification~\cite{Haas:2017:BWU:3062341.3062363} (the table fields are elided as they are only used by \KK{call\_indirect}):
$$
\small
\begin{array}{l@{\;}c@{\;}l@{\;}rl}
\textit{s} &::=& \{ & \text{funcs:} & \textit{func list} \\
&&&\text{mems:} & \textit{mem list} \\
&&&\text{globs:} & \textit{glob list} ~ \}\\
\end{array}
\qquad
\begin{array}{l@{\;}c@{\;}l@{\;}rl}
\textit{inst} &::=& \{ & \text{faddrs:} & \textit{nat list} \\
&&& \text{maddr:} & \textit{nat option} \\
&&& \text{gaddrs:} & \textit{nat list} ~ \}\\
\end{array}
\qquad
\begin{array}{rcl}
\textit{locs} &::=& \textit{$\consts$ list} \\
\textit{labs} &::=& \textit{nat list} \\
\textit{ret} &::=& \textit{nat option} \\
\end{array}
$$

The reification stage further relates abstract heaps to WebAssembly stores, giving the concrete WebAssembly memories that are consistent with the \textbf{size} resource, such that all owned locations exist. Store reification is defined between a WebAssembly store, instance, abstract heap, abstract variable store, and function list, as follows:
$$
\small
\prftree[r]{rei$_{sto}$}
{
\pbox{0.55\textwidth}{
\centering
\mbox{$\forall i.~F!i = \text{funcs}(s)!((\text{faddrs}(inst))!i)$}
\mbox{$\forall (i,c) \in \text{fst}(h).~c = (\text{mems}(s)!(\text{maddr}~inst))!i$}
\mbox{$\text{snd}(h) \neq \bullet \Longrightarrow \textit{pages}((\text{mems}(s)!(\text{maddr}~inst))) = \text{snd}(h)$}
\mbox{$\forall (g_i, c) \in \store.~ c = \text{globs}(s)!((\text{gaddrs}(inst))!i)$}
}
}
{
\textit{reifies}_{sto}(\textit{s, inst, h, $\store$, F})
}
$$

We also define reification for local variables, labels, and returns:
$$
\small
\prftree[r]{rei$_{loc}$}
{
\forall (l_i, v) \in \store.~ v = \textit{locs}!i
}
{
\textit{reifies}_{loc}(\textit{locs, $\store$})
}
~~
\prftree[r]{rei$_{lab}$}
{
\forall i.~(L!i = P_n) \Longleftrightarrow (\textit{labs}!i = n)
}
{
\textit{reifies}_{lab}(\textit{labs, L})
}
~~
\prftree[r]{rei$_{ret}$}
{
(R = R_n) \Longleftrightarrow (\textit{ret} = n)
}
{
\textit{reifies}_{ret}(\textit{ret, R})
}
$$

%

%
%
%
%

\subparagraph{Semantic Interpretation} We define the semantic interpretation of \logicname triples in Figure~\ref{fig:semanticinterp}.
We say that a triple $(s,\x{locs},v^\ast)$ satisfies an assertion $P$ if its members can be reified from a member of the interpretation of $P$.
The judgement \mbox{$F,L,R \vDash \speclineold{\{P\}}~e^\ast~\speclineold{\{Q\}}$} means, intuitively, that for all triples $(s,\x{locs},v^\ast_e)$ that satisfy $P$, executing $(s,\x{locs}, (v\!^{f\ast}_e) (v^\ast_e) e^\ast)$ to completion will result in a triple $(s',\x{locs}',\x{res})$ with the following properties:
if \x{res} is of the form  $\KK{Normal}~v^\ast$, then $(s',\x{locs}', {v}^\ast)$ satisfies $Q$;
if \x{res} is of the form $\KK{Break}~i~v^\ast$, then $(s',\x{locs}', {v}^\ast)$ satisfies $L!{i}$;
if \x{res} is of the form $\KK{Return}~v^\ast$, then $(s',\x{locs}', {v}^\ast)$ satisfies $R$.

Note that framing is featured in three places in the definition: in the heap~$(h^f)$; in the stack~$(v^{f\ast})$; and in the labels $(\x{labs}^f)$. 
The heap frame is treated in the standard way. 
The stack frame remains in the case of a \KK{Normal} result, but is discarded in case of the \KK{Break} and \KK{Return} results automatically, by WebAssembly's semantics. Finally, the labels frame encodes that the full label context during reduction may be arbitrarily large, but that only the initial labels \x{labs} will be targeted by the \KK{br} instructions present in $e^\ast$.


\begin{figure}[!t]
\small
$$
\begin{array}{l}
F,L,R \vDash \speclineold{\{P\}}~e^\ast~\speclineold{\{Q\}} \triangleq \frall {s, \textit{locs}, v^\ast, \textit{labs}, \textit{labs}^f, v^{f\ast}, h, h^f, \store, \textit{ret}, s', \textit{locs}', \textit{res}}  (v^\ast,h) \in \interp{P}(\store)~\wedge\\
\quad \textit{reifies}_s({s, \textit{inst}, h \uplusbullet h^f, \store, F}) \wedge \textit{reifies}_{\textit{loc}}({\textit{locs}, \store}) \wedge \textit{reifies}_{lab}({labs, L}) \wedge \textit{reifies}_\textit{ret}(\textit{ret}, R)~\wedge \\
\qquad (s, \textit{locs}, v^{f\ast}_\x{e}v^\ast_\x{e}e^\ast) \Downarrow^{(\textit{labs};\textit{labs}^f), \textit{ret}}_{\textit{inst}}~({s', \textit{locs}', \textit{res}}) \Longrightarrow \\
 \qquad \quad \textit{res} \neq \KK{Trap}~\wedge \\
 \qquad \quad \exsts {h', \store'} \textit{reifies}_s(\textit{$s'$, inst, $h' \uplusbullet h^f$, $\store'$, F}) \wedge \textit{reifies}_{loc}(\textit{locs$'$, $\store'$})~\wedge \\
 \qquad\qquad(\textit{res} = \KK{Normal}~v^\ast \Rightarrow \exists {v'}^\ast.~{v}^\ast = v^{f\ast}{v'}^\ast \wedge ({v'}^\ast,h') \in \interp{Q}(\store'))~\wedge \\
 \qquad\qquad(\textit{res} = \KK{Break}~i~v^\ast \Rightarrow ({v}^\ast, h') \in \interp{L!{i}}(\store'))~\wedge \\
 \qquad\qquad(\textit{res} = \KK{Return}~v^\ast \Rightarrow ({v}^\ast, h') \in \interp{R}(\store')) \\[0.2cm]
 F,L,R ~{\footnotesize \ddtstile{}{}}~ \textit{specs} \triangleq (\forall (\speclineold{\{P\}}~e^\ast~\speclineold{\{Q\}}) \in \textit{specs}.~F,L,R \vDash \speclineold{\{P\}}~e^\ast~\speclineold{\{Q\}})
 \\[0.2cm]
F,A,L,R~{\footnotesize \ddtstile{}{}}~\textit{specs} \triangleq (F,[],\epsilon ~{\footnotesize \ddtstile{}{}}~ A \Rightarrow F,L,R ~{\footnotesize \ddtstile{}{}}~ \textit{specs})
\end{array}
$$
\caption{Semantic interpretation of the specification triple.}
\label{fig:semanticinterp}
\end{figure}

\subparagraph{Soundness} We now state our soundness result, fully mechanised in Isabelle/HOL.
\begin{theorem1}[inference\_rules\_sound]
\mbox{$ \Gamma\Vdash \textit{specs} \Longrightarrow \Gamma~{\footnotesize \ddtstile{}{}}~\textit{specs}$}
\end{theorem1}


\section{Related Work}
\label{sec:relwork}

WebAssembly's official specification is given as a pen-and-paper formal semantics~\cite{Haas:2017:BWU:3062341.3062363, Rossberg:2018:BWU:3293542.3282510}, a large core of which has been mechanised in Isabelle~\cite{Watt:2018:MVW:3176245.3167082}.
Our mechanised soundness results build on this existing mechanisation.
CT-Wasm~\cite{Watt:2019:CTS:3302515.3290390} is a proposed cryptographic extension to WebAssembly's type system that protects against side-channel and information flow leaks.
Aside from this, research on WebAssembly has focussed mainly on dynamic analysis.
Wasabi~\cite{Lehmann:2019:WFD:3297858.3304068} is a general purpose framework for dynamic analysis.
Other work has focussed on taint tracking and binary instrumentation~\cite{1802.01050,1807.08349}; and the detection of unauthorised WebAssembly-based cryptocurrency miners~\cite{10.1007/978-3-319-98989-1_7, 1808.09474}.

\subparagraph{Control Flow}
Our proof rules for \logicname's break/continue-to-block-style semi-structured control flow take inspiration from the program logic for ``structured \KK{goto}'' proposed by Clint and Hoare~\cite{Clint1972} and first proven sound by de Bruin~\cite{deBruin1981}.
These works use a traditional Hoare Logic based on first-order logic; we have adapted their approach to our \logicname. In doing so, we have observed that the existential elimination and consequence rules of Hoare logic, and the frame rule of separation logic, require modification, as detailed in \S\ref{sec:proof_system}.

Huisman and Jacobs~\cite{10.1007/3-540-46428-X_20} describe an early Hoare logic for Java, and their treatment of Java's \KK{break} and \KK{continue} statements in their operational semantics is similar to our use of the \KK{Break} and \KK{Return} execution results.
However, their specifications must explicitly track in the post-condition that a statement terminates via \KK{break} or \KK{continue}, leading to unwieldy proof rules for loops, since separate specifications must be proven for each possible kind of termination of the loop body.

It is common for program logics which handle unstructured control flow, such as \KK{goto} or continuations, to include a context of target assumptions in the semantics of the triple~\cite{Benton:2005:TCL:2099708.2099741, deBruin1981, Tan:2006:CLC:2146228.2146234,Saabas:2006:CTS:2523791.2523798}.
Separation logics for such languages require a ``higher-order frame rule'', which distributes the frame across all such assumptions~\cite{Jensen:2013:HSL:2429069.2429105,Birkedal:2007:RPS:1760037.1760047,Yang:2005:SST:1078035.1079700,O'Hearn:2004:SIH:964001.964024,neelthesis}.
Similarly, our adaptions to the ``structured \KK{goto}'' approach result in rules akin to a higher-order frame rule, despite the first-order nature of our logic.


\subparagraph{Stack-Based Logics} Two existing program logics are defined over languages which are close to WebAssembly in their typed treatments of the stack: Benton~\cite{Benton:2005:TCL:2099708.2099741}, and Bannwart and M{\"u}ller~\cite{Bannwart:2005:PLB:1705549.1706041}.
However, unlike \logicname, these works does not propose a structured assertion syntax for the stack, instead using unstructured assertions about the values of individual stack positions.
This means that assertions must be re-written with a \textit{shift} operation whenever the shape of the stack changes due to the execution of an instruction, and irrelevant portions of the assertions cannot be framed off during local proofs without keeping track of the necessary resulting shift.
Saabas and Uustalu~\cite{Saabas:2006:CTS:2523791.2523798} give a program logic for a low-level stack-based language with no heap.
Their stack assertion is related to ours in that it has a list structure, but their proof rules rely on a global style of term substitution, and their discussion of compositionality does not appear to extend to generalising existing specifications to larger stacks.
This means that one cannot conduct local proofs over just the portion of the stack that is changing in the program fragment, which we permit thanks to our [extension] rule.
There has been other previous work on program logics for low-level, assembly-like languages, often incorporating a stack~\cite{Myreen:2007:HLR:1763507.1763565,Beringer:2006:BLJ:2172767.2172802,5254213,
Myreen:2007:HLA:1775223.1775241,10.1007/978-3-319-06200-6_8,Jensen:2013:HSL:2429069.2429105}.
These languages do not have type system restrictions on the stack that are as strong as WebAssembly's, and must therefore find other, less structured ways to represent the stack formally.


\section{Conclusions and Future Work}
\label{futureconclusions}

We have presented \logicname, a sound program logic for first-order, encapsulated WebAssembly, and proven the soundness result in Isabelle/HOL. Using \logicname, we have specified and verified a simple WebAssembly B-tree library, giving abstract specifications independent of the underlying implementation.

In designing \logicname, we have found the properties of WebAssembly's type system helpful for streamlining the assertions of \logicname.
The restrictions placed on the runtime behaviour of the WebAssembly stack by the type system are mirrored in the structured nature of our logic's stack assertions.
To account for WebAssembly's uncommon control flow, we have adapted the standard separation logic triple and proof rules, inspired by the early approach of Clint and Hoare~\cite{Clint1972} for ``structured \KK{goto}''.

%
%
%
%

We plan to extend \logicname to handle programs made up of multiple WebAssembly modules composed together.
To do this, we must extend \logicname with the ability to reason about multiple, disjoint memories.
Moreover, we would need to account for the JavaScript ``glue code'', mandatory for module interoperability.
This is part of our broader goal of integrating JavaScript and WebAssembly reasoning.
To achieve this, however, we will need to support some higher-order reasoning, as WebAssembly modules and functions are first-class entities in JavaScript.
We also plan to extend \logicname to be able to reason about higher-order pure WebAssembly code and the \KK{call\_indirect} instruction. For both of these goals, we will refer to existing work on higher-order separation logics~\cite{Varming:2008:HSL:1454792.1455072,Krebbers:2017:EHC:3089361.3089371}.
Although WebAssembly's higher-order constructs are not entirely standard, we believe that it is possible to map WebAssembly's use of the \textit{table} as a higher-order store to the more traditional program states of other higher-order logics, and hence take direct inspiration from their proof rules and soundness approaches. Again, we would also need to account for the  JavaScript component required to mutate the table.
%

Our long-term goal is to be able to reason, in a single formalism, about integrated JavaScript/WebAssembly programs as they will appear on the Web. 
We ultimately hope to integrate our work on \logicname with existing work on program analysis for JavaScript~\cite{Gardner:2012:TPL:2103656.2103663,FragosoSantos:2017:JJV:3177123.3158138,FragosoSantos:2019:JCS:3302515.3290379} to provide a combined proof system, as well as a verification tool.

We expect WebAssembly to be extended with threads and concurrency primitives in the near future~\cite{wasm-threads}.
Because there is no sharing of stacks in the WebAssembly threads proposal, we believe that many of our proof rules will be fully transferrable to a hypothetical concurrent separation logic for WebAssembly with threads, although proof rules for the (now shared) heap will need revising, as will the semantic interpretation. For this, we will take inspiration from various modern concurrent separation logics~\cite{Brookes:2016:CSL:2984450.2984457,Vafeiadis:2013:RSL:2509136.2509532,10.1007/978-3-662-46669-8_30}.

\bibliography{bibliography}

\newpage
\appendix

\section{Full Big-Step Semantics}
\label{app:bigstep}

The definition of the Wasm AST remains identical to that of~\cite{Haas:2017:BWU:3062341.3062363} Figs. 1 and 2. Our big-step judgement, $\Downarrow^{labs,ret}_{inst}$, is parameterised by a list of labels (nat list), a return (nat option), and the current instance.

Our judgement has a structure almost identical to that of the original small-step judgement. The only difference is that the right-hand side results in a $res$ object with the following structure, rather than an intermediate stack.

Note that for a given list of values $v^\ast$, by convention we write $v_e^\ast$ to represent the same values wrapped by the approprate \KK{const} operation.

$$
\begin{array}{lcl}
\textit{res} &::=&  v^\ast~\pmb{|}~\KK{Break}~n~v^\ast~\pmb{|}~\KK{Return}~v^\ast~\pmb{|}~\KK{Trap}\\
\end{array}
$$

$$
\begin{array}{lclrl}
\textit{inst} &::=& \{ & \text{faddrs:} & \textit{nat list} \\
&&& \text{taddr:}  & \textit{nat option} \\
&&& \text{maddr:} & \textit{nat option} \\
&&& \text{gaddrs:} & \textit{nat list} ~ \}\\
\end{array}
$$

\paragraph*{arithmetic operations}
$$
\begin{array}{rcl}
s,~locs,  (\textit{t}.\KK{const}~c) (\textit{t}.\KK{unop}~\textit{op}) &\Downarrow^{labs,ret}_{inst}& s,~locs, ~(\textit{op}(c)) \\
\\
s,~locs,~  (\textit{t}.\KK{const}~c_1) (\textit{t}.\KK{const}~c_2) (\textit{t}.\KK{binop}~\textit{op}) &\Downarrow
^{labs,ret}_{inst}& s,~locs,~ (\textit{op}(c_1,c_2)) \\
\end{array}
$$

\paragraph*{control operations}

$$
\prftree[r]{}
{
s,~locs,  (\KK{label}_m\{[]\}~ (v_e^n~\textit{es}))\Downarrow^{labs,ret}_{inst} s',~locs',~res
}
{
s,~locs, ~v_e^n~(\KK{block } t^n \rightarrow t^m~\textit{es}) \Downarrow^{labs,ret}_{inst} s',~locs',~res
}
$$

$$
\prftree[r]{}
{
s,~locs,  (\KK{label}_n\{[(\KK{loop } t^n \rightarrow t^m~\textit{es})]\}~ (v_e^n~\textit{es}))\Downarrow^{labs,ret}_{inst} s',~locs',~res
}
{
s,~locs, ~v_e^n~(\KK{loop } t^n \rightarrow t^m~\textit{es}) \Downarrow^{labs,ret}_{inst} s',~locs',~res
}
$$

$$
\prftree[r]{}
{
s,~locs, ~v_e^n~(\KK{block } \textit{tf}~\textit{es}) \Downarrow^{labs,ret}_{inst} s',~locs',~res\qquad
}
{
c \neq 0_\textit{i32}
}
{
s,~locs, ~v_e^n~(\KK{i32.const}~c)~(\KK{if } \textit{tf}~\textit{es}~\textit{es}') \Downarrow^{labs,ret}_{inst} s',~locs',~res
}
$$

$$
\prftree[r]{}
{
s,~locs, ~v_e^n~(\KK{block } \textit{tf}~\textit{es}') \Downarrow^{labs,ret}_{inst} s',~locs',~res\qquad
}
{
c = 0_\textit{i32}
}
{
s,~locs, ~v_e^n~(\KK{i32.const}~c)~(\KK{if } \textit{tf}~\textit{es}~\textit{es}') \Downarrow^{labs,ret}_{inst} s',~locs',~res
}
$$

$$
\prftree[r]{}
{
labs!n = k
}
{
s,~locs, ~v_e^k~(\KK{br } n) \Downarrow^{labs,ret}_{inst} s,~locs,~\KK{Break }n~v^k
}
$$

$$
\prftree[r]{}
{
s,~locs, ~v_e^k~(\KK{br } n) \Downarrow^{labs,ret}_{inst} s',~locs',~res\qquad
}
{
c \neq 0_\textit{i32}
}
{
s,~locs, v_e^k~(\KK{i32.const}~c) (\KK{br\_if } n) \Downarrow^{labs,ret}_{inst} s',~locs',~res
}
$$

$$
\prftree[r]{}
{
c = 0_\textit{i32}
}
{
s,~locs, ~(\KK{i32.const}~c) (\KK{br\_if } n) \Downarrow^{labs,ret}_{inst} s,~locs,~\epsilon
}
$$

$$
\prftree[r]{}
{
s,~locs, ~v_e^k~(\KK{br } k') \Downarrow^{labs,ret}_{inst} s',~locs',~res\qquad
}
{
ns!c = k'
}
{
s,~locs, ~v_e^k~(\KK{i32.const}~c) (\KK{br\_table } ns~n') \Downarrow^{labs,ret}_{inst} s',~locs',~res
}
$$

$$
\prftree[r]{}
{
s,~locs, ~v_e^k~(\KK{br } n') \Downarrow^{labs,ret}_{inst} s',~locs',~res
}
{
c \geq length(ns)
}
{
s,~locs, ~v_e^k~(\KK{i32.const}~c) (\KK{br\_table } ns~n') \Downarrow^{labs,ret}_{inst} s',~locs',~res
}
$$

$$
\prftree[r]{}
{
k = ret
}
{
s,~locs, ~v_e^k~\KK{return} \Downarrow^{labs,ret}_{inst} s,~locs,~\KK{Return}~v^k
}
$$

\paragraph*{stack operations}
$$
\begin{array}{rcl}
s,~locs, ~v_e~(\KK{drop}) &\Downarrow^{labs,ret}_{inst}& s,~locs,~\epsilon \\
\\

s,~locs,~v_{1e}~v_{2e}~(\KK{i32.const}~c)~(\KK{select}) &\Downarrow^{labs,ret}_{inst}& s,~locs,~v_1 \quad \text{if $c \neq 0_\textit{i32}$}\\
\\

s,~locs,~v_{1e}~v_{2e}~(\KK{i32.const}~c)~(\KK{select}) &\Downarrow^{labs,ret}_{inst}& s,~locs, ~v_2 \hfill \text{if $c = 0_\textit{i32}$}\\
\end{array}
$$

\paragraph*{local operations}
$$
\begin{array}{rcl}
s,~locs, (\KK{get\_local } n) &\Downarrow^{labs,ret}_{inst}& s,~locs,  v \quad \text{if $locs!n = v$}\\
\\

s,~locs, v~(\KK{set\_local } n) &\Downarrow^{labs,ret}_{inst}& s,~locs[n := v],~\epsilon \\
\\

s,~locs, v~(\KK{tee\_local } n) &\Downarrow^{labs,ret}_{inst}& s,~locs[n := v],~v \\
\end{array}
$$

\paragraph*{global operations}
$$
\begin{array}{rcl}
s,~locs, (\KK{get\_global } n) &\Downarrow^{labs,ret}_{inst}& s,~locs, ~v \\
&&\text{if}\\
&& \text{gaddrs($inst$)$!n = k$}\\
&& \text{$\text{globs}(s)!k = v$} \\
\\
s,~locs, v~(\KK{set\_global } n) &\Downarrow^{labs,ret}_{inst}& s[\text{globs}!k := v],~locs, \epsilon \\
&&\text{if}\\
&& \text{gaddrs($inst$)$!n = k$}\\
\end{array}
$$

\paragraph*{memory operations}
$$
\begin{array}{rcl}
s,~locs, ~(\KK{i32.const}~n)(t.\KK{load}\textit{ off}) &\Downarrow^{labs,ret}_{inst}& s,~locs,  (t.\KK{const}~c) \\
&\multicolumn{2}{@{\qquad}l}{\text{if}}\\
&\multicolumn{2}{@{\qquad}l}{\text{maddr($inst$)$ = k$}}\\
&\multicolumn{2}{@{\qquad}l}{\text{$\text{mems}(s)!k = m$}} \\
&\multicolumn{2}{@{\qquad}l}{m[(n+\textit{off})...(n+\textit{off} + |t|)] = \textit{bytes}}\\
&\multicolumn{2}{@{\qquad}l}{\textit{from\_bytes}(t, \textit{bytes}) = c}\\
\\
s,~locs, ~(\KK{i32.const}~n)(\KK{i32.const}~c)(t.\KK{store}\textit{ off}) &\Downarrow^{labs,ret}_{inst}& s[\text{mems}!k := m'],~locs, ~\epsilon \\
&\multicolumn{2}{@{\hspace{-0.3cm}}l}{\text{if}}\\
&\multicolumn{2}{@{\hspace{-0.3cm}}l}{\text{maddr($inst$)$ = k$}}\\
&\multicolumn{2}{@{\hspace{-0.3cm}}l}{\text{$\text{mems}(s)!k = m$}} \\
&\multicolumn{2}{@{\hspace{-0.3cm}}l}{\textit{to\_bytes}(t, c) = \textit{bytes}}\\
&\multicolumn{2}{@{\hspace{-0.3cm}}l}{m[(n+\textit{off})...(n+\textit{off} + |t|) := \textit{bytes}] = m'}\\
\\
\end{array}
$$

$$
\begin{array}{rcl}
s,~locs, ~(\KK{i32.const}~c)(\KK{mem.grow}) &\Downarrow^{labs,ret}_{inst}& s[\text{mems}!k := m'],~locs,  (\KK{i32.const}~n)  \\
&&\text{if}\\
&& \text{maddr($inst$)$ = k$}\\
&& \text{$\text{mems}(s)!k = m$} \\
&& pages(m) = n \\
&& add\_pages(m,c) = m' \\
&& n \leq 2^{16}\\
\\
s,~locs, ~(\KK{i32.const}~c)(\KK{mem.grow}) &\Downarrow^{labs,ret}_{inst}& s,~locs, ~(\KK{i32.const -1}) \\
&&\text{if}\\
&& \text{maddr($inst$)$ = k$}\\
&& \text{$\text{mems}(s)!k = m$} \\
\\
s,~locs, ~(\KK{mem.size}) &\Downarrow^{labs,ret}_{inst}& s,~locs,  (\KK{i32.const}~n) \\
&&\text{if}\\
&& \text{maddr($inst$)$ = k$}\\
&& \text{$\text{mems}(s)!k = m$} \\
&& pages(m) = n \\
\end{array}
$$

\paragraph*{call operations}

$$
\prftree[r]{}
{
\pbox{\textwidth}{
\mbox{$\text{faddrs($inst$)$!n = k$}$ \qquad $\text{$\text{funcs}(s)!k = cl$}$}\\[1ex]
\mbox{$s,~locs, v_e^\ast~ (\KK{callcl } cl) \Downarrow^{labs,ret}_{inst} s',~locs',~res$}\\[-1ex]
}
}
{
s,~locs, v_e^\ast~(\KK{call } n) \Downarrow^{labs,ret}_{inst} s',~locs',~res
}
$$

$$
\prftree[r]{}
{
\pbox{\textwidth}{
\mbox{$\text{(tables(s)!(taddr($inst$)))$!n = cl$}$}\\[1ex]
\mbox{$s,~locs, v_e^\ast~ (\KK{callcl } cl) \Downarrow^{labs,ret}_{inst} s',~locs',~res$}\\[-1ex]
}
}
{
s,~locs, v_e^\ast~(\KK{i32.const}~n)~(\KK{call\_indirect}~\x{tf}) \Downarrow^{labs,ret}_{inst} s',~locs',~res
}
$$

$$
\prftree[r]{}
{
\pbox{\textwidth}{
\mbox{$\text{(tables(s)!(taddr($inst$)))$!n = cl$}$\qquad $\textit{cl} = \KK{f}~\x{tf} \ldots$}\\[1ex]
\mbox{$s,~locs, ~ (\KK{callcl } cl) \Downarrow^{labs,ret}_{inst} s',~locs',~res$}\\[-1ex]
}
}
{
s,~locs, v_e^\ast~(\KK{i32.const}~n)~(\KK{call\_indirect}~\x{tf}) \Downarrow^{labs,ret}_{inst} s',~locs',~res
}
$$

\paragraph*{callcl operation}

$$
\prftree[r]{}
{
\pbox{\textwidth}{
\mbox{$\textit{cl} = \KK{f}~t^n \rightarrow t^m ~\textnormal{locals}~l_{n}..l_{n+n'}~es$ \qquad $z^{n'} = zerovals(l_{n}..l_{n+n'})$}\\[1ex]
\mbox{$s,~locs, ~ (\KK{local}_m\{(v^n~z^{n'})\}~(\KK{block }[] \rightarrow t^m~es)) \Downarrow^{labs,ret}_{inst} s',~locs',~res$}\\[-1ex]
}
}
{
s,~locs, v_e^n~(\KK{callcl } cl) \Downarrow^{labs,ret}_{inst} s',~locs',~res
}
$$

\paragraph*{value congruence}

$$
\prftree[r]{}
{
s,~locs, [] \Downarrow^{labs, ret}_{inst} s,~locs, ~ []
}
\qquad
\prftree[r]{}
{
s,~locs, ~ es \Downarrow^{labs, ret}_{inst} s',~locs',  ~ \KK{v}^\ast
}
{
s,~locs, ~v_e^\ast ~ es \Downarrow^{labs, ret}_{inst} s',~locs',  ~ v^\ast ~ \KK{v}^\ast
}
$$

$$
\prftree[r]{}
{
s,~locs, ~ es \Downarrow^{labs,ret}_{inst} s'',~locs'',  ~ v^\ast
\qquad
}
{
s'',~locs'', ~ v_e^\ast~es' \Downarrow^{labs,ret}_{inst} s',~locs',  ~ {v'}^\ast
}
{
s,~locs, ~(es~es') \Downarrow^{labs,ret}_{inst} s',~locs', ~ {v'}^\ast
}
$$

$$
\prftree[r]{}
{
s,~locs, ~ es \Downarrow^{n:labs, ret}_{inst} s',~locs',  ~ v^\ast
}
{
s,~locs, ~ \KK{label}_n\{e^\ast\}~es \Downarrow^{labs,ret}_{inst} s',~locs',  ~ v^\ast
}
$$

$$
\prftree[r]{}
{
s,~llocs, ~ es \Downarrow^{[],r}_{i} s',~llocs',  ~ v^\ast
}
{
s,~locs, ~ \KK{local}_r\{llocs\}~es \Downarrow^{labs,ret}_{inst} s',~locs, ~v^\ast
}
$$

\paragraph*{break congruence}

$$
\prftree[r]{}
{
s,~locs, ~ es \Downarrow^{labs,ret}_{inst} s',~locs',  ~ \KK{Break}~n~vs
\qquad
}
{
s,~locs, ~ v_e^\ast~(es~es') \Downarrow^{labs,ret}_{inst} s',~locs',~\KK{Break}~n~vs
}
$$

$$
\prftree[r]{}
{
s,~locs, ~ es \Downarrow^{n':labs,ret}_{inst} s',~locs',  ~\KK{Break}~(n+1)~vs
}
{
s,~locs, ~ \KK{label}_{n'}\{e^\ast\}~es \Downarrow^{labs,ret}_{inst} s',~locs',  ~\KK{Break}~n~vs
}
$$

$$
\prftree[r]{}
{
\pbox{\textwidth}{
\mbox{$s,~locs, ~ es \Downarrow^{n':labs,r}_{inst} s'',~locs'',  ~\KK{Break}~0~vs$}\\[1ex]
\mbox{$s'',~locs'', ~ v_e^\ast~vs_e~e^\ast \Downarrow^{labs,ret}_{inst} s',~locs',  res$}\\[-1ex]
}
}
{
s,~locs, ~ v_e^\ast~\KK{label}_{n'}\{e^\ast\}~es \Downarrow^{labs,ret}_{inst} s',~locs', ~  res
}
$$

\paragraph*{return congruence}

$$
\prftree[r]{}
{
s,~locs, ~ es \Downarrow^{labs,ret}_{inst} s',~locs',  ~ \KK{Return}~vs
\qquad
}
{
s,~locs, ~ v_e^\ast~(es~es') \Downarrow^{labs,ret}_{inst} s',~locs',~\KK{Return}~vs
}
$$

$$
\prftree[r]{}
{
s,~locs, ~ es \Downarrow^{n':labs,ret}_{inst} s',~locs',  ~\KK{Return}~vs
}
{
s,~locs, ~ \KK{label}_{n'}\{e^\ast\}~es \Downarrow^{labs,ret}_{inst} s',~locs',  ~\KK{Return}~vs
}
$$

$$
\prftree[r]{}
{
s,~llocs, ~ es \Downarrow^{[],r}_{inst} s',~llocs',  ~\KK{Return}~vs
}
{
s,~locs, ~ \KK{local}_r\{llocs\}~es \Downarrow^{labs,ret}_{inst} s',~locs, ~vs
}
$$

\paragraph*{trap congruence}

$$
\prftree[r]{}
{
s,~locs, ~ es \Downarrow^{labs,ret}_{inst} s',~locs',  ~ \KK{Trap}
\qquad
}
{
s,~locs, ~ v_e^\ast~(es~es') \Downarrow^{labs,ret}_{inst} s',~locs',~\KK{Trap}
}
$$

$$
\prftree[r]{}
{
s,~locs, ~ es \Downarrow^{n':labs,ret}_{inst} s',~locs',  ~\KK{Trap}
}
{
s,~locs, ~ \KK{label}_{n'}\{e^\ast\}~es \Downarrow^{labs,ret}_{inst} s',~locs',  ~\KK{Trap}
}
$$

$$
\prftree[r]{}
{
s,~llocs, ~ es \Downarrow^{[],ret}_{i} s',~llocs',  ~ \KK{Trap}
}
{
s,~locs, ~ \KK{local}_r\{llocs\}~es \Downarrow^{labs,ret}_{inst} s',~locs, ~\KK{Trap}
}
$$



\newpage

\section{Verification: B-Trees}

%
%

\subsection{Ordered, Bounded Arrays} 
\label{app:oba}

\begin{align*}
\obaseg{x}{\alpha} := \Star{0 \leq i < \llen \alpha}(\ittcell{x + 4 \cdot i}{\alpha \lidx i});
\end{align*}

\begin{align*}
\ba{x}{n}{\alpha} :=&~\left(\ittcell{x}{\llen{\alpha}} \slstar \obaseg{x + 4}{\alpha} \slstar \Star{\llen{\alpha} < i \leq n}(\ittcell{x + 4 \cdot i}{-})\right)\wedge \\ 
&~(\llen{\alpha} \leq n~\wedge (x + 4 \cdot (n + 1) \leq \ittmax)_\mathbb{Z}),
\end{align*}

\begin{align*}
\ordered{\alpha} := \frall{i, i'} 0 \leq i < \llen{\alpha} \rightarrow 0 \leq i' < i \rightarrow \alpha \lidx i' \leq \alpha \lidx i;
\end{align*}

\begin{align*}
\oba{x}{n}{\alpha} :=~& \ba x n \alpha \wedge \ordered{\alpha}
\end{align*}

\subparagraph{OBAGet} The $\pred{OBAGet}(x, k)$ function, specified and verified below, retrieves the $k$-th element of the OBA which starts from memory location $x$. \\

{
\footnotesize\noindent
$\specline{\stackheap{x, k}{\oba{x}{n}{\alpha} \wedge 0 \leq k < \llen \alpha}}$ \\
\begin{Wfunce}[]{\pred{OBAGet}}{[\Witype, \Witype] \rightarrow [\Witype]}
  $\specline{\stackheap{}{\oba{x}{n}{\alpha} \wedge 0 \leq k < \llen \alpha \wedge l_0 = x \wedge l_1 = k}}$ \\
  $\begin{leftvruled}{frame}
  \specline{\stackheap{}{\emp}} \\
  \Wgetlocal{0} \\
  \specline{\stackheap{l_0}{\emp}} \\
  \begin{leftvruled}{extension}
  \specline{\stackheap{}{\emp}} \\
  \Wgetlocal{1} \\
  \specline{\stackheap{l_1}{\emp}} \\
  \end{leftvruled} \\
  \specline{\stackheap{l_0, l_1}{\emp}} \\
  \Wiconst{4} \\
  \specline{\stackheap{l_0, l_1, 4}{\emp}} \\
  \Wimul 
  \Wiadd \\
  \specline{\stackheap{l_0 + 4 \cdot l_1}{\emp}} 
  \end{leftvruled}$ \\
  $\specline{\stackheap{l_0 + 4 \cdot l_1}{\oba{x}{n}{\alpha} \wedge 0 \leq k < \llen \alpha \wedge l_0 = x \wedge l_1 = k}}$ \\
  $\specline{\stackheap{x + 4 \cdot k}{\oba{x}{n}{\alpha} \wedge 0 \leq k < \llen \alpha \wedge l_0 = x \wedge l_1 = k}}$ \annot{(by consequence)} \\
  \annot{[[ Unfold $\oba{x}{n}{\alpha}$ ]]} \\
  $\specline{\stackheap{x + 4 \cdot k}{(\ittcell{x}{\llen{\alpha}} \slstar \Star{0 \leq i < \llen \alpha}(\ittcell{x + 4 + 4 \cdot i}{\alpha \lidx i}) \slstar \Star{\llen{\alpha} < i \leq n}(\ittcell{x + 4 \cdot i}{-}))\wedge \\ \hspace*{1.55cm}(\ordered{\alpha} \wedge \llen{\alpha} \leq n~\wedge (x + 4 \cdot (n + 1) \leq \ittmax)_\mathbb{Z}) \wedge 0 \leq k < \llen \alpha~\wedge \\ \hspace*{1.57cm}l_0 = x \wedge l_1 = k}}$ \\
  $\begin{leftvruled}{frame}
  \specline{\stackheap{x + 4 \cdot k}{\ittcell{x + 4 + 4 \cdot k}{\alpha \lidx k}}} \\
  \Wiloadoff{4} \\
  \specline{\stackheap{\alpha \lidx k}{\ittcell{x + 4 + 4 \cdot k}{\alpha \lidx k}}} \\
  \end{leftvruled}$ \\
  $\specline{\stackheap{\alpha \lidx k}{(\ittcell{x}{\llen{\alpha}} \slstar \Star{0 \leq i < \llen \alpha}(\ittcell{x + 4 + 4 \cdot i}{\alpha \lidx i}) \slstar \Star{\llen{\alpha} < i \leq n}(\ittcell{x + 4 \cdot i}{-}))\wedge \\ \hspace*{0.95cm}(\ordered{\alpha} \wedge \llen{\alpha} \leq n~\wedge (x + 4 \cdot (n + 1) \leq \ittmax)_\mathbb{Z}) \wedge 0 \leq k < \llen \alpha~\wedge \\ \hspace*{0.98cm}l_0 = x \wedge l_1 = k}}$ \\
  \annot{[[ Fold $\oba{x}{n}{\alpha}$ ]]} \\
  $\specline{\stackheap{\alpha \lidx k}{\oba{x}{n}{\alpha} \wedge 0 \leq k < \llen \alpha \wedge l_0 = x \wedge l_1 = k}}$ \\
$\specline{\stackheap{\alpha \lidx k}{\oba{x}{n}{\alpha} \wedge 0 \leq k < \llen \alpha}}$ \annot{(by consequence)}
\end{Wfunce} \\
$\specline{\stackheap{\alpha \lidx k}{\oba{x}{n}{\alpha} \wedge 0 \leq k < \llen \alpha}}$
}

\subparagraph{OBAFind} The $\pred{OBAFind}(x, e)$ function, specified and verified below, finds the appropriate index for an element in an OBA list. It takes two parameters: $x$~denotes the memory address at which the OBA is allocated, while $e$ denotes the element that is being searched for. The function returns an integer $i$ with the following properties: if $e$ is in the OBA, $i$ equals the index of its first occurrence; if $e$ is not in the OBA, $i$ is equal to the index of the first element of the OBA larger than $e$, if such an element exists, and to the OBA length otherwise. \\

{\footnotesize\noindent
$\specline{\stackheap{x, e}{\oba{x}{n}{\alpha}}}$ \\
\begin{Wfunce}[\\(\KK{locals} \Witype)]{OBAFind}{[\Witype, \Witype] \rightarrow [\Witype]}
  $\specline{\stackheap{}{\oba{x}{n}{\alpha} \wedge l_0 = x \wedge l_1 = e \wedge l_2 = 0}}$ \\
  \annot{$P_{inv} : \oba{x}{n}{\alpha} \wedge l_0 = x \wedge l_1 = e \wedge 0 \leq l_2 \leq \llen\alpha \wedge \left(\frall{j} 0 \leq j < l_2 \rightarrow \alpha \lidx j < e \right)$ \\
  $\specline{\stackheap{}{P_{inv}}}$} \annot{(by consequence)} \\
  \begin{Wloope}[\\$(\stackheap{}{P_{inv}}) \vdash$]{}
    $\specline{\stackheap{}{P_{inv}}}$ \\
    $\begin{leftvruled}{frame}
    {\color{purple} (\stackheap{}{\emp}) \vdash}~\specline{\stackheap{}{\emp}} \\
    \Wgetlocal{2} \\
    {\color{purple} (\stackheap{}{\emp}) \vdash}~\specline{\stackheap{l_2}{\emp}} \\
    \end{leftvruled}$ \\
    $\specline{\stackheap{l_2}{P_{inv}}}$ \\
    \Wgetlocal{0} 
    \Wiload \\
    $\specline{\stackheap{l_2, \llen \alpha}{P_{inv}}}$ \\
    \Wilt \\
    $\specline{\exsts{v}\stackheap{v}{P_{inv} \wedge C_1}}$ \\
    ${\color{purple} (\exsts v \stackheap{}{P_{inv}}) \vdash~} \specline{\exsts{v}\stackheap{v}{P_{inv} \wedge C_1}}$ \annot{(by consequence)}\\
      $\begin{leftvruled}{exists}
      \begin{minipage}{0.7\textwidth}
    ${\color{purple} (\stackheap{}{P_{inv}}) \vdash~} \specline{\stackheap{v}{P_{inv} \wedge C_1}}$ \\
    \begin{Wife}[\\$(\stackheap{}{P_{inv} \wedge C_3}), (\stackheap{}{P_{inv}}) \vdash$]{}
      $\specline{\stackheap{}{P_{inv} \wedge l_2 < \llen{\alpha}}}$ \\
      \Wgetlocal{0}
      \Wgetlocal{2} 
      \Wcall{\text{OBAGet}} \\
      $\specline{\stackheap{\alpha!l_2}{P_{inv} \wedge l_2 < \llen{\alpha}}}$ \\
      \Wgetlocal{1} 
      \Wilt \\
      $\specline{\exsts{v}\stackheap{v}{P_{inv} \wedge l_2 < \llen{\alpha} \wedge C_2}}$ \\
    \begin{Wife}[\\$(\stackheap{}{P_{inv} \wedge C_3}), (\stackheap{}{P_{inv} \wedge C_3}), (\stackheap{}{P_{inv}})  \vdash$]{}
         $\specline{\stackheap{}{P_{inv} \wedge l_2 < \llen{\alpha} \wedge \alpha \lidx l_2 < e}}$ \\
        \Wgetlocal{2} 
        \Wiconst{1} 
        \Wiadd \\
        $\specline{\stackheap{l_2 + 1}{P_{inv} \wedge l_2 < \llen{\alpha} \wedge \alpha \lidx l_2 < e}}$ \\
        \Wsetlocal{2} \\
        $\specline{\stackheap{}{\oba{x}{n}{\alpha} \wedge l_0 = x \wedge l_1 = e \wedge l_2{-}1 < \llen{\alpha}~\wedge \\ \tab~~ \left(\frall{j} 0 \leq j < l_2{-}1 \rightarrow \alpha \lidx j < e \right) \wedge \alpha \lidx (l_2{-}1) < e}}$ \\
        $\specline{\stackheap{}{\oba{x}{n}{\alpha} \wedge l_0 = x \wedge l_1 = e \wedge  \left(\frall{j} 0 \leq j < l_2 \rightarrow \alpha \lidx j < e \right) \wedge l_2 \leq \llen{\alpha}}}$ \\
        $\specline{\stackheap{}{P_{inv}}}$ \\
        \Wbr{2} \\
        $\specline{\stackheap{}{P_{inv} \wedge C_3}}$ 
      \end{Wife} \\
      $\specline{\stackheap{}{P_{inv} \wedge C_3}}$ 
    \end{Wife} \\
    $\specline{\stackheap{}{P_{inv} \wedge C_3}}$ 
    \end{minipage} 
    \end{leftvruled}$ \\
    ${\color{purple} (\exsts v \stackheap{}{P_{inv}}) \vdash~} \specline{\exsts v \stackheap{}{P_{inv} \wedge C_3}}$ \\
    ${\color{purple} (\stackheap{}{P_{inv}}) \vdash~} \specline{\stackheap{}{P_{inv} \wedge C_3}}$ \annot{(by consequence)} 
  \end{Wloope} \\
  $\specline{\stackheap{}{P_{inv} \wedge C_3}}$ \\
  \Wgetlocal{2} \\
  $\specline{\stackheap{l_2}{\oba{x}{n}{\alpha} \wedge l_0 = x \wedge l_1 = e \wedge 0 \leq l_2 \leq \llen\alpha \wedge \left(\frall{j} 0 \leq j < l_2 \rightarrow \alpha \lidx j < e \right) \wedge C_3}}$ \\
  $\specline{\exsts{i} \stackheap{i}{\oba{x}{n}{\alpha} \wedge l_0 = x \wedge l_1 = e \wedge l_2 = i \wedge 0 \leq i \leq \llen\alpha~\wedge \\ \tab\tab\tab~\left(\frall{j} 0 \leq j < i \rightarrow \alpha \lidx j < e \right) \wedge \left(\frall{j} i \leq j < \llen{\alpha}\rightarrow e \leq \alpha \lidx j\right)}}$ \\
\end{Wfunce} \\
$\specline{\exsts{i}\stackheap{i}{\oba{x}{n}{\alpha} \wedge 0 \leq i \leq \llen\alpha \wedge \left(\frall{j} 0 \leq j < i \rightarrow \alpha \lidx j < e \right) \wedge \left(\frall{j} i \leq j < \llen{\alpha}\rightarrow e \leq \alpha \lidx j\right) }}$} \\

\noindent where:
\begin{align*}
C_1 \equiv~& (v = 0 \rightarrow l_2 = \llen{\alpha}) \wedge (v > 0 \rightarrow l_2 < \llen{\alpha}); \\
C_2 \equiv~& (v = 0 \rightarrow \alpha \lidx l_2 \geq e) \wedge (v \neq 0 \rightarrow \alpha \lidx l_2 < e); \text{ and} \\
C_3 \equiv~& ((l_2 < \llen{\alpha} \wedge \alpha \lidx l_2 \geq e) \vee l_2 = \llen\alpha).
\end{align*}

\subparagraph{AsegShl} The $\pred{AsegShl}(x, n)$ function, specified and verified below, shifts an array segment to the left. It takes two parameters: $x$~denotes the address in memory at which the array segment is allocated, while $n$~denotes the length of the segment that is to be shifted to the left. The resulting OBA segment contains all of its previous elements except the one at the front, and the last element of the original array segment is forgotten. \\

{
\footnotesize\noindent
$\specline{\stackheap{x, n}{\obaseg{x}{a \lcons \alpha} \wedge n = \llen\alpha}}$ \\
\begin{Wfunce}[\\(\KK{locals} \Witype, \Witype)]{AsegShl}{[\Witype, \Witype] \rightarrow []}
  $\specline{\stackheap{}{\obaseg{x}{a \lcons \alpha} \wedge l_0 = x \wedge l_1 = \llen\alpha \wedge l_2 = 0 \wedge l_3 = 0}}$ \\
  \Wgetlocal{0} \Wteelocal{2} \Wiconst{4} \Wiadd \Wsetlocal{3} \Wiconst{0} \Wsetlocal{0} \\
  $\specline{\stackheap{}{\obaseg{x}{a \lcons \alpha} \wedge l_0 = 0 \wedge l_1 = \llen\alpha \wedge l_2 = x \wedge l_3 = x + 4}}$ \\
  $P_{inv}: 0 \leq l_0 \leq l_1 \wedge l_1 = \llen\alpha \wedge l_2 = x + 4 \cdot l_0 \wedge l_3 = x + 4 \cdot (l_0 + 1) \wedge \\ 
  \hspace*{0.85cm} \obaseg{x}{\sublist\alpha{0}{l_0}} \slstar \obaseg{l_2}{\sublist{a \lcons \alpha}{l_0}{\llen{a \lcons \alpha} - l_0}}$ \\
  $\specline{\stackheap{}{P_{inv}}}$ \\
  \begin{Wloope}[\\$(\stackheap{}{P_{inv}}) \vdash$]{}
    $\specline{\stackheap{}{P_{inv}}}$ \\
  	\Wgetlocal{0}\Wgetlocal{1}\Wilt \\
    $\specline{\exsts{v}\stackheap{v}{P_{inv} \wedge 
    (v = 0 \rightarrow l_0 = l_1) \wedge 
    (v > 0 \rightarrow l_0 < l_1)}}$ \\
	\begin{Wife}[\\$-, (\stackheap{}{P_{inv}}) \vdash$]{}
	        $\specline{\stackheap{}{P_{inv} \wedge l_0 < l_1)}}$ \\
		\Wgetlocal{2} \Wgetlocal{3} \Wiload \Wistore \\
		$\specline{\stackheap{}{0 \leq l_0 < l_1 \wedge l_1 = \llen\alpha \wedge l_2 = x + 4 \cdot l_0 \wedge l_3 = x + 4 \cdot (l_0 + 1) \wedge \\ 
  \tab~~ \obaseg{x}{\sublist\alpha{0}{l_0}} \slstar \ittcell{l_2}{\alpha \lidx l_0}~\slstar \\ \tab~~ \obaseg{l_3}{\sublist{a \lcons \alpha}{l_0+1}{\llen{a \lcons \alpha} - l_0 - 1}}}}$ \\
		\tab~~ \Wgetlocal{2} \Wiconst{4} \Wiadd \Wteelocal{2} \Wiconst{4} \Wiadd \Wsetlocal{3} \\
		\tab~~ \Wgetlocal{0} \Wiconst{1} \Wiadd \Wsetlocal{0} \\
		\tab~~ $\specline{\stackheap{}{0 \leq l_0 \leq l_1 \wedge l_1 = \llen\alpha \wedge l_2 = x + 4 \cdot l_0 \wedge l_3 = x + 4 \cdot (l_0 + 1) \wedge \\ 
  \tab~~ \obaseg{x}{\sublist\alpha{0}{l_0-1}} \slstar \ittcell{l_2 - 4}{\alpha \lidx l_0-1} \slstar \obaseg{l_2}{\sublist{a \lcons \alpha}{l_0}{\llen{a \lcons \alpha} - l_0}}}}$ \\
		$\specline{\stackheap{}{0 \leq l_0 \leq l_1 \wedge l_1 = \llen\alpha \wedge l_2 = x + 4 \cdot l_0 \wedge l_3 = x + 4 \cdot (l_0 + 1) \wedge \\ 
  \tab~~ \obaseg{x}{\sublist\alpha{0}{l_0}} \slstar \obaseg{l_2}{\sublist{a \lcons \alpha}{l_0}{\llen{a \lcons \alpha} - l_0}}}}$ \\
		\Wbr{1} \\
		$\specline{\stackheap{}{P_{inv} \wedge l_0 = l_1)}}$ 
	\end{Wife} \\
	$\specline{\stackheap{}{P_{inv} \wedge l_0 = l_1)}}$ 
  \end{Wloope} \\
  $\specline{\stackheap{}{P_{inv} \wedge l_0 = l_1)}}$ \\
		$\specline{\stackheap{}{ 
  \obaseg{x}{\sublist\alpha{0}{\llen\alpha}} \slstar \obaseg{x + 4 \cdot \llen\alpha}{\sublist{a \lcons \alpha}{\llen\alpha}{1}}}}$ \\
		$\specline{\stackheap{}{ 
  \obaseg{x}{\alpha} \slstar \ittcell{x + 4 \cdot \llen\alpha}{-}}}$ 
    \end{Wfunce} \\
$\specline{\stackheap{}{\obaseg{x}{\alpha} \slstar \ittcell{x+4\cdot\llen\alpha}{-}}}$}

\subparagraph{AsegShr} The $\pred{AsegShr}(x, n)$ function, specified below and verified analogously to $\pred{AsegShl}$, shifts an array segment to the right. It takes two parameters: $x$~denotes the memory address at which the (non-empty) array segment is allocated, while $n$ denotes the length of the segment to be shifted to the right. We also require ownership of one element past the length of the list. The resulting array segment contains an additional, duplicated element at the front. \\ 

{\footnotesize\noindent
$\specline{\stackheap{x, n}{\obaseg{x}{a \lcons \alpha} \slstar \ittcell{x + 4\cdot\llen{a \lcons \alpha}}{-} \wedge n = \llen{a \lcons \alpha}}}$ \\
$(\KK{func}~[\Witype, \Witype] \rightarrow~ []~\text{AsegShr} \ldots~\KK{end})$ \\
$\specline{\stackheap{}{\obaseg{x}{a \lcons {(a \lcons \alpha)}}}}$}

\subsection{Ordered, Bounded Arrays without Duplication} 
\label{app:OBAnd}
\begin{align*}
\oband{x}{n}{\alpha} :=~\oba{x}{n}{\alpha} \wedge \llen\alpha = \card{\listToSet\alpha}
\end{align*}

\subparagraph{OBAInsert} The $\pred{OBAInsert}(x, e)$ function, specified and verified below, inserts an element into a given $\pred{OBA_{nd}}$. If the element already is in the OBA, the OBA is not modified. \\

{\footnotesize\noindent
$\specline{\stackheap{x, e}{\oband{x}{n}{\alpha} \wedge \llen\alpha < n}}$ \\
\begin{Wfunce}[\\(\KK{locals} \Witype)]{OBAInsert}{[\Witype, \Witype] \rightarrow []}
  $\specline{\stackheap{}{\oband{x}{n}{\alpha} \wedge l_0 = x \wedge l_1 = e \wedge l_2 = 0 \wedge \llen\alpha < n \wedge e \notin \listToSet\alpha}}$ \\
  \Wgetlocal{0} \Wteelocal{1} \Wcall{OBAFind} \\
  $\specline{\exsts i \stackheap{i}{\oband{x}{n}{\alpha} \wedge l_0 = x \wedge l_1 = e \wedge l_2 = 0 \wedge  \llen\alpha < n \wedge e \notin \listToSet\alpha \wedge 0 \leq i \leq \llen\alpha~\wedge \\ \tab\tab\tab~ \left(\frall{j} 0 \leq j < i \rightarrow \alpha \lidx j < e \right) \wedge \left(\frall{j} i \leq j < \llen{\alpha}\rightarrow e < \alpha \lidx j\right) }}$ \\
  \Wteelocal{2} \Wgetlocal{0} \Wiload \Wilt \\
  $\specline{\exsts v \stackheap{v}{\oband{x}{n}{\alpha} \wedge l_0 = x \wedge l_1 = e \wedge  \llen\alpha < n \wedge e \notin \listToSet\alpha \wedge 0 \leq l_2 \leq \llen\alpha~\wedge \\ \tab\tab\tab~~ \left(\frall{j} 0 \leq j < l_2 \rightarrow \alpha \lidx j < e \right) \wedge \left(\frall{j} l_2 \leq j < \llen{\alpha}\rightarrow e < \alpha \lidx j\right) \wedge \\ \tab\tab\tab~~ (v = 0 \rightarrow \l_2 = \llen\alpha) \wedge (v \neq 0 \rightarrow \l_2 < \llen\alpha) }}$ \\
  \begin{Wife}[\\$- \vdash$]{}
  $\specline{\stackheap{}{\oband{x}{n}{\alpha} \wedge l_0 = x \wedge l_1 = e \wedge  \llen\alpha < n \wedge e \notin \listToSet\alpha \wedge 0 \leq l_2 < \llen\alpha~\wedge \\ \tab~~ \left(\frall{j} 0 \leq j < l_2 \rightarrow \alpha \lidx j < e \right) \wedge \left(\frall{j} l_2 \leq j < \llen{\alpha}\rightarrow e \leq \alpha \lidx j\right)}}$ \\  
  \Wgetlocal{0} \Wgetlocal{2} \Wcall{AsegShr} \Wgetlocal 1 \Wine \\
  $\specline{\exsts v \stackheap{v}{\oband{x}{n}{\alpha} \wedge l_0 = x \wedge l_1 = e \wedge  \llen\alpha < n \wedge e \notin \listToSet\alpha \wedge 0 \leq l_2 < \llen\alpha~\wedge \\ \tab\tab\tab~~ \left(\frall{j} 0 \leq j < l_2 \rightarrow \alpha \lidx j < e \right) \wedge \left(\frall{j} l_2 \leq j < \llen{\alpha}\rightarrow e \leq \alpha \lidx j\right)~\wedge \\
  \tab\tab\tab~~ (v = 0 \rightarrow \alpha \lidx l_2 = e) \wedge (v \neq 0 \rightarrow \alpha \lidx l_2 \neq e) }}$ \\
    \begin{Wife}[\\$-, - \vdash$]{}
	    $\specline{\stackheap{}{\oband{x}{n}{\alpha} \wedge l_0 = x \wedge l_1 = e \wedge  \llen\alpha < n \wedge e \notin \listToSet\alpha \wedge 0 \leq l_2 < \llen\alpha~\wedge \\ \tab~~ \left(\frall{j} 0 \leq j < l_2 \rightarrow \alpha \lidx j < e \right) \wedge \left(\frall{j} l_2 \leq j < \llen{\alpha}\rightarrow e < \alpha \lidx j\right)}}$ \\
		\Wgetlocal{0} \Wgetlocal{2} \Wcall{AsegShr} \\
		\Wgetlocal{0} \Wgetlocal{2} \Wiconst{4} \Wimul \Wiadd \Wgetlocal{1} \Wistoreoff{4} \\
		\Wgetlocal{0} \Wgetlocal{0} \Wiload \Wiconst{1} \Wisub \Wistore \\
		$\specline{\exsts{\alpha'} \stackheap{}{\oband{x}{n}{\alpha'} \wedge \listToSet{\alpha'} = \listToSet{\alpha} \cup \{e\}}}$ \\
		\Welse \\
		$\specline{\stackheap{}{\oband{x}{n}{\alpha} \wedge l_0 = x \wedge l_1 = e \wedge  \llen\alpha < n \wedge e \notin \listToSet\alpha \wedge l_2 = \llen\alpha~\wedge \\ \tab\tab\tab~~ \left(\frall{j} 0 \leq j < \llen\alpha \rightarrow \alpha \lidx j < e \right)}}$ \\
		\Wgetlocal{0} \Wgetlocal{2} \Wiconst{4} \Wimul \Wiadd \Wgetlocal{1} \Wistoreoff{4} \\
		\Wgetlocal{0} \Wgetlocal{0} \Wiload \Wiconst{1} \Wisub \Wistore \\
		$\specline{\exsts{\alpha'} \stackheap{}{\oband{x}{n}{\alpha'} \wedge \listToSet{\alpha'} = \listToSet{\alpha} \cup \{e\}}}$
	\end{Wife} \\
		$\specline{\exsts{\alpha'} \stackheap{}{\oband{x}{n}{\alpha'} \wedge \listToSet{\alpha'} = \listToSet{\alpha} \cup \{e\}}}$
	\end{Wife} \\
	$\specline{\exsts{\alpha'} \stackheap{}{\oband{x}{n}{\alpha'} \wedge \listToSet{\alpha'} = \listToSet{\alpha} \cup \{e\}}}$
\end{Wfunce} \\
$\specline{\exsts{\alpha'} \stackheap{}{\oband{x}{n}{\alpha'} \wedge \listToSet{\alpha'} = \listToSet{\alpha} \cup \{e\}}}$}

\subparagraph{OBADelete} The $\pred{OBADelete}(x, e)$ function, specified and verified below, deletes an element from a given $\pred{OBA_{nd}}$. If the element is not in the OBA, the OBA is not modified. \\

{
\footnotesize\noindent
$\specline{\stackheap{x, e}{\oband{x}{n}{\alpha}}}$ \\
\begin{Wfunce}[\\(\KK{locals} \Witype)]{OBADelete}{[\Witype, \Witype] \rightarrow []}
  $\specline{\stackheap{}{\oband{x}{n}{\alpha} \wedge l_0 = x \wedge l_1 = e \wedge l_2 = 0}}$ \\
  \Wgetlocal{0} \Wgetlocal{0} \Wgetlocal{1} \Wcall{OBAFind} \\
  $\specline{\exsts i \stackheap{x, i}{\oband{x}{n}{\alpha} \wedge l_0 = x \wedge l_1 = e \wedge 0 \leq i < \llen\alpha~\wedge \\ \tab\tab\tab\tab~ \left(\frall{j} 0 \leq j < i \rightarrow \alpha \lidx j < e \right) \wedge \left(\frall{j} i \leq j < \llen{\alpha}\rightarrow e \leq \alpha \lidx j\right) }}$ \\
  \Wteelocal{2} \Wcall{OBAGet} \Wgetlocal{2} \Wieq \\
  $\specline{\exsts v \stackheap{v}{\oband{x}{n}{\alpha} \wedge l_0 = x \wedge l_1 = e \wedge 0 \leq i < \llen\alpha~\wedge \\ \tab\tab\tab~~ \left(\frall{j} 0 \leq j < l_2 \rightarrow \alpha \lidx j < e \right) \wedge \left(\frall{j} l_2 \leq j < \llen{\alpha}\rightarrow e \leq \alpha \lidx j\right)~\wedge \\ 
  \tab\tab\tab~~ (v = 0 \rightarrow \alpha \lidx l_2 \neq e) \wedge (v \neq 0 \rightarrow \alpha \lidx l_2 = e)}}$ \\
  \begin{Wife}[\\$- \vdash$]{}
  $\specline{\stackheap{}{\oband{x}{n}{\alpha} \wedge l_0 = x \wedge l_1 = e \wedge 0 \leq i < \llen\alpha \wedge \left(\frall{j} 0 \leq j < l_2 \rightarrow \alpha \lidx j < e \right)~\wedge \\ \tab~~\left(\frall{j} l_2 < j < \llen{\alpha}\rightarrow e < \alpha \lidx j\right) \wedge \alpha \lidx l_2 = e}}$ \\
  \Wgetlocal 0 \Wgetlocal 2 \Wcall{AsegShl} \\ 
  \Wgetlocal{0} \Wgetlocal{0} \Wiload \Wiconst{1} \Wisub \Wistore \\
$\specline{\exsts{\alpha'} \stackheap{}{\oband x n {\alpha'} \wedge e \in \listToSet\alpha \wedge \listToSet{\alpha'} = \listToSet{\alpha} \setminus \{e\} \wedge l_0 = x \wedge l_1 = e}}$
  \end{Wife} \\
  $\specline{\exsts{\alpha'} \stackheap{}{\oband{x}{n}{\alpha'} \wedge e \in \listToSet\alpha \wedge \listToSet{\alpha'} = \listToSet{\alpha} \setminus \{e\} \wedge l_0 = x \wedge l_1 = e}}$
    \end{Wfunce} \\
$\specline{\exsts{\alpha'} \stackheap{}{\oband{x}{n}{\alpha'} \wedge \listToSet{\alpha'} = \listToSet{\alpha} \setminus \{e\}}}$
}

\subsection{B-Tree Metadata Page}
\begin{align*}
\page{a} := \Star{(a \cdot 64k) \leq i < ((a+1) \cdot 64k)}(\ittcell{i}{-}) \land 0 \leq a \land ((a+1) \cdot 64k) \leq \ittmax
\end{align*}

\begin{align*}
\free{\alpha} := \oband{8}{16381}{\alpha} \Star{0 \leq i < \llen \alpha}(\page{\alpha \lidx i});
\end{align*}

\begin{align*}
\meta t r l \alpha := \ittcell{0}{t} \slstar \ittcell{4}{r} \slstar \Wsize{l} \slstar \free{\alpha}.
\end{align*}

\subsection{B-Tree Nodes}
\label{app:bnodes}
\begin{align*}
\nkeys{x}{\kappa} :=~& \oband{x \cdot 64k + 4}{4095}{\kappa}; \\
\nptrs{x}{\pi} :=~& \ba{x \cdot 64k + 32k}{4096}{\pi}; \\
\node x \lambda \kappa \pi :=~& \ittcell{x \cdot 64k}{\lambda} \slstar \nkeys{x}{\kappa} \slstar \nptrs{x}{\pi}.
\end{align*}

\subparagraph{InitNode} The $\pred{InitNode}$ function, specified and verified below, initialises a given WebAssembly memory page to represent a B-tree leaf node. \\

{\footnotesize\noindent
$\specline{\stackheap{x}{\page{x}}}$ \\
\begin{Wfunce}{InitNode}{[\Witype] \rightarrow []}
$\specline{\stackheap{x}{\page{x}} \land l_0 = x}$ \\
\Wgetlocal 0 \Wiconst {64k} \Wimul \Wsetlocal 0 \quad \annot{[Store $x \cdot 64k$ in $l_0$]} \\
$\specline{\stackheap{x}{\page{x}} \land l_0 = x \cdot 64k}$ \\
\Wgetlocal 0 \Wiconst{1} \Wistore \quad \annot{(Set leaf information)}\\
\Wgetlocal 0 \Wiconst{0} \Wistoreoff{4} \quad \annot{(Keys are empty)}\\
\Wgetlocal 0 \Wiconst{0} \Wistoreoff{32k} \quad \annot{(Pointers are empty)}\\
$\specline{\stackheap{}{\node x 1 \lemp \lemp \land l_0 = x \cdot 64k} }$ \\
\annot{[[Fold $\nkeys x \lemp$]; Fold $\nptrs x \lemp$; Fold $\node x 1 \lemp \lemp$]]} \\
$\specline{\stackheap{}{\node x 1 \lemp \lemp \wedge l_0 = x \cdot 64k} }$
\end{Wfunce} \\
$\specline{\stackheap{}{\node x 1 \lemp \lemp} }$
}

\subparagraph{FreeNode} The $\pred{FreeNode}$ function, specified and verified below, frees the memory page belonging to a given B-tree node. If the free page set is not full, the page is added to the free page set. Otherwise, the function does not terminate. \\

{\footnotesize\noindent
$\specline{\stackheap{x}{\free{\alpha} \wedge \llen \alpha < 16381 \slstar \node x - - -}}$ \\
\begin{Wfunce}{FreeNode}{[] \rightarrow []}
$\specline{\stackheap{}{\free{\alpha} \slstar \node x - - - \wedge l_0 = x}}$ \\
$\begin{leftvruled}{frame}
\begin{minipage}{0.95\textwidth}
$\specline{\stackheap{}{\ittcell{8}{\llen \alpha}}}$ \\
\begin{Wloope}[\\$- \vdash$]{}
$\specline{\stackheap{}{\ittcell{8}{\llen \alpha}}}$ \\
\Wiconst {16381} \Wiconst{8} \Wiload  \Wile \Wbrif{0} \quad \annot{(Ensure we can add a free page)} \\ 
$\specline{\stackheap{}{\ittcell{8}{\llen \alpha} \wedge \llen\alpha < 16381}}$ 
\end{Wloope} \\
$\specline{\stackheap{}{\ittcell{8}{\llen \alpha} \wedge \llen\alpha < 16381}}$ \\
\end{minipage}
\end{leftvruled}$ \\
$\specline{\stackheap{}{\free{\alpha} \slstar \node x - - - \wedge l_0 = x \wedge \llen\alpha < 16381}}$ \\
\Wiconst{8}
\Wgetlocal{0}
\Wcall{OBAInsert} \quad \annot{(Insert the freed page into the list of free pages)} \\
$\specline{\exsts {\alpha'} \stackheap{}{\free{\alpha'} \land \listToSet {\alpha'} = \listToSet \alpha \uplus \{x \}} }$
\end{Wfunce} \\
$\specline{\exsts {\alpha'} \stackheap{}{\free{\alpha'} \land \listToSet {\alpha'} = \listToSet \alpha \uplus \{x \}} }$
}

\subparagraph{AllocNode} The $\pred{AllocNode}$ function, specified and verified below, allocates a B-tree leaf node. The node's address is chosen from
the list of free pages, if that list is non-empty, Otherwise, a new page is allocated, if possible. If not, the function does not terminate. \\

{\footnotesize\noindent
$\specline{\stackheap{}{\KK{size}(l) \slstar \free{\alpha}}}$ \\
\begin{Wfunce}[\\(\KK{locals} \Witype)]{AllocNode}{[] \rightarrow []}
$\specline{\stackheap{}{\KK{size}(l) \slstar \free{\alpha}}}$ \\
\Wiconst 8 \Wiload \Wiconst 0 \Wigt \quad \annot{(Check if we have freed pages)} \\
$\specline{\exsts{v}\stackheap{v}{\Wsize l \slstar \free \alpha \land 
(v = 0 \rightarrow \alpha = \lemp) \wedge 
(v \neq 0 \rightarrow \exsts{l', \alpha'} \alpha = l' \lcons \alpha')}}$ \\
\begin{Wife} ~
$\specline{\stackheap{}{\exsts{l', \alpha'}  \Wsize l \slstar \free{l' \lcons \alpha'}}}$ \\
\Wiconst{12} \Wiload \Wcall{InitNode} \quad \annot{(We do, initialise node on first free page)} \\
\Wiconst {8} \Wiconst{12} \Wiload \\ \Wcall{OBADelete} \quad \annot{(Remove the re-allocated page from the list of free pages)} \\
$\specline{POST}$ \\
\Welse \\
\begin{Wloope}{}
$\specline{\stackheap{}{\KK{size}(l)}}$ \\
\Wiconst{1}
\Wgrowmemory \quad \annot{(We do not, attempt to allocate another page)} \\
\Wteelocal{0} \\
\Wiconst{-1}
\Wieq
\Wbrif{0} \quad \annot{(Loop forever if we cannot allocate)}\\
$\specline{\stackheap{}{\page{l} \slstar \KK{size}(l+1) \wedge l_0 = l}}$
\end{Wloope} \\
\Wgetlocal{0} \quad \annot{(Get the address of the newly allocated page)} \\
$\specline{\stackheap{l}{\page{l}} \slstar \KK{size}(l+1) \slstar \free{\alpha} \wedge l_0 = l}$ \\
\Wcall{InitNode} \quad \annot{(Initialise node on the newly allocated page)} \\
$\specline{POST}$ 
\end{Wife} \\
$\specline{POST}$ 
\end{Wfunce} \\
$\specline{POST}$}

where {\footnotesize $\specline{POST} = \specline{
\exsts l' \stackheap{l'}{\node {l'} \bot \lemp \lemp \slstar ((\alpha = \lemp \wedge l' = l \wedge \KK{size}(l+1) \slstar \free{\alpha}~\lor) \\
\hspace*{3.95cm} (\exsts{\alpha'} \alpha = l' \lcons {\alpha'} \slstar \Wsize l \slstar \free{\alpha'}))
}}$} \\

Onward, when calling $\pred{AllocNode}$, we will assume the set of free pages to be empty, for simplicity, and use the following specification: \\

{\footnotesize
$\begin{array}{l}
\specline{\stackheap{}{\KK{size}(l) \slstar \free{\lemp}}} \\
(\KK{func}~\pred{AllocNode}~[] \rightarrow [\Witype] \ldots \KK{end}) \\
\specline{
\stackheap{l}{\node l 1 \lemp \lemp \slstar \KK{size}(l+1) \slstar \free{\lemp}}}
\end{array}$}

\subparagraph{GetNodeLeaf} The $\pred{GetNodeLeaf}(x) $ function, specified and verified below, returns the leaf information of a given B-tree node at address $x$. \\

{\footnotesize\noindent
$\specline{\stackheap{x}{\node x \lambda \kappa \pi} }$ \\
\begin{Wfunce}{GetNodeLeaf}{[\Witype] \rightarrow [\Witype]}
$\specline{\stackheap{}{\node x \lambda \kappa \pi \land l_0 = x}}$ \\
\Wgetlocal 0 \Wiconst {64k} \Wimul \Wiload \\
$\specline{\stackheap{\lambda}{\node x \lambda \kappa \pi} \land l_0 = x }$
\end{Wfunce} \\
$\specline{\stackheap{\lambda}{\node x \lambda \kappa \pi} }$
}

\subparagraph{SetNodeLeaf} The $\pred{SetNodeLeaf}$ function, specified and verified below, stores the provided leaf information $\lambda$ in a given B-tree node at address $x$. \\

{\footnotesize\noindent
$\specline{\stackheap{x, \lambda}{\node x - \kappa \pi} }$ \\
\begin{Wfunce}{SetNodeLeaf}{[\Witype, \Witype] \rightarrow []}
$\specline{\stackheap{}{\node x \lambda \kappa \pi \land l_0 = x \land l_1 = \lambda}}$ \\
\Wgetlocal 0 \Wiconst {64k} \Wimul \Wgetlocal 1 \Wistore \\
$\specline{\stackheap{}{\node x \lambda \kappa \pi} \land l_0 = x \land l_1 = \lambda }$
\end{Wfunce} \\
$\specline{\stackheap{}{\node x \lambda \kappa \pi} }$
}

\subparagraph{GetNodeKey} The $\pred{GetNodeKey}$ function, specified and verified below, retrieves the $i$-th key of a B-tree node at address $x$. \\

{\footnotesize\noindent
$\specline{\stackheap{x, i}{\node x \lambda \kappa \pi} \wedge 0 \leq i < \llen \kappa }$ \\
\begin{Wfunce}{GetNodeKey}{[\Witype, \Witype] \rightarrow [\Witype]}
$\specline{\stackheap{}{\node x \lambda \kappa \pi \wedge 0 \leq i < \llen \kappa \land l_0 = x \land l_1 = i}}$ \\
\Wgetlocal 0 \Wiconst {64k} \Wimul \Wiconst 4 \Wiadd \Wgetlocal 1 \\
$\specline{\stackheap{x + 64k + 4, i}{\node x \lambda \kappa \pi \wedge 0 \leq i < \llen \kappa \land l_0 = x \land l_1 = i}}$ \\
\Wcall{OBAGet} \\
$\specline{\stackheap{\kappa \lidx i}{\node x \lambda \kappa \pi} \wedge 0 \leq i < \llen \kappa \land l_0 = x \land l_1 = i }$
\end{Wfunce} \\
$\specline{\stackheap{\kappa \lidx i}{\node x \lambda \kappa \pi \wedge 0 \leq i < \llen \kappa} }$
}

\subparagraph{GetNodePtr} The $\pred{GetNodePtr}$ function, specified and verified below, retrieves the $i$-th pointer of a non-leaf B-tree node at address $x$. \\

{\footnotesize\noindent
$\specline{\stackheap{x, i}{\node x 0 \kappa \pi} \wedge 0 \leq i < \llen \pi }$ \\
\begin{Wfunce}{GetNodePtr}{[\Witype, \Witype] \rightarrow [\Witype]}
$\specline{\stackheap{}{\node x 0 \kappa \pi \wedge 0 \leq i < \llen \pi \land l_0 = x \land l_1 = i}}$ \\
\Wgetlocal 0 \Wiconst {64k} \Wimul \Wiconst {32k} \Wiadd \Wgetlocal 1 \\
$\specline{\stackheap{x + 64k + 32k, i}{\node x 0 \kappa \pi \wedge 0 \leq i < \llen \pi \land l_0 = x \land l_1 = i}}$ \\
\Wcall{OBAGet} \\
$\specline{\stackheap{\pi \lidx i}{\node x 0 \kappa \pi} \wedge 0 \leq i < \llen \pi \land l_0 = x \land l_1 = i }$
\end{Wfunce} \\
$\specline{\stackheap{\pi \lidx i}{\node x 0 \kappa \pi \wedge 0 \leq i < \llen \pi} }$
}

\subparagraph{InsertNodeKey} The $\pred{InsertNodeKey}$ function, specified and verified below, inserts the key~$k$ into the keys a of B-tree node at address $x$, potentially extending the keys. \\

{\footnotesize\noindent
$\specline{\stackheap{x, k}{\node x \lambda \kappa \pi} \wedge \llen \kappa < 4095 }$ \\
\begin{Wfunce}{InsertNodeKey}{[\Witype, \Witype] \rightarrow []}
$\specline{\stackheap{}{\node x \lambda \kappa \pi \wedge \llen \kappa < 4095 \land l_0 = x \land l_1 = k}}$ \\
\Wgetlocal 0 \Wiconst {64k} \Wimul \Wiconst {4} \Wiadd \Wgetlocal 1 \\
$\specline{\stackheap{x + 64k + 4, k}{\node x \lambda \kappa \pi \wedge \llen \kappa < 4095 \land l_0 = x \land l_1 = k}}$ \\
\Wcall{OBAInsert} \\
$\specline{\exsts {\kappa'} \stackheap{}{\node x \lambda {\kappa'} \pi \land  \listToSet{\kappa'} =  \listToSet{\kappa} \cup \{ k \} \land l_0 = x \land l_1 = k }}$
\end{Wfunce} \\
$\specline{\exsts {\kappa'} \stackheap{}{\node x \lambda {\kappa'} \pi \land  \listToSet{\kappa'} =  \listToSet{\kappa} \cup \{ k \} }}$} \\

\subparagraph{SetNodePtr} The $\pred{SetNodePtr}$ function, specified and verified below, sets the $i$-th pointer of a given B-tree non-leaf node at address $x$ to value $p$, potentially extending the pointers. \\

{\footnotesize\noindent
$\specline{\stackheap{x, i, p}{\node x 0 \kappa \pi} \wedge ((0 \leq i < \llen \pi) \vee (i = \llen \pi < 4096)) }$ \\
\begin{Wfunce}{SetNodePtr}{[\Witype, \Witype, \Witype] \rightarrow []}
$\specline{\stackheap{}{\node x 0 \kappa \pi \wedge ((0 \leq i < \llen \pi) \vee (i = \llen \pi < 4096)) \land l_0 = x \land l_1 = i \land l_2 = p}}$ \\
\Wgetlocal 0 \Wiconst {64k} \Wimul \Wiconst {32k} \Wiadd \Wgetlocal 1 \Wgetlocal 2\\
$\specline{\stackheap{x + 64k + 32k, i, p}{\node x 0 \kappa \pi \wedge ((0 \leq i < \llen \pi) \vee (i = \llen \pi < 4096))}}$ \\
\Wcall{BASet} \\
$\specline{\exsts {\pi'} \stackheap{}{\node x 0 \kappa {\pi'} \land 0 \leq i \leq \llen \pi \wedge \pi' =  \sublist \pi 0 i \lcat [ e ] \lcat \sublist \pi {i + 1} {\llen \pi - i - 1} }}$
\end{Wfunce} \\
$\specline{\exsts {\pi'} \stackheap{}{\node x 0 \kappa {\pi'} \land 0 \leq i \leq \llen \pi \wedge \pi' =  \sublist \pi 0 i \lcat [ p ] \lcat \sublist \pi {i + 1} {\llen \pi - i - 1} }}$} \\

\noindent where the specification of the $\pred{BASet}(x, i, p)$ function, which sets an element of a given bounded array to a given value, is: \\

{\footnotesize
$\begin{array}{l}
\specline{\stackheap{x, i, e}{\ba x n \alpha \wedge ((0 \leq i < \llen \alpha) \vee (i = \llen \alpha < n)) }} \\
(\KK{func}~\pred{BASet}~[\Witype, \Witype, \Witype] \rightarrow [] \ldots \KK{end}) \\
\specline{
\exsts {\alpha'} \stackheap{}{\ba x n {\alpha'} \wedge 0 \leq i \leq \llen \alpha \wedge \alpha' =  \sublist \alpha 0 i \lcat [ e ] \lcat \sublist \alpha {i + 1} {\llen \alpha - i - 1}}}
\end{array}$}

\subsection{B-Trees}
\label{app:btrees}

\subparagraph{Definition} We define an abstract predicate describing what it means to be a WebAssembly B-Tree with branching factor $t$ and keys $\kappa$ as follows:
\begin{align*}
\btree t \kappa \triangleq~& \exsts{r, l, \alpha, \lambda, \phi} \meta t r l \alpha \slstar \btreerec r \kappa \lambda \phi,
\end{align*}
where:
$$
\begin{array}{l}
\btreerec x \kappa \lambda \phi \triangleq~ x < l~\wedge\\ 
\quad (\exsts{\kappa_x, \pi_x} (\node x {\lambda} {\kappa_x} {\pi_x} \wedge \llen {\kappa_x} \leq 2t-1 \wedge (x \neq r \rightarrow t - 1 \leq \llen {\kappa_x})~\wedge \\
\qquad (\llen {\kappa_x} < 2t-1 \rightarrow \phi = 0) \wedge (\llen {\kappa_x} = 2t-1 \rightarrow \phi \neq 0)) \slstar  \\
\qquad (\lambda \neq 0 \rightarrow \pi_x = \lemp \wedge \listToSet {\kappa_x} = \kappa \wedge \emp) \slstar \\
\qquad (\lambda = 0 \rightarrow 0 < \llen {\kappa_x} = \llen {\pi_x} - 1 \wedge \exsts{\lambda', \overline{\kappa}, \overline\phi} (\llen {\overline \kappa} = \llen {\overline \phi} = \llen {\pi_x}~\wedge \\
\qquad \quad \kappa = (\bigcup_{0 \leq i < \llen {\overline \kappa}} {\overline \kappa} \lidx i) \cup \kappa_x~\wedge \Star{0 \leq i < \llen {\pi_x}} \btreerec {\pi_x \lidx i}{\overline \kappa \lidx i}{\lambda'}{\overline \phi \lidx i}~\wedge \\
\qquad \qquad (\bigwedge_{0 \leq i < \llen {\overline\kappa} - 1} \frall{k, k'} k \in  {\overline \kappa \lidx i} \rightarrow k' \in {\overline \kappa \lidx (i + 1)} \rightarrow 
k < \kappa_x \lidx i < k')~\wedge \\
\qquad \qquad \quad (\bigwedge_{0 \leq i < \llen {\pi_x}} {\pi_x} \lidx i < l)))).
\end{array}
$$

\subsubsection{B-Tree Creation}

A B-Tree is always created from an empty memory. We first allocate the metadata page, then allocate the first, root node, which is set to be a leaf and left empty. The branching factor of the B-Tree, $t$ is given as the only parameter. \\

{\footnotesize\noindent
$\specline{\stackheap{t}{\Wsize 0} \wedge 2 \leq t \leq 2048}$ \\
\begin{Wfunce}{BTreeCreate}{[\Witype] \rightarrow []}
$\specline{\stackheap{}{\Wsize 0} \wedge 2 \leq t \leq 2048 \wedge l_0 = t}$ \\
\begin{Wloope}[\\$- \vdash$]{}
$\specline{\stackheap{}{\Wsize 0} \wedge 2 \leq t \leq 2048 \wedge l_0 = t}$ \\
\Wiconst{1} \Wgrowmemory  \Wiconst{-1} \Wieq \Wbrif{0} \quad \annot{(Allocate metadata page)} \\
$\specline{\stackheap{}{\page{0} \slstar \Wsize 1 \wedge 2 \leq t \leq 2048 \wedge l_0 = t}}$
\end{Wloope} \\
$\specline{\stackheap{}{\page{0} \slstar \Wsize 1 \wedge 2 \leq t \leq 2048 \wedge l_0 = t}}$ \\
\Wiconst 0 \Wgetlocal 0 \Wistore \quad \annot{(Set the branching factor)}\\
\Wiconst 4 \Wiconst 0 \Wistore \quad \annot{(Set the root node)}\\
\Wiconst 8 \Wiconst 0 \Wistore \quad \annot{(Set the free page list to empty)}\\
\annot{[[Fold $\free \lemp$; Fold $\meta t 0 1 \lemp$]]} \\
\Wcall{AllocNode} \Wdrop \\
$\specline{\stackheap{}{\meta t 0 2 \lemp \slstar \node l 1 \lemp \lemp \wedge 2 \leq t \leq 2048}}$ \\
\annot{[[Fold $\pred{BTreeRec}^{t, 0, 2}(0, 1, \emptyset)$]]} \\
$\specline{\stackheap{}{\meta t 0 2 \lemp \slstar \pred{BTreeRec}^{t, 0, 2}(0, 1, \emptyset) \wedge 2 \leq t \leq 2048}} $ \\
\annot{[[Fold $\btree t \emptyset $]]} \\
$\specline{\stackheap{}{\btree t \emptyset \wedge 2 \leq t \leq 2048}}$ 
\end{Wfunce} \\
$\specline{\stackheap{}{\btree t \emptyset} \wedge 2 \leq t \leq 2048 }$}

\subsubsection{B-Tree Search}
The B-tree search function, $\pred{BTreeSearch}$, takes a key $k$ that is being searched for and returns a non-zero result if the B-tree contains the tree $k$, and zero if it does not. It uses an auxiliary function, $\pred{BTreeSearchRec}$, which traverses the B-tree recursively. \\

{\footnotesize\noindent
$\specline{\stackheap{x, k}{\btreerec x \kappa \lambda \phi}}$ \\
\begin{Wfunce}[\\(\KK{locals} \Witype)]{BTreeSearchRec}{[\Witype, \Witype] \rightarrow [\Witype]}
$\specline{\stackheap{}{\btreerec x \kappa \lambda \phi \wedge l_0 = x \wedge l_1 = k \wedge l_2 = 0}}$ \\
\Wgetlocal 0 \Wiconst {64k} \Wimul \Wiconst 4 \Wiadd \Wgetlocal 1 \\
$\specline{\stackheap{x\cdot64k + 4, k}{\btreerec x \kappa \lambda \phi \wedge l_0 = x \wedge l_1 = k \wedge l_2 = 0}}$ \\
$\specline{\stackheap{x\cdot64k + 4, k}{\node x \lambda {\kappa_x} {\pi_x} \slstar \ldots}}$ \\
\Wcall {OBAFind} \\
$\specline{\exsts{i}\stackheap{i}{\node{x}{\lambda}{\kappa_x}{\pi_x} \wedge 0 \leq i \leq \llen{\kappa_x} \wedge \left(\frall{j} 0 \leq j < i \rightarrow \kappa_x \lidx j < k \right) \wedge \\ 
\hspace*{1cm}\left(\frall{j} i \leq j < \llen{\kappa_x}\rightarrow k \leq \kappa_x \lidx j\right) \slstar \ldots }}$ \\
\Wteelocal 2 \Wgetlocal 0 \Wiload \\
$\specline{\stackheap{l_2, {\llen{\kappa_x}}}{\node{x}{\lambda}{\kappa_x}{\pi_x} \wedge 0 \leq l_2 \leq \llen{\kappa_x} \wedge \left(\frall{j} 0 \leq j < l_2 \rightarrow \kappa_x \lidx j < k \right) \wedge \\ 
\hspace*{1.9cm}\left(\frall{j} l_2 \leq j < \llen{\kappa_x}\rightarrow k \leq \kappa_x \lidx j\right) \slstar \ldots }}$ \\
\Wilt \\
\begin{Wife}{}
$\specline{\stackheap{}{\node{x}{\lambda}{\kappa_x}{\pi_x} \wedge 0 \leq l_2 < \llen{\kappa_x} \wedge \left(\frall{j} 0 \leq j < l_2 \rightarrow \kappa_x \lidx j < k \right) \wedge \\ 
\hspace*{0.5cm}\left(\frall{j} l_2 \leq j < \llen{\kappa_x}\rightarrow k \leq \kappa_x \lidx j\right) \wedge \slstar \ldots }}$ \\
\Wgetlocal 0 \Wgetlocal 2 \Wcall {GetNodeKey} \Wgetlocal 1 \\
$\specline{\stackheap{\kappa_x \lidx l_2, k}{\node{x}{\lambda}{\kappa_x}{\pi_x} \wedge 0 \leq l_2 < \llen{\kappa_x} \wedge \left(\frall{j} 0 \leq j < l_2 \rightarrow \kappa_x \lidx j < k \right) \wedge \\ 
\hspace*{1.3cm}\left(\frall{j} l_2 \leq j < \llen{\kappa_x}\rightarrow k \leq \kappa_x \lidx j\right) \wedge \slstar \ldots }}$ \\
\Wieq \\
\begin{Wife}{}
$\specline{\stackheap{}{\node{x}{\lambda}{\kappa_x}{\pi_x} \wedge 0 \leq l_2 < \llen{\kappa_x} \wedge \kappa_x \lidx l_2 = k \wedge \left(\frall{j} 0 \leq j < l_2 \rightarrow \kappa_x \lidx j < k \right) \wedge \\ 
\hspace*{0.5cm}\left(\frall{j} l_2 < j < \llen{\kappa_x}\rightarrow k < \kappa_x \lidx j\right) \wedge \slstar \ldots }}$ \\
\Wiconst 1 \\
$\specline{\stackheap{1}{\node{x}{\lambda}{\kappa_x}{\pi_x} \wedge 0 \leq l_2 < \llen{\kappa_x} \wedge \kappa_x \lidx l_2 = k \wedge \left(\frall{j} 0 \leq j < l_2 \rightarrow \kappa_x \lidx j < k \right) \wedge \\ 
\hspace*{0.65cm}\left(\frall{j} l_2 < j < \llen{\kappa_x}\rightarrow k < \kappa_x \lidx j\right) \wedge \slstar \ldots }}$ \\
$\specline{\stackheap{1}{\btreerec r \kappa \lambda \phi \wedge k \in \kappa}}$ \\
$\specline{POST}$ \\
\Welse \\
$\specline{\stackheap{}{\node{x}{\lambda}{\kappa_x}{\pi_x} \wedge 0 \leq l_2 < \llen{\kappa_x} \wedge \left(\frall{j} 0 \leq j < l_2 \rightarrow \kappa_x \lidx j < k \right) \wedge \\ 
\hspace*{0.5cm}\left(\frall{j} l_2 \leq j < \llen{\kappa_x}\rightarrow k < \kappa_x \lidx j\right) \wedge \slstar \ldots }}$ \\
\Wgetlocal 0 \Wcall{GetNodeLeaf} 
\Wiconst 0 \\ 
$\specline{\stackheap{\lambda, 0}{\node{x}{\lambda}{\kappa_x}{\pi_x} \wedge 0 \leq l_2 < \llen{\kappa_x} \wedge \left(\frall{j} 0 \leq j < l_2 \rightarrow \kappa_x \lidx j < k \right) \wedge \\ 
\hspace*{1cm}\left(\frall{j} l_2 \leq j < \llen{\kappa_x}\rightarrow k < \kappa_x \lidx j\right) \wedge \slstar \ldots }}$ \\
\Wine \\
\begin{Wife}{}
$\specline{\stackheap{}{\node{x}{\lambda}{\kappa_x}{\pi_x} \wedge 0 \leq l_2 < \llen{\kappa_x} \wedge \lambda \neq 0 \wedge \left(\frall{j} 0 \leq j < l_2 \rightarrow \kappa_x \lidx j < k \right) \wedge \\ 
\hspace*{0.5cm}\left(\frall{j} l_2 \leq j < \llen{\kappa_x}\rightarrow k < \kappa_x \lidx j\right) \wedge \listToSet {\kappa_x} = \kappa \slstar \ldots }}$ \\
\Wiconst 0 \\
$\specline{\stackheap{0}{\node{x}{\lambda}{\kappa_x}{\pi_x} \wedge 0 \leq l_2 < \llen{\kappa_x} \wedge \lambda \neq 0 \wedge \left(\frall{j} 0 \leq j < l_2 \rightarrow \kappa_x \lidx j < k \right) \wedge \\ 
\hspace*{0.65cm}\left(\frall{j} l_2 \leq j < \llen{\kappa_x}\rightarrow k < \kappa_x \lidx j\right)  \wedge \listToSet {\kappa_x} = \kappa \slstar \ldots }}$ \\
$\specline{\stackheap{0}{\btreerec r \kappa \lambda \phi \wedge k \notin \kappa}}$ \\
$\specline{POST}$ \\
\Welse \\
$\specline{\stackheap{}{\node{x}{\lambda}{\kappa_x}{\pi_x} \wedge 0 \leq l_2 < \llen{\kappa_x} \wedge \lambda = 0 \wedge \left(\frall{j} 0 \leq j < l_2 \rightarrow \kappa_x \lidx j < k \right) \wedge \\ 
\hspace*{0.5cm}\left(\frall{j} l_2 \leq j < \llen{\kappa_x}\rightarrow k < \kappa_x \lidx j\right) \wedge \kappa = (\bigcup_{0 \leq i < \llen {\overline \kappa}} {\overline \kappa} \lidx i) \cup \kappa_x \wedge k \notin \bigcup_{0 \leq j < \llen {\overline \kappa}, j \neq i} {\overline \kappa} \lidx j \slstar \ldots}}$ \\
\Wiconst 0 \Wgetlocal 2 \Wcall{GetNodePtr} \Wgetlocal 1 \\
$\specline{\stackheap{\pi_x \lidx {l_2}, k}{\btreerec{\pi_x \lidx {l_2}}{\overline\kappa \lidx {l_2}}{\lambda'}{\overline\phi \lidx {l_2}} \wedge 0 \leq l_2 < \llen{\kappa_x} \wedge \lambda = 0 \wedge \left(\frall{j} 0 \leq j < l_2 \rightarrow \kappa_x \lidx j < k \right) \wedge \\ 
\hspace*{0.65cm}\left(\frall{j} l_2 \leq j < \llen{\kappa_x}\rightarrow k < \kappa_x \lidx j\right) \wedge \kappa = (\bigcup_{0 \leq i < \llen {\overline \kappa}} {\overline \kappa} \lidx i) \cup \kappa_x \wedge k \notin \bigcup_{0 \leq j < \llen {\overline \kappa}, j \neq l_2} {\overline \kappa} \lidx j \slstar \ldots}}$ \\
\Wcall{BTreeRec} \\
$\specline{\exsts {b} \stackheap{b}{\btreerec{\pi_x \lidx {l_2}}{\overline\kappa \lidx {l_2}}{\lambda'}{\overline\phi \lidx {l_2}} \wedge 0 \leq l_2 < \llen{\kappa_x} \wedge \lambda = 0 \wedge \left(\frall{j} 0 \leq j < l_2 \rightarrow \kappa_x \lidx j < k \right) \wedge \\ 
\hspace*{1cm}\left(\frall{j} l_2 \leq j < \llen{\kappa_x}\rightarrow k < \kappa_x \lidx j\right) \wedge \kappa = (\bigcup_{0 \leq i < \llen {\overline \kappa}} {\overline \kappa} \lidx i) \cup \kappa_x \wedge k \notin \bigcup_{0 \leq j < \llen {\overline \kappa}, j \neq l_2} {\overline \kappa} \lidx j~\wedge \\ \hspace*{1cm}(k \in \overline\kappa \lidx {l_2} \rightarrow b = 1) \wedge (k \notin \overline\kappa \lidx {l_2} \rightarrow b = 0) \slstar \ldots}}$ \\
$\specline{POST}$
\end{Wife} 
\end{Wife} \\
\Welse \\
$\specline{\stackheap{}{\node{x}{\lambda}{\kappa_x}{\pi_x} \wedge l_2 = \llen{\kappa_x} \wedge \left(\frall{j} 0 \leq j < l_2 \rightarrow \kappa_x \lidx j < k \right) \wedge \slstar \ldots }}$ \\
\Wgetlocal 0 \Wcall{GetNodeLeaf}
\Wiconst 0 \\
$\specline{\stackheap{\lambda, 0}{\node{x}{\lambda}{\kappa_x}{\pi_x} \wedge l_2 = \llen{\kappa_x} \wedge \left(\frall{j} 0 \leq j < l_2 \rightarrow \kappa_x \lidx j < k \right) \wedge \slstar \ldots }}$ \\
\Wine \\
\begin{Wife}{}
$\specline{\stackheap{}{\node{x}{\lambda}{\kappa_x}{\pi_x} \wedge l_2 = \llen{\kappa_x} \wedge \lambda \neq 0 \wedge \left(\frall{j} 0 \leq j < l_2 \rightarrow \kappa_x \lidx j < k \right) \wedge \listToSet{\kappa_x} = \kappa \slstar \ldots }}$ \\
\Wiconst 0 \\
$\specline{\stackheap{0}{\node{x}{\lambda}{\kappa_x}{\pi_x} \wedge l_2 = \llen{\kappa_x} \wedge \lambda \neq 0 \wedge \left(\frall{j} 0 \leq j < l_2 \rightarrow \kappa_x \lidx j < k \right) \wedge \listToSet{\kappa_x} = \kappa \slstar \ldots }}$ \\
$\specline{\stackheap{0}{\btreerec r \kappa \lambda \phi \wedge k \notin \kappa}}$ \\
$\specline{POST}$ \\
\Welse \\
$\specline{\stackheap{}{\node{x}{\lambda}{\kappa_x}{\pi_x} \wedge l_2 = \llen{\kappa_x} \wedge \lambda = 0 \wedge \left(\frall{j} 0 \leq j < l_2 \rightarrow \kappa_x \lidx j < k \right) \wedge~\\ 
\hspace*{0.5cm} \kappa = (\bigcup_{0 \leq i < \llen {\overline \kappa}} {\overline \kappa} \lidx i) \cup \kappa_x \wedge k \notin \bigcup_{0 \leq j < \llen {\overline \kappa}, j \neq i} {\overline \kappa} \lidx j \slstar \ldots}}$ \\
\Wiconst 0 \Wgetlocal 2 \Wcall{GetNodePtr} \Wgetlocal 1 \\
$\specline{\stackheap{\pi_x \lidx {l_2}, k}{\btreerec{\pi_x \lidx {l_2}}{\overline\kappa \lidx {l_2}}{\lambda'}{\overline\phi \lidx {l_2}} \wedge l_2 = \llen{\kappa_x} \wedge \lambda = 0 \wedge \left(\frall{j} 0 \leq j < l_2 \rightarrow \kappa_x \lidx j < k \right) \wedge~\\ 
\hspace*{1.48cm} \kappa = (\bigcup_{0 \leq i < \llen {\overline \kappa}} {\overline \kappa} \lidx i) \cup \kappa_x \wedge k \notin \bigcup_{0 \leq j < \llen {\overline \kappa}, j \neq l_2} {\overline \kappa} \lidx j \slstar \ldots}}$ \\
\Wcall{BTreeRec} \\
$\specline{\exsts {b} \stackheap{b}{\btreerec{\pi_x \lidx {l_2}}{\overline\kappa \lidx {l_2}}{\lambda'}{\overline\phi \lidx {l_2}} \wedge l_2 = \llen{\kappa_x} \wedge \lambda = 0 \wedge \left(\frall{j} 0 \leq j < l_2 \rightarrow \kappa_x \lidx j < k \right) \wedge~\\ 
\hspace*{1.1cm} \kappa = (\bigcup_{0 \leq i < \llen {\overline \kappa}} {\overline \kappa} \lidx i) \cup \kappa_x \wedge k \notin \bigcup_{0 \leq j < \llen {\overline \kappa}, j \neq l_2} {\overline \kappa} \lidx j~\wedge \\ \hspace*{1.1cm}(k \in \overline\kappa \lidx {l_2} \rightarrow b = 1) \wedge (k \notin \overline\kappa \lidx {l_2} \rightarrow b = 0) \slstar \ldots}}$ \\
$\specline{POST}$
\end{Wife} \\
$\specline{POST}$
\end{Wife} \\
$\specline{POST}$
\end{Wfunce} \\
$\specline{POST}$} \\

\noindent where {\footnotesize $\specline{POST} = \specline{\exsts {b} \stackheap{b}{\btreerec r \kappa \lambda \phi  \wedge (k \in \kappa \rightarrow b = 1) \wedge (k \notin \kappa \rightarrow b = 0)}}$}. \\

{\footnotesize\noindent
$\specline{\stackheap{k}{\btree t \kappa}}$ \\
\begin{Wfunce}{BTreeSearch}{[\Witype] \rightarrow [\Witype]}
$\specline{\stackheap{}{\btree t \kappa \wedge l_0 = k}}$ \\
\annot{[[Unfold $\btree t \kappa $]]} \\
$\specline{\stackheap{}{\exsts{r, l, \alpha, \lambda} \meta t r l \alpha \slstar \btreerec r \kappa \lambda \phi \wedge l_0 = k}}$ \\
$\specline{\exsts{r, l, \alpha, \lambda}\stackheap{}{\meta t r l \alpha \slstar \btreerec r \kappa \lambda \phi \wedge l_0 = k}}$ \\
$\begin{leftvruled}{exists}
\specline{\stackheap{}{\meta t r l \alpha \slstar \btreerec r \kappa \lambda \phi \wedge l_0 = k}} \\
\Wiconst 0 \Wiload \Wgetlocal 0 \\
\specline{\stackheap{r, k}{\meta t r l \alpha \slstar \btreerec r \kappa \lambda \phi \wedge l_0 = k}} \\
\begin{leftvruled}{frame}
\specline{\stackheap{r, k}{\btreerec r \kappa \lambda \phi}} \\
\Wcall{\text{~BTreeSearchRec}} \\
\specline{\exsts {b} \stackheap{b}{\btreerec r \kappa \lambda \phi \wedge (k \in \kappa \rightarrow b = 1) \wedge (k \notin \kappa \rightarrow b = 0)}} 
\end{leftvruled} \\
\specline{\exsts {b} \stackheap{b}{\meta t r l \alpha \slstar \btreerec r \kappa \lambda \phi\wedge l_0 = k \wedge (k \in \kappa \rightarrow b = 1) \wedge (k \notin \kappa \rightarrow b = 0 \wedge l_0 = k)}} 
\end{leftvruled}$ \\
$\specline{\exsts {r, l, \alpha, \lambda, b} \stackheap{b}{\meta t r l \alpha \slstar \btreerec r \kappa \lambda \phi \wedge l_0 = k \wedge (k \in \kappa \rightarrow b = 1) \wedge (k \notin \kappa \rightarrow b = 0) \wedge l_0 = k}}$ \\
$\specline{\exsts {b} \stackheap{b}{\exsts {r, l, \alpha, \lambda} \meta t r l \alpha \slstar \btreerec r \kappa \lambda \phi \wedge l_0 = k \wedge (k \in \kappa \rightarrow b = 1) \wedge (k \notin \kappa \rightarrow b = 0 \wedge l_0 = k)}}$ \\
\annot{[[Fold $\btree t \kappa $]]} \\
$\specline{\exsts {b} \stackheap{b}{\btree t \kappa \wedge (k \in \kappa \rightarrow b = 1) \wedge (k \notin \kappa \rightarrow b = 0) \wedge l_0 = k}}$ 
\end{Wfunce} \\
$\specline{\exsts {b} \stackheap{b}{\btree t \kappa \wedge (k \in \kappa \rightarrow b = 1) \wedge (k \notin \kappa \rightarrow b = 0) }}$}

\subsubsection{B-Tree Insertion}

\subparagraph{B-Tree Child Splitting} 
The auxiliary function $\pred{BTreeSplitChild}$ splits the (full) $i$-th child of a non-full, non-leaf node at address $x$ into two nodes with $t-1$ keys each and moves its median key into the node at address $x$. The set of keys is effectively left unchanged. We note that the $\pred{SubList}$ predicate returns an empty list instead if its arguments do not make sense in the context of the given list. \\

{\footnotesize\noindent
$\specline{\stackheap{x, i} {\ittcell{0}{t} \slstar \node x 0 {\kappa_x} {\pi_x} \slstar \node {\pi_x \lidx i} {\kappa_y} {\lambda_y} {\pi_y} \slstar \Wsize l \wedge 0 \leq i < \llen {\pi_x}~\wedge\\ 
\hspace*{0.95cm} \llen {\kappa_x} = \llen {\pi_x} - 1 \wedge \llen {\kappa_i} = \llen {\pi_i} - 1 = 2t - 1 \wedge C(i, \kappa_x, \kappa_y) }}$ \\
\begin{Wfunce}[\\(\KK{locals} \Witype, \Witype, \Witype)]{BTreeSplitChild}{[\Witype, \Witype] \rightarrow []}
$\specline{\stackheap{}{\ittcell{0}{t} \slstar \node x {0} {\kappa_x} {\pi_x} \slstar \node {\pi_x \lidx i} {\lambda_y} {\kappa_y} {\pi_y} \slstar \Wsize l \wedge 0 \leq i < \llen {\pi_x}~\wedge\\ 
\hspace*{0.5cm} \llen {\kappa_x} = \llen {\pi_x} - 1 \wedge \llen {\kappa_i} = \llen {\pi_i} - 1 = 2t - 1 \wedge C(i, \kappa_x, \kappa_i)~\wedge\\
\hspace*{0.5cm}l_0 = x \wedge l_1 = i \wedge l_2 = 0 \wedge l_3 = 0 \wedge l_4 = 0}}$ \\
\Wcall{AllocNode} \Wsetlocal 3 \\
$\specline{\stackheap{}{\ittcell{0}{t} \slstar \node x {0} {\kappa_x} {\pi_x} \slstar \node {\pi_x \lidx i} {\lambda_y} {\kappa_y} {\pi_y} \slstar \node {l_3} 1 \lemp \lemp \slstar \Wsize{l+1} \wedge 0 \leq i < \llen {\pi_x}~\wedge\\ 
\hspace*{0.5cm} \llen {\kappa_x} = \llen {\pi_x} - 1 \wedge \llen {\kappa_i} = \llen {\pi_i} - 1 = 2t - 1 \wedge C(i, \kappa_x, \kappa_i)~\wedge\\
\hspace*{0.5cm}l_0 = x \wedge l_1 = i \wedge l_2 = 0 \wedge l_3 = l \wedge l_4 = 0}}$ \\
\Wgetlocal 0 \Wgetlocal 1 \Wcall{GetNodePtr} \Wsetlocal 2 \\
$\specline{\stackheap{}{\ittcell{0}{t} \slstar \node x {0} {\kappa_x} {\pi_x} \slstar \node {l_2} {\lambda_y} {\kappa_y} {\pi_y} \slstar \node {l_3} 1 \lemp \lemp \slstar \Wsize{l+1} \wedge 0 \leq i < \llen {\pi_x}~\wedge\\ 
\hspace*{0.5cm} \llen {\kappa_x} = \llen {\pi_x} - 1 \wedge \llen {\kappa_i} = \llen {\pi_i} - 1 = 2t - 1 \wedge C(i, \kappa_x, \kappa_i)~\wedge\\
\hspace*{0.5cm}l_0 = x \wedge l_1 = i \wedge l_2 = \pi_x\lidx i \wedge l_3 = l \wedge l_4 = 0}}$ \\
\Wgetlocal 4 \Wgetlocal 2 \Wcall{GetNodeLeaf} \Wcall{SetNodeLeaf} \\
$\specline{\stackheap{}{\ittcell{0}{t} \slstar \node x {0} {\kappa_x} {\pi_x} \slstar \node {l_2} {\lambda_y} {\kappa_y} {\pi_y} \slstar \node {l_3} {\lambda_y} \lemp \lemp \slstar \Wsize{l+1} \wedge 0 \leq i < \llen {\pi_x}~\wedge\\ 
\hspace*{0.5cm} \llen {\kappa_x} = \llen {\pi_x} - 1 \wedge \llen {\kappa_i} = \llen {\pi_i} - 1 = 2t - 1 \wedge C(i, \kappa_x, \kappa_i)~\wedge\\
\hspace*{0.5cm}l_0 = x \wedge l_1 = i \wedge l_2 = \pi_x\lidx i \wedge l_3 = l \wedge l_4 = 0}}$ \\
%
%
\Wiconst 0 \Wiload \Wsetlocal 4 \\
$\specline{\stackheap{}{\ittcell{0}{t} \slstar \node x 0 {\kappa_x} {\pi_x} \slstar \node {l_2} {\lambda_y} {\kappa_y} {\pi_y} \slstar \node {l_3} {\lambda_y} \lemp \lemp \slstar \Wsize{l+1} \wedge 0 \leq i < \llen {\pi_x}~\wedge\\ 
\hspace*{0.5cm} \llen {\kappa_x} = \llen {\pi_x} - 1 \wedge \llen {\kappa_i} = \llen {\pi_i} - 1 = 2t - 1 \wedge C(i, \kappa_x, \kappa_i)~\wedge\\
\hspace*{0.5cm}l_0 = x \wedge l_1 = i \wedge l_2 = \pi_x\lidx i \wedge l_3 = l \wedge l_4 = t}}$ \\
\begin{Wloope}{}
	\Wgetlocal 3 \Wgetlocal 2 \Wgetlocal 4 \Wcall{GetNodeKey} \Wcall{InsertNodeKey} \\
	\Wgetlocal 4 \Wiconst 0 \Wiload \Wiconst 2 \Wimul \Wiconst 2 \Wisub \Wilt \\
	\begin{Wife}{}
		\Wgetlocal 5 \Wiconst 1 \Wiadd \Wbr 1 
	\end{Wife}
\end{Wloope} \\
$\specline{\stackheap{}{\ittcell{0}{t} \slstar\node x 0 {\kappa_x} {\pi_x} \slstar \node {l_2} {\lambda_y} {\kappa_y} {\pi_y}~\slstar\\
\hspace*{0.5cm}\node {l_3} {\lambda_y} {\sublist {\kappa_y} t {t-1}} \lemp \slstar \Wsize{l+1} \wedge 0 \leq i < \llen {\pi_x}~\wedge \\ \hspace*{0.5cm} \llen {\kappa_x} = \llen {\pi_x} - 1 \wedge \llen {\kappa_i} = \llen {\pi_i} - 1 = 2t - 1 \wedge C(i, \kappa_x, \kappa_i)~\wedge\\
\hspace*{0.5cm}l_0 = x \wedge l_1 = i \wedge l_2 = \pi_x\lidx i \wedge l_3 = l \wedge l_4 = 2t-2}}$ \\
\Wgetlocal 0 \Wgetlocal 2 \Wiconst 0 \Wiload \Wiconst 1 \Wisub \\
\Wcall{GetNodeKey} \Wcall{InsertNodeKey} \\
$\specline{\stackheap{}{\ittcell{0}{t} \slstar \node x 0 {\sublist{\kappa_x} 0 i \lcat [ \kappa_y \lidx (t-1) ] \lcat \sublist{\kappa_x} i {\llen{\kappa_x} - i}} {\pi_x}~\slstar\\
\hspace*{0.5cm} \node {l_2} {\lambda_y} {\kappa_y} {\pi_y}~\slstar\\
\hspace*{0.5cm}\node {l_3} {\lambda_y} {\sublist {\kappa_y} t {t-1}} \lemp \slstar \Wsize{l+1} \wedge 0 \leq i < \llen {\pi_x}~\wedge \\ \hspace*{0.5cm} \llen {\kappa_x} = \llen {\pi_x} - 1 \wedge \llen {\kappa_i} = \llen {\pi_i} - 1 = 2t - 1 \wedge C(i, \kappa_x, \kappa_i)~\wedge\\
\hspace*{0.5cm}l_0 = x \wedge l_1 = i \wedge l_2 = \pi_x\lidx i \wedge l_3 = l \wedge l_4 = 2t-2}}$ \\
\Wgetlocal 0 \Wiconst {64k} \Wimul \Wiconst {32k} \Wiadd \\
\Wgetlocal 1 \Wiconst 1 \Wiadd \Wcall{AsegShr} \\
\Wgetlocal 0 \Wiconst {64k} \Wimul \Wiconst {32k} \Wiadd \\
\Wgetlocal 1 \Wiconst 1 \Wiadd \Wiconst 4 \Wimul \\
\Wgetlocal 3 \Wistore \\
$\specline{\stackheap{} {\ittcell{0}{t} \slstar \node x 0 {\sublist{\kappa_x} 0 i \lcat [ \kappa_y \lidx (t-1) ] \lcat \sublist{\kappa_x} i {\llen{\kappa_x} - i}} {\\\hspace*{3.35cm} \sublist{\pi_x} 0 {i + 1} \lcat [ l ] \lcat \sublist{\pi_x} {i + 1} {\llen {\pi_x} - i - 1}}  \slstar \\
\hspace*{0.5cm} \node {l_2} {\lambda_y} {\kappa_y} {\pi_y}~\slstar\\
\hspace*{0.5cm}\node {l_3} {\lambda_y} {\sublist {\kappa_y} t {t-1}} \lemp \slstar \Wsize{l+1} \wedge 0 \leq i < \llen {\pi_x}~\wedge \\ \hspace*{0.5cm} \llen {\kappa_x} = \llen {\pi_x} - 1 \wedge \llen {\kappa_i} = \llen {\pi_i} - 1 = 2t - 1 \wedge C(i, \kappa_x, \kappa_i)~\wedge\\
\hspace*{0.5cm}l_0 = x \wedge l_1 = i \wedge l_2 = \pi_x\lidx i \wedge l_3 = l \wedge l_4 = 2t-2}}$ \\
\Wgetlocal 2 \Wcall{GetNodeLeaf} \Wiconst 0 \Wieq \\
\begin{Wife}{}
$\specline{\stackheap{} {\ittcell{0}{t} \slstar \node x 0 {\sublist{\kappa_x} 0 i \lcat [ \kappa_y \lidx (t-1) ] \lcat \sublist{\kappa_x} i {\llen{\kappa_x} - i}} {\\\hspace*{3.35cm} \sublist{\pi_x} 0 {i + 1} \lcat [ l ] \lcat \sublist{\pi_x} {i + 1} {\llen {\pi_x} - i - 1}}  \slstar \\
\hspace*{0.5cm} \node {l_2} {\lambda_y} {\kappa_y} {\pi_y}~\slstar\\
\hspace*{0.5cm}\node {l_3} {\lambda_y} {\sublist {\kappa_y} t {t-1}} \lemp \slstar \Wsize{l+1} \wedge 0 \leq i < \llen {\pi_x}~\wedge \\ \hspace*{0.5cm} \llen {\kappa_x} = \llen {\pi_x} - 1 \wedge \llen {\kappa_i} = \llen {\pi_i} - 1 = 2t - 1 \wedge C(i, \kappa_x, \kappa_i)~\wedge\\
\hspace*{0.5cm}l_0 = x \wedge l_1 = i \wedge l_2 = \pi_x\lidx i \wedge l_3 = l \wedge l_4 = 2t-2 \wedge \lambda_y = 0}}$ \\
\Wiconst 0 \Wiconst 0 \Wsetlocal 1 \Wiload \\
$\specline{\stackheap{} {\ittcell{0}{t} \slstar \node x 0 {\sublist{\kappa_x} 0 i \lcat [ \kappa_y \lidx (t-1) ] \lcat \sublist{\kappa_x} i {\llen{\kappa_x} - i}} {\\\hspace*{3.35cm} \sublist{\pi_x} 0 {i + 1} \lcat [ l ] \lcat \sublist{\pi_x} {i + 1} {\llen {\pi_x} - i - 1}}  \slstar \\
\hspace*{0.5cm} \node {l_2} {\lambda_y} {\kappa_y} {\pi_y}~\slstar\\
\hspace*{0.5cm}\node {l_3} {\lambda_y} {\sublist {\kappa_y} t {t-1}} \lemp \slstar \Wsize{l+1} \wedge 0 \leq i < \llen {\pi_x}~\wedge \\ \hspace*{0.5cm} \llen {\kappa_x} = \llen {\pi_x} - 1 \wedge \llen {\kappa_i} = \llen {\pi_i} - 1 = 2t - 1 \wedge C(i, \kappa_x, \kappa_i)~\wedge\\
\hspace*{0.5cm}l_0 = x \wedge l_1 = 0 \wedge l_2 = \pi_x\lidx i \wedge l_3 = l \wedge l_4 = t \lambda_y = 0}}$ \\
\begin{Wloope}{}
	\Wgetlocal 3 \Wgetlocal 1 \Wgetlocal 2 \Wgetlocal 4 \\
	\Wcall{GetNodePtr} \Wcall{SetNodePtr} \\
	\Wgetlocal 4 \Wiconst 1 \Wiadd \Wsetlocal 4 \\
	\Wgetlocal 1 \Wiconst 1 \Wiadd \Wteelocal 1 \\
	\Wiconst 0 \Wiload \Wine \Wbrif 0
\end{Wloope} \\
$\specline{\stackheap{} {\ittcell{0}{t} \slstar \node x 0 {\sublist{\kappa_x} 0 i \lcat [ \kappa_y \lidx (t-1) ] \lcat \sublist{\kappa_x} i {\llen{\kappa_x} - i}} {\\\hspace*{3.35cm} \sublist{\pi_x} 0 {i + 1} \lcat [ l ] \lcat \sublist{\pi_x} {i + 1} {\llen {\pi_x} - i - 1}}  \slstar \\
\hspace*{0.5cm} \node {l_2} {\lambda_y} {\kappa_y} {\pi_y}~\slstar\\
\hspace*{0.5cm}\node {l_3} {\lambda_y} {\sublist {\kappa_y} t {t-1}} {\sublist{\pi_i} t t} \slstar \Wsize{l+1} \wedge 0 \leq i < \llen {\pi_x}~\wedge \\ \hspace*{0.5cm} \llen {\kappa_x} = \llen {\pi_x} - 1 \wedge \llen {\kappa_i} = \llen {\pi_i} - 1 = 2t - 1 \wedge C(i, \kappa_x, \kappa_i)~\wedge\\
\hspace*{0.5cm}l_0 = x \wedge l_1 = t \wedge l_2 = \pi_x\lidx i \wedge l_3 = l \wedge l_4 = 2t \wedge \lambda_y = 0}}$ 
\end{Wife} \\
$\specline{\stackheap{} {\ittcell{0}{t} \slstar \node x 0 {\sublist{\kappa_x} 0 i \lcat [ \kappa_y \lidx (t-1) ] \lcat \sublist{\kappa_x} i {\llen{\kappa_x} - i}} {\\\hspace*{3.35cm} \sublist{\pi_x} 0 {i + 1} \lcat [ l ] \lcat \sublist{\pi_x} {i + 1} {\llen {\pi_x} - i - 1}}  \slstar \\
\hspace*{0.5cm} \node {l_2} {\lambda_y} {\kappa_y} {\pi_y}~\slstar\\
\hspace*{0.5cm}\node {l} {\lambda_y} {\sublist {\kappa_y} t {t-1}} {\sublist{\pi_i} t t} \slstar \Wsize{l+1} \wedge 0 \leq i < \llen {\pi_x}~\wedge \\ \hspace*{0.5cm} \llen {\kappa_x} = \llen {\pi_x} - 1 \wedge \llen {\kappa_i} = \llen {\pi_i} - 1 = 2t - 1 \wedge C(i, \kappa_x, \kappa_i)~\wedge\\
\hspace*{0.5cm}l_0 = x \wedge l_2 = \pi_x\lidx i}}$ \\
\Wgetlocal 2 \Wiconst {64k} \Wimul \Wiconst 4 \Wiadd \\
\Wiconst 0 \Wiload \Wiconst 1 \Wisub \Wistore \\
\Wgetlocal 2 \Wiconst {64k} \Wimul \Wiconst {32k} \Wiadd \\
\Wiconst 0 \Wiload \Wistore \\
$\specline{\stackheap{} {\ittcell{0}{t} \slstar \node x 0 {\sublist{\kappa_x} 0 i \lcat [ \kappa_y \lidx (t-1) ] \lcat \sublist{\kappa_x} i {\llen{\kappa_x} - i}} {\\\hspace*{3.35cm} \sublist{\pi_x} 0 {i + 1} \lcat [ l ] \lcat \sublist{\pi_x} {i + 1} {\llen {\pi_x} - i - 1}}  \slstar \\ 
\hspace*{0.5cm} \node {\pi_x \lidx i} {\lambda_y} {\sublist {\kappa_y} 0 {t-1}} {\sublist{\pi_i} 0 t} \slstar \\ 
\hspace*{0.5cm} \node {l} {\lambda_y} {\sublist {\kappa_y} t {t-1}} {\sublist{\pi_i} t t} \wedge \Wsize{l + 1} \wedge 0 \leq i < \llen {\pi_x}~\wedge\\
\hspace*{0.5cm}  \llen {\kappa_x} = \llen {\pi_x} - 1 \wedge C(i, \kappa_x, \kappa_y)}}$ 
\end{Wfunce} \\
$\specline{\stackheap{} {\ittcell{0}{t} \slstar \node x 0 {\sublist{\kappa_x} 0 i \lcat [ \kappa_y \lidx (t-1) ] \lcat \sublist{\kappa_x} i {\llen{\kappa_x} - i}} {\\\hspace*{3.35cm} \sublist{\pi_x} 0 {i + 1} \lcat [ l ] \lcat \sublist{\pi_x} {i + 1} {\llen {\pi_x} - i - 1}}  \slstar \\ 
\hspace*{0.5cm} \node {\pi_x \lidx i} {\lambda_y} {\sublist {\kappa_y} 0 {t-1}} {\sublist{\pi_i} 0 t} \slstar \\ 
\hspace*{0.5cm} \node {l} {\lambda_y} {\sublist {\kappa_y} t {t-1}} {\sublist{\pi_i} t t} \wedge \Wsize{l + 1} \wedge 0 \leq i < \llen {\pi_x}~\wedge\\
\hspace*{0.5cm}  \llen {\kappa_x} = \llen {\pi_x} - 1 \wedge C(i, \kappa_x, \kappa_y)}}$} \\

\noindent where
$C(i, \kappa_x, \kappa_y) \equiv \frall{k'} k' \in \listToSet{\kappa_y} \rightarrow C'(i, \kappa_x, k')$
and
$C'(i, \kappa_x, k) \equiv 
	 (i = 0 \rightarrow \llen {\kappa_x} > 0 \rightarrow k < \kappa_x \lidx i) \wedge
	 (0 < i < \llen {\kappa_x} \rightarrow \kappa_x \lidx (i - 1) < k < \kappa_x \lidx i) \wedge
	 (0 < i = \llen {\kappa_x} \rightarrow \kappa_x \lidx i < k)$.

\subparagraph{B-Tree Insertion into Non-Full Nodes} 
The auxiliary function $\pred{BTreeInsertNonFull}$ recursively traverses a subtree with a non-full root to insert a key $k$, which is not already in the subtree, into it. This function may  extend the allocated memory by adding new nodes, as described in its post-condition. \\

{\footnotesize\noindent
$\specline{\stackheap{x, k} {\ittcell{0}{t} \slstar \btreerec x {\kappa_x} {\lambda_x} 0 \slstar \Wsize l \wedge k \notin \kappa_x}}$ \\
\begin{Wfunce}[\\(\KK{locals} \ldots)]{BTreeInsertNonFull}{[\Witype, \Witype] \rightarrow []}
	$\specline{\stackheap{} {\ittcell{0}{t} \slstar \btreerec x {\kappa_x} {\lambda_x} 0 \slstar \Wsize l \wedge k \notin \kappa_x \wedge l_0 = x \wedge l_1 = k}}$ \\
	$\specline{\stackheap{} {\ittcell{0}{t} \slstar \node x {\lambda_x} {\kappa_x} {\pi_x} \slstar \ldots \slstar \Wsize l \wedge k \notin \kappa_x \wedge l_0 = x \wedge l_1 = k}}$ \\
	\Wgetlocal 0 \Wcall{GetNodeLeaf} \Wiconst 0 \Wine \\
	$\specline{\stackheap{\lambda_x} {\ittcell{0}{t} \slstar \node x {\lambda_x} {\kappa_x} {\pi_x} \slstar \ldots \slstar \Wsize l \wedge k \notin \kappa_x \wedge l_0 = x \wedge l_1 = k}}$ \\
	\begin{Wife}{}
		$\specline{\stackheap{} {\node x {\lambda_x} {\kappa_x} {\pi_x} \slstar \ldots \slstar \Wsize l \wedge k \notin \kappa_x \wedge l_0 = x \wedge l_1 = k \wedge \lambda_x \neq 0}}$ \\
		\Wgetlocal 0 \Wgetlocal 0 \Wgetlocal 1 \\
		\Wcall{OBAFind} \Wteelocal 2 \\
		\Wcall{GetNodePtr} \Wsetlocal 3 \\
		$\specline{\stackheap{} { \ittcell{0}{t} \slstar  \node x {\lambda_x} {\kappa_x} {\pi_x} \slstar \ldots \slstar \Wsize l \wedge k \notin \kappa_x \wedge l_0 = x \wedge l_1 = k \wedge l_3 = \pi_x \lidx l_2 \wedge \lambda_x \neq 0~\slstar\\
		\hspace*{0.5cm} \btreerec {\pi_x \lidx l_2}{\overline \kappa \lidx l_2}{\lambda'}{\overline \phi \lidx l_2} \wedge C(l_2, \kappa_x, \setToList{\overline\kappa \lidx l_2}) \\}}$ \\
		\Wgetlocal 3 \Wiload \\
		\Wiconst 0 \Wiload \Wiconst 2 \Wimul \Wiconst 1 \Wisub \Wieq \\
		$\specline{\stackheap{\overline\phi \lidx l_2} { \ittcell{0}{t} \slstar  \node x {\lambda_x} {\kappa_x} {\pi_x} \slstar \ldots \slstar \Wsize l \wedge k \notin \kappa_x \wedge l_0 = x \wedge l_1 = k \wedge l_3 = \pi_x \lidx l_2 \wedge \lambda_x \neq 0~\slstar\\
		\hspace*{1cm} \btreerec {\pi_x \lidx l_2}{\overline \kappa \lidx l_2}{\lambda'}{\overline \phi \lidx l_2} \wedge C(l_2, \kappa_x, \setToList{\overline\kappa \lidx l_2}) \\}}$ \\
		\begin{Wife} {}
		$\specline{\stackheap{} { \ittcell{0}{t} \slstar  \node x {\lambda_x} {\kappa_x} {\pi_x} \slstar \ldots \slstar \Wsize l \wedge k \notin \kappa_x \wedge l_0 = x \wedge l_1 = k \wedge l_3 = \pi_x \lidx l_2 \wedge \lambda_x \neq 0~\slstar\\
		\hspace*{0.5cm} \btreerec {\pi_x \lidx l_2}{\overline \kappa \lidx l_2}{\lambda'}{\overline \phi \lidx l_2} \wedge C(l_2, \kappa_x, \setToList{\overline\kappa \lidx l_2}) \wedge \overline\phi \lidx l_2 \neq 0 \\}}$ \\
		\Wgetlocal 0 \Wgetlocal 2 \\
		$\specline{\stackheap{x, l_2} { \ittcell{0}{t} \slstar  \node x {\lambda_x} {\kappa_x} {\pi_x} \slstar \ldots \slstar \Wsize l \wedge k \notin \kappa_x \wedge l_0 = x \wedge l_1 = k \wedge l_3 = \pi_x \lidx l_2 \wedge \lambda_x \neq 0~\slstar\\
		\hspace*{1.05cm} \btreerec {\pi_x \lidx l_2}{\overline \kappa \lidx l_2}{\lambda'}{\overline \phi \lidx l_2} \wedge C(l_2, \kappa_x, \setToList{\overline\kappa \lidx l_2}) \wedge \overline\phi \lidx l_2 \neq 0 \\}}$ \\
		$\begin{leftvruled}{frame, cons}
		\specline{\stackheap{x, l_2} {\ittcell{0}{t} \slstar \node x 0 {\kappa_x} {\pi_x} \slstar \node {\pi_x \lidx l_2} {\lambda'} {\kappa'}  {\pi'} \slstar \Wsize l \wedge 0 \leq i < \llen {\pi_x}~\wedge\\ 
\hspace*{0.95cm} \llen {\kappa_x} = \llen {\pi_x} - 1 \wedge \llen {\kappa'} = \llen {\pi'} - 1 = 2t - 1 \wedge C({l_2}, \kappa_x, \kappa')~\wedge\\
\hspace*{0.95cm} C'(l_2, \kappa_x, k) \wedge
	(\frall {k'} k' \in \listToSet{\kappa'} \rightarrow C'(l_2, \kappa_x, k')) }}\\
		\Wcall{~\text{BTreeSplitChild}} \\
        \specline{\stackheap{} {\ittcell{0}{t} \slstar \node x 0 {\sublist{\kappa_x} 0 {l_2} \lcat [ \kappa' \lidx (t-1) ] \lcat \sublist{\kappa_x} {l_2} {\llen{\kappa_x} - {l_2}}} {\\\hspace*{3.35cm} \sublist{\pi_x} 0 {{l_2} + 1} \lcat [ l ] \lcat \sublist{\pi_x} {{l_2} + 1} {\llen {\pi_x} - {l_2} - 1}}  \slstar \\ 
\hspace*{0.5cm} \node {\pi_x \lidx {l_2}} {\lambda'} {\sublist {\kappa'} 0 {t-1}}  {\sublist{\pi'} 0 t} \slstar \\ 
\hspace*{0.5cm} \node {l} {\sublist {\kappa'} t {t-1}} {\lambda'} {\sublist{\pi'} t t} \slstar \Wsize{l + 1} \wedge 0 \leq i < \llen {\pi_x}~\wedge\\
\hspace*{0.5cm}  \llen {\kappa_x} = \llen {\pi_x} - 1 \wedge C(l_2, \kappa_x, \kappa') \wedge (\frall {k'} k' \in \listToSet{\kappa'} \rightarrow C'(l_2, \kappa_x, k')) }}\\
	\end{leftvruled}$\\
        $\specline{\stackheap{} {\ittcell{0}{t} \slstar \node x 0 {\sublist{\kappa_x} 0 {l_2} \lcat [ \kappa' \lidx (t-1) ] \lcat \sublist{\kappa_x} {l_2} {\llen{\kappa_x} - {l_2}}} {\\\hspace*{3.35cm} \sublist{\pi_x} 0 {{l_2} + 1} \lcat [ l ] \lcat \sublist{\pi_x} {{l_2} + 1} {\llen {\pi_x} - {l_2} - 1}}  \slstar \\ 
\hspace*{0.5cm} \pred{BTreeRec}^{r, t, l+1}({\pi_x \lidx l_2},{\kappa''},{\lambda'},{0}) \slstar 
                \pred{BTreeRec}^{r, t, l+1}({l},{\kappa'''},{\lambda'},{0}) \slstar \Wsize{l + 1}~\slstar \ldots~\wedge  \\
  \hspace*{0.5cm}  \overline\kappa \lidx l_2 = \kappa'' \cup \{ \kappa' \lidx l_2 \} \cup \kappa'''  \wedge 0 \leq i < \llen {\pi_x} ~\wedge (C(l_2, \kappa_x, \setToList{\kappa''}) \vee C(l_2, \kappa_x, \setToList{\kappa'''})~\wedge\\
  \hspace*{0.5cm} k \notin \kappa_x \wedge l_0 = x \wedge l_1 = k \wedge l_3 = \pi_x \lidx l_2 \wedge \lambda_x \neq 0 }}$ \\
		\Wgetlocal 0 \Wgetlocal 2 \\
		\Wcall{GetNodeKey} 
		\Wgetlocal 2 \\
		\Wigt \\
        $\specline{\exsts v \stackheap{v} {\ittcell{0}{t} \slstar \node x 0 {\sublist{\kappa_x} 0 {l_2} \lcat [ \kappa' \lidx (t-1) ] \lcat \sublist{\kappa_x} {l_2} {\llen{\kappa_x} - {l_2}}} {\\\hspace*{3.95cm} \sublist{\pi_x} 0 {{l_2} + 1} \lcat [ l ] \lcat \sublist{\pi_x} {{l_2} + 1} {\llen {\pi_x} - {l_2} - 1}}  \slstar \\ 
\hspace*{1.1cm} \pred{BTreeRec}^{r, t, l+1}({\pi_x \lidx l_2},{\kappa''},{\lambda'},{0}) \slstar 
                \pred{BTreeRec}^{r, t, l+1}({l},{\kappa'''},{\lambda'},{0}) \slstar \Wsize{l + 1}~\slstar \ldots~\wedge  \\
  \hspace*{1.1cm}  \overline\kappa \lidx l_2 = \kappa'' \cup \{ \kappa' \lidx l_2 \} \cup \kappa'''  \wedge 0 \leq i < \llen {\pi_x} ~\wedge (C(l_2, \kappa_x, \setToList{\kappa''}) \vee C(l_2, \kappa_x, \setToList{\kappa'''})~\wedge\\
  \hspace*{1.1cm} (v = 0 \rightarrow k \leq \kappa' \lidx l_2) \wedge (v \neq 0 \rightarrow k > \kappa' \lidx l_2) \wedge k \notin \kappa_x \wedge l_0 = x \wedge l_1 = k \wedge l_3 = \pi_x \lidx l_2 \wedge \lambda_x \neq 0 }}$ \\
		\begin{Wife}{}
			\Wgetlocal 2 \Wiconst 1 \Wiadd \Wsetlocal 2
		\end{Wife} \\
		$\specline{\stackheap{} { \ittcell{0}{t} \slstar  \node x {0} {\kappa'_x} {\pi'_x} \slstar \ldots \slstar \Wsize {l + 1} \wedge k \notin \kappa'_x \wedge l_0 = x \wedge l_1 = k \wedge l_3 = \pi_x \lidx l_2~\slstar\\
		\hspace*{0.5cm} \pred{BTreeRec}^{r, t, l+1}({\pi'_x \lidx l_2},{\kappa''''},{\lambda'},0) \wedge C(l_2, \kappa_x, \kappa'''') \\}}$ \\
		\Wgetlocal 0 \Wgetlocal 2 \\
		\Wcall{GetNodePtr} \\
		\Wgetlocal 1 \\
		$\specline{\stackheap{\pi'_x \lidx l_2, k} {  \ittcell{0}{t} \slstar  \node x {0} {\kappa'_x} {\pi'_x} \slstar \ldots \slstar \Wsize {l + 1} \wedge k \notin \kappa'_x \wedge l_0 = x \wedge l_1 = k \wedge l_3 = \pi_x \lidx l_2~\slstar\\
		\hspace*{1.5cm} \pred{BTreeRec}^{r, t, l+1}({\pi'_x \lidx l_2},{\kappa''''},{\lambda'},0) \wedge C(l_2, \kappa_x, \kappa'''') }}$
$\begin{leftvruled}{frame, cons}
		\specline{\stackheap{\pi'_x \lidx l_2, k} { \ittcell{0}{t} \slstar  \pred{BTreeRec}^{r, t, l+1}({\pi'_x \lidx l_2},{\kappa''''},{\lambda'},0) \slstar \Wsize {l+1} \wedge k \notin \kappa''''}} \\
		\Wcall{~\text{BTreeInsertNonFull}} \\
		\specline{\exsts {l', \phi'} \stackheap{} { \ittcell{0}{t} \slstar \pred{BTreeRec}^{t, r, l'}(\pi_x \lidx l_2, \kappa'''' \cup \{ k \}}, \lambda_x, \phi') \slstar \Wsize {l'} \wedge k \notin \kappa'''' \wedge l' \geq l}
		\end{leftvruled}$ \\
		$\specline{\exsts {l', \phi'} \stackheap{} { \ittcell{0}{t} \slstar  \node x 0 {\kappa'_x} {\pi'_x} \slstar \ldots \slstar \Wsize {l'} \wedge k \notin \kappa_x \wedge l_0 = x \wedge l_1 = k \wedge l_3 = \pi_x \lidx l_2 \wedge \lambda_x \neq 0~\slstar\\
		\hspace*{1.35cm} \pred{BTreeRec}^{t, r, l'}(\pi'_x \lidx l_2, \kappa'''' \cup \{ k \}}, \lambda', \phi') \wedge C(l_2, \kappa_x, \kappa'''')  \\}$ \\
		$\specline{\exsts {l'} \stackheap{} { \ittcell{0}{t} \slstar \pred{BTreeRec}^{t, r, l'}(x, \kappa_x \cup \{ k \}}, \lambda', 0) \slstar \Wsize {l'} \wedge k \notin \kappa_x \wedge l' \geq l}$ \\
		$\specline{\exsts {l', \phi_x} \stackheap{} { \ittcell{0}{t} \slstar \pred{BTreeRec}^{t, r, l'}(x, \kappa_x \cup \{ k \}}, \lambda', \phi_x) \slstar \Wsize {l'} \wedge k \notin \kappa_x \wedge l' \geq l}$ \\
		\Welse \\
		$\specline{\stackheap{} { \ittcell{0}{t} \slstar  \node x {\lambda_x} {\kappa_x} {\pi_x} \slstar \ldots \slstar \Wsize l \wedge k \notin \kappa_x \wedge l_0 = x \wedge l_1 = k \wedge l_3 = \pi_x \lidx l_2 \wedge \lambda_x \neq 0~\slstar\\
		\hspace*{0.5cm} \btreerec {\pi_x \lidx l_2}{\overline \kappa \lidx l_2}{\lambda'}{\overline \phi \lidx l_2} \wedge C(l_2, \kappa_x, \overline\kappa \lidx l_2) \wedge \overline\phi \lidx l_2 = 0 \\}}$ \\
			\Wgetlocal 3 \Wgetlocal 1 \\
		$\specline{\stackheap{\pi_x \lidx l_2, k} { \ittcell{0}{t} \slstar  \node x {\lambda_x} {\kappa_x} {\pi_x} \slstar \ldots \slstar \Wsize l \wedge k \notin \kappa_x \wedge l_0 = x \wedge l_1 = k \wedge l_3 = \pi_x \lidx l_2 \wedge \lambda_x \neq 0~\slstar\\
		\hspace*{1.45cm} \btreerec {\pi_x \lidx l_2}{\overline \kappa \lidx l_2}{\lambda'}{\overline \phi \lidx l_2} \wedge C(l_2, \kappa_x, \overline\kappa \lidx l_2) \wedge \overline\phi \lidx l_2 = 0 \\}}$ \\
		$\begin{leftvruled}{frame, cons}
		\specline{\stackheap{\pi_x \lidx l_2, k} { \ittcell{0}{t} \slstar  \btreerec {\pi_x \lidx l_2}{\overline \kappa \lidx l_2}{\lambda'}{0} \slstar \Wsize l \wedge k \notin \overline\kappa \lidx l_2}} \\
		\Wcall{~\text{BTreeInsertNonFull}} \\
		\specline{\exsts {l', \phi'} \stackheap{} { \ittcell{0}{t} \slstar \pred{BTreeRec}^{t, r, l'}(\pi_x \lidx l_2, \overline\kappa \lidx l_2 \cup \{ k \}}, \lambda_x, \phi') \slstar \Wsize {l'} \wedge k \notin \overline\kappa \lidx l_2 \wedge l' \geq l}
		\end{leftvruled}$
		\end{Wife} \\
		$\specline{\exsts {l', \phi'} \stackheap{} { \ittcell{0}{t} \slstar  \node x {\lambda_x} {\kappa_x} {\pi_x} \slstar \ldots \slstar \Wsize {l'} \wedge k \notin \kappa_x \wedge l_0 = x \wedge l_1 = k \wedge l_3 = \pi_x \lidx l_2 \wedge \lambda_x \neq 0~\slstar\\
		\hspace*{1.35cm} \pred{BTreeRec}^{t, r, l'}(\pi_x \lidx l_2, \overline\kappa \lidx l_2 \cup \{ k \}}, \lambda_x, \phi') \wedge C(l_2, \kappa_x, \overline\kappa \lidx l_2) \wedge \overline\phi \lidx l_2 = 0 \\}$ \\
		$\specline{\exsts {l'} \stackheap{} { \ittcell{0}{t} \slstar \pred{BTreeRec}^{t, r, l'}(x, \kappa_x \cup \{ k \}}, \lambda', 0) \slstar \Wsize {l'} \wedge k \notin \kappa_x \wedge l' \geq l}$ \\
		$\specline{\exsts {l', \phi_x} \stackheap{} { \ittcell{0}{t} \slstar \pred{BTreeRec}^{t, r, l'}(x, \kappa_x \cup \{ k \}}, \lambda', \phi_x) \slstar \Wsize {l'} \wedge k \notin \kappa_x \wedge l' \geq l}$ \\
		\Welse \\
		$\specline{\stackheap{} {\ittcell{0}{t} \slstar \node x 0 {\kappa} {\lemp} \slstar \ldots \slstar \Wsize l \wedge k \notin \kappa \wedge l_0 = x \wedge l_1 = k}}$ \\
		\Wgetlocal 0 \Wgetlocal 1 \\
		$\specline{\stackheap{x, k} {\ittcell{0}{t} \slstar \node x 0 {\kappa} {\lemp} \slstar \ldots \slstar \Wsize l \wedge k \notin \kappa \wedge l_0 = x \wedge l_1 = k}}$ \\
		\Wcall{InsertNodeKey} \\
		$\specline{\stackheap{} {\ittcell{0}{t} \slstar \node x 0 {\kappa \cup \{ k \}} {\lemp} \slstar \ldots \slstar \Wsize l \wedge k \notin \kappa \wedge l_0 = x \wedge l_1 = k}}$ \\	
		$\specline{\exsts {\phi_x} \stackheap{} { \ittcell{0}{t} \slstar \pred{BTreeRec}^{t, r, l}(x, \kappa_x \cup \{ k \}}, \lambda_x, \phi_x) \slstar \ldots \slstar \Wsize l \wedge k \notin \kappa \wedge l_0 = x \wedge l_1 = k}$ \\	
		$\specline{\exsts {l', \phi_x} \stackheap{} { \ittcell{0}{t} \slstar \pred{BTreeRec}^{t, r, l'}(x, \kappa_x \cup \{ k \}}, \lambda_x, \phi_x) \slstar \Wsize {l'} \wedge k \notin \kappa_x \wedge l' \geq l}$  
	\end{Wife}
\end{Wfunce} \\
$\specline{\exsts {l', \phi_x} \stackheap{} { \ittcell{0}{t} \slstar \pred{BTreeRec}^{t, r, l'}(x, \kappa_x \cup \{ k \}}, \lambda_x, \phi_x) \slstar \Wsize {l'} \wedge k \notin \kappa_x \wedge l' \geq l}$} \\

\subparagraph{B-Tree Insertion} 
The function $\pred{BTreeInsert}$ inserts a key $k$ into the B-tree. If the key already exists, the B-tree is not modified. \\

{\footnotesize\noindent
$\specline{\stackheap{k} {\btree x {\kappa}}}$ \\
\begin{Wfunce}[\\(\KK{locals} \Witype, \Witype)]{BTreeInsert}{[\Witype] \rightarrow []}
	$\specline{\stackheap{} {\btree x {\kappa} \wedge l_0 = k \wedge l_1 = 0 \wedge l_2 = 0}}$ \\
	\Wgetlocal 0 \Wcall{BTreeSearch} \Wiconst 0 \Wieq \\
	$\specline{\exsts v \stackheap{v}{\btree x {\kappa} \wedge l_0 = k \wedge l_1 = 0 \wedge l_2 = 0 \wedge (v = 0 \rightarrow k \in \kappa) \wedge (v \neq 0 \rightarrow k \notin \kappa)}}$ \\
	\begin{Wife}{}
	$\specline{\stackheap{} {\btree x {\kappa}} \wedge k \notin \kappa \wedge l_0 = k \wedge l_1 = 0 \wedge l_2 = 0}$ \\
	$\specline{\stackheap{}{\meta t r l \alpha \slstar \btreerec r \kappa \lambda \phi \wedge l_0 = k \wedge l_1 = 0 \wedge l_2 = 0 \wedge k \notin \kappa}}$ \\
	\Wiconst 4 \Wiload \Wsetlocal 1 \\
	\Wiconst 0 \Wiload \Wiconst 2 \Wimul \Wiconst 1 \Wisub \Wsetlocal 2 \\
	$\specline{\stackheap{}{\meta t r l \alpha \slstar \btreerec r \kappa \lambda \phi \wedge l_0 = k \wedge l_1 = r \wedge l_2 = 2t - 1 \wedge k \notin \kappa}}$ \\
	\Wgetlocal 1 \Wiload \Wgetlocal 2 \Wieq \\
	\begin{Wife}{}
		$\specline{\stackheap{}{\meta t r l \alpha \slstar \btreerec r \kappa \lambda \phi \wedge l_0 = k \wedge l_1 = r \wedge l_2 = 2t - 1 \wedge k \notin \kappa \wedge \phi \neq 0}}$ \\
		\Wcall{AllocateNode} \Wteelocal 2 \\
		$\specline{\stackheap{l}{\meta t r {l + 1} \alpha \slstar \pred{BTreeRec}^{t, r, l+1}(r, \kappa, \lambda, \phi) \slstar \node l 1 \lemp \lemp~\wedge \\
	 	\hspace*{0.65cm}l_0 = k \wedge l_1 = r \wedge l_2 = l \wedge k \notin \kappa \wedge \phi \neq 0}}$ \\
		\Wiconst 0 \Wcall{SetNodeLeaf} \\
		\Wgetlocal 2 \Wiconst 0 \Wgetlocal 1 \Wiconst 0 \\
		$\specline{\stackheap{l, 0}{\meta t r {l + 1} \alpha \slstar \pred{BTreeRec}^{t, r, l+1}(r, \kappa, \lambda, \phi) \slstar \node l 0 \lemp {[ r ]}~\wedge \\		
		\hspace*{0.9cm}l_0 = k \wedge l_1 = r \wedge l_2 = l \wedge k \notin \kappa \wedge \phi \neq 0}}$ \\
		$\begin{leftvruled}{frame, cons}
			\specline{\stackheap{l, 0} {0 \rightarrow r \slstar \node l 0 {\lemp} {[ r ]} \slstar \btreerec  {r} {\kappa} \lambda {\phi} \slstar \Wsize{l + 1}~\wedge\\ 
\hspace*{0.95cm}0 \leq 0 < \llen {[r]} \wedge \llen {\lemp} = \llen {[r]} - 1 \wedge C(0, \lemp, \kappa) \wedge \phi \neq 0}} \\
		\Wcall{~\text{BTreeSplitChild}} \\
		\specline{\exsts k' \stackheap{} {0 \rightarrow r \slstar \node x 0 {[ k' ]} {[r, l + 1]}  \slstar \pred{BTreeRec}^{t, r, l + 2}({r}, {\kappa \setminus \{ k' \}},  {\lambda}, 0) \slstar \Wsize{l + 2} \wedge k' \in \kappa}} \\
		\Wiconst 0 \Wgetlocal 2 \Wistore \\
		\specline{\exsts k' \stackheap{} {0 \rightarrow l \slstar \node x 0 {[ k' ]} {[r, l + 1]}  \slstar \pred{BTreeRec}^{t, r, l + 2}({r}, {\kappa \setminus \{ k' \}},  {\lambda}, 0) \slstar \Wsize{l + 2} \wedge k' \in \kappa}} \\
		\end{leftvruled}$ \\
		$\specline{\stackheap{}{\meta t l {l + 2} \alpha \slstar \pred{BTreeRec}^{t, l, l+2}(l, \kappa, 0, 0)  \wedge l_0 = k \wedge l_1 = r \wedge l_2 = l \wedge k \notin \kappa}}$ \\
		\Wgetlocal 2 \Wgetlocal 0 \\
		$\specline{\stackheap{l, k}{\meta t l {l + 2} \alpha \slstar \pred{BTreeRec}^{t, l, l+2}(l, \kappa, 0, 0) \wedge l_0 = k \wedge l_1 = r \wedge l_2 = l \wedge k \notin \kappa}}$ \\
		\Wcall{BTreeInsertNonFull} \\
		$\specline{\stackheap{}{\meta t l {l + 2} \alpha \slstar \pred{BTreeRec}^{t, l, l+2}(l, \kappa \cup \{ k \}, 0, \phi') \wedge l_0 = k \wedge l_1 = r \wedge l_2 = l \wedge k \notin \kappa}}$ \\
		$\specline{\stackheap{}{\btree x {\kappa \cup \{ k \}}}}$ \\		
		\Welse \\
		$\specline{\stackheap{}{\meta t r l \alpha \slstar \btreerec r \kappa \lambda \phi \wedge l_0 = k \wedge l_1 = r \wedge l_2 = 2t - 1 \wedge k \notin \kappa \wedge \phi = 0}}$ \\
		\Wgetlocal 1 \Wgetlocal 0 \\
		$\specline{\stackheap{r, k}{\meta t r l \alpha \slstar \btreerec r \kappa \lambda \phi \wedge l_0 = k \wedge l_1 = r \wedge l_2 = 2t - 1 \wedge k \notin \kappa \wedge \phi = 0}}$ \\
		\Wcall{BTreeInsertNonFull} \\
		$\specline{\stackheap{}{\meta t r {l'} \alpha \slstar \pred{BTreeRec}^{t, r, l'}(r, \kappa \cup \{ k \}, \lambda, \phi') \wedge l_0 = k \wedge l_1 = r \wedge l_2 = 2t - 1 \wedge k \notin \kappa}}$ \\
		$\specline{\stackheap{}{\btree x {\kappa \cup \{ k \}}}}$
	\end{Wife}
	\end{Wife} \\
	$\specline{\stackheap{}{\btree x {\kappa \cup \{ k \}}}}$
\end{Wfunce} \\
$\specline{\stackheap{}{\btree x {\kappa \cup \{ k \}}}}$ \\

\end{document}